\documentclass[12pt]{book}
\usepackage{epsfig}

\def\beq{\begin{equation}}
\def\eeq{\end{equation}}
\def\ba{\begin{eqnarray}}
\def\er{\end{array}}
\def\br{\begin{array}}
\def\ea{\end{eqnarray}}
\def\bfg{\begin{figure}}
\def\efg{\end{figure}}
\def\etal{{\it et al. }}
\def\cxhy{C$_x$H$_y$}
\def\ctht{C$_2$H$_2$}
\def\cht{CH$_3$}

\begin{document}

\pagestyle{empty}

\title{\vspace{-3em}\large DOCTORAL DISSERTATION\\
\vspace{4em} \LARGE \bf Computer simulations of structural and hopping
conduction properties of disordered solids
\vspace{0em}}
\author{\bf \Large Kriszti\'an Koh\'ary\\
\\
\\
\\
Thesis advisors:\\
\\
\fontshape{sc}\selectfont Dr. S\'andor Kugler\\
\it Department of Theoretical Physics,\\
\it Budapest University of Technology and Economics,\\
\it H-1521 Budapest, Hungary\\
\\
and
\\
\\
\fontshape{sc}\selectfont PD Dr. Sergei Baranovskii \\
\it Institut f\"ur Physikalische Chemie und Zentrum f\"ur\\
\it Materialwissenschaften der Philipps--Universit\"at Marburg,\\
\it Philipps--Universit\"at Marburg,\\
\it D-35032 Marburg, Germany\\
\date{2001}
}
\maketitle

\newpage

\null\vspace{18.0truecm}
\noindent%
Vom Fachbereich Physik der Philipps-Universit\"at\\
als Dissertation angenommen am 18.06.2001\\
Erstgutachter: Prof. Dr. Koichi Shimakawa\\
Zweitgutachter: Prof. Dr. Peter Thomas\\
Tag der m\"undlichen Pr\"ufung: 02.07.2001

\null
\pagebreak

\baselineskip=1.3em

\pagenumbering{roman}

\pagestyle{plain}
\tableofcontents

\mainmatter
\pagenumbering{arabic}
\pagestyle{headings}

\chapter{Introduction}


{\it Disordered materials} have been the subject of numerous
experimental and theoretical works for a significantly long period of
time. There are several forms of disorder and they can be classified
in the following way.  Topological disorder is a structural
arrangement where the long range order of a crystal is missing. This
can be considered as the definition of {\it amorphous materials},
which are non-crystalline materials.  Other types of
disorder also exist.  Substitutional disorder is where the units of the
elementary cell change randomly with other units in the lattice.
Vibrational disorder is where the positions of the atoms around their
equilibrium positions are random due to temperature.  Finally, some
materials have magnetic disorder. Here the underlying crystal can posses
randomly oriented spins at lattice points.  Materials which are
topologically disordered and posses a randomly oriented spin are
called spin glasses.  They must be not confused with "real" glasses.
The latter are prepared by quick quenching from the melt. The glass
transition appears when the supercooled liquid has a sudden change in
derivative thermodynamic properties. Henceforth, we will
define the glasses, which exhibit this glass transition. The glass
transition temperature is always lower than the melting temperature
for crystallines and it partly depends on thermal history.

It is highly important that one is aware of the detailed structure of
amorphous materials, as this determines the main physical properties.
The structure of amorphous covalently bonded materials is still
however not known in detail and lots of physical phenomena are yet to
be explained. The amorphous structure cannot be understood in terms of
a unit cell consisting of a few atoms.  The lack of periodicity in the
material does not allow for the use of elegant theorems as used for
crystalline structures. The complex nature of the systems also cannot
be described by the means of theories used for the periodic structures
of solids.  In amorphous materials the structural information can only
be determined in a statistical sense.  Diffraction experiments are
employed to retrieve information about atomic arrangements, which are
to yield the radial distribution functions containing integrated
information about the amorphous structures. Continuous random network
models were proposed to describe the structure of covalent amorphous
semiconductors on an atomic scale.  In the past two decades, however,
vast advances in information technology has allowed scientists to use
computers in a more effective way to study disordered materials on the
atomic scale. 

Transport and energy relaxation of charge carriers in disordered
solids has been the subject of intensive experimental and theoretical
research for the past few decades. Detailed knowledge of the real
underlying structure is not necessary in order to understand the
electronic properties. Instead, simple models (extended Hamiltonians)
were proposed, which are able to give reasonable descriptions for
these physical properties. However, the model parameters used in these
Hamiltonians could only be determined by some minimal information
about the structure.  It is possible to describe conductivity,
mobility, thermopower, and other electronical properties of disordered
materials with these models. 

This thesis focuses on two physical properties of disordered
materials: {\it (i)} on the {\it structure} of amorphous carbon and
amorphous silicon, and {\it (ii)} on the {\it mobility} of the quasi
one-dimensional disordered organic solids.  In chapter\ \ref{aC}, the
description of the molecular dynamics simulation growth of amorphous
carbon with low energy carbon bombardment is given.  The atomistic
simulation of the bombardment process during the bias enhanced
nucleation phase of diamond chemical vapor deposition is presented in
chapter\ \ref{CxHy}. Chapter\ \ref{Si} deals with the study of the
growth of amorphous silicon structures by the same preparation method
as it was applied for the amorphous carbon structures (chapter\
\ref{aC}). Finally, in chapter\ \ref{hopping}, the mobility of quasi
one-dimensional organic semiconductors will be examined via Monte
Carlo simulations. The thesis concludes with a summary in English,
German and Hungarian. 
\chapter{Molecular dynamics simulation of amorphous carbon structures}
\markboth{MD SIMULATION OF a$-$C STRUCTURES}{MD SIMULATION OF a$-$C STRUCTURES}
\label{aC}
 
Amorphous carbon (a$-$C) has been the subject of numerous experimental
and theoretical works in recent decades. The ability of carbon atoms
to form $sp^3$, $sp^2$ and $sp^1$  bonding configurations leads to a
large variation in structure.  Various methods of preparation produce
a wide range of macroscopic parameters for a$-$C.  Tetrahedrally bonded
amorphous carbon~(ta$-$C) is found to be optically transparent,
extremely hard and chemically inert whereas a$-$C which mostly consists
of threefold coordinated atoms has a small gap in the electronic
density of states.  In amorphous forms of carbon the most elementary
parameter is the ratio of fourfold, threefold, and twofold coordinated
atomic sites which determine electronic and mechanical properties of
the film.

In this chapter the growth of amorphous carbon thin film on a [111]
diamond surface is studied with the tight-binding molecular dynamics
technique. Six different three-dimensional networks were constructed
with periodic boundary conditions in two dimensions.  Time-dependent
non-equilibrium growth was studied and densities, radial distribution
functions, coordination numbers,  bond angle distributions, and ring
statistics were analyzed.

\section{Introduction}

\subsection{Experimental background}

At the early stage of structure investigations the amorphous carbon
was studied using diffraction techniques. The obtained results show a
wide range of structures.  One of the first experiments was carried
out by a Japanese group on an evaporated a$-$C sample using electron
diffraction \cite{kakinoki_ac60}.  From the measured data they
proposed a microcrystalline model of amorphous carbon structures,
consisting of graphite-like and diamond-like regions.  The first
neutron diffraction measurement was carried out by Mildner and
Carpenter \cite{mildner_jncs82}.  They estimated a 5\% diamond-like
atomic configuration in their sample.  Li and Lannin published another
neutron diffraction measurement on a sputtered a$-$C sample
\cite{li_prl90}.  The obtained radial distribution functions differed
in detail from the theoretical models. A~small fraction of $sp^3$
atoms were found and the chainlike configurations are also present in the
disordered a$-$C network. The absence of a significant fraction of
ordered sixfold rings was also shown, which was observed in glassy
carbon.  A~diffraction study of a small volume of amorphous carbon
materials prepared by filtered vacuum-arc method proposed nearly 100\%
of the $sp^3$ atomic configuration \cite{gaskell_prl91}.  Another
neutron diffraction study was employed on an anomalous a$-$C sample
\cite{shimakawa_pml91} in order to determine the atomic arrangement
and to decide whether the structure has temperature dependence or not
\cite{kugler_jncs93}.  The sample contained mostly $sp^2$ atomic
arrangements.  One of the most recent neutron diffraction measurement
was carried out at the Rutherford Appleton Laboratory
\cite{gilkes_prb95}.  About half a gram of tetrahedral amorphous
carbon was prepared by deposition on glass substrates.

In the experiments amorphous carbon films are usually prepared by
evaporation growth techniques. During the evaporation, carbon atoms
are directed towards the target.  If they reach the surface they could
{\it(i)}~scatter backwards, {\it(ii)}~penetrate under the surface
atoms, {\it(iii)}~start collision-cascade, or {\it(iv)}~chemically
bond on the surface, producing the growth of amorphous carbon. The
above mentioned processes ({\it(i)$-$(iv)}) depend mainly on the
environmental conditions.  The most important parameters are substrate
temperature T$_{sub}$ and the kinetic energy $E_{beam}$ of bombarding
carbon atoms.  The diamond-like amorphous carbon is usually prepared
at low pressure (10$^{-5}$~mbar), with the substrate temperature lower
than 80~$^\circ$C. In the experiments, negative bias is usually
applied to accelerate the carbon ions from the plasma, causing an
extreme increase of initial kinetic energy (usually around
300$-$500~eV from 5$-$10~eV) \cite{milne}.  However, there are some
neutral atoms that mainly contribute to the surface growth because
their kinetic energy~(1$-$10~eV) is too low to penetrate under the
surface.

\subsection{Proposed models for tetrahedrally bonded amorphous carbon}

The diamond-like amorphous carbon, otherwise frequently referred to as
tetrahedral amorphous carbon~(ta$-$C), has been the subject of several
experiments and theoretical works in the last decade. This is so, as
the material has lots of very good applicable parameters. The ta$-$C
itself has  a high percentage of $sp^3$ carbon atoms~(nearly to even
100\%) and the density of states has a wide gap, similar to that of a
diamond.  There are two main models which try to describe how the
C$^+$ ions change the $sp^2$ rich matrix target to $sp^3$ rich film.
These two models were proposed empirically from the results of
experiments and computer simulations.  

The so-called {\it subplantation model} is based on experimental
observation and physical intuition.  It was proposed in 1989 to
describe the film deposition technology for semiconductors, metals,
and ceramics with low energy E = 1$-$1000~eV hyperthermal species
\cite{lifshitz_prl89, lifshitz_prb90}.  Later similar models were
proposed by Davis \cite{davis_tsf93} and Robertson
\cite{robertson_drm93}. In these models the ions are implanted into
the subsurface of the film, producing densification and compressive
stress in the process. The ion impact affects processes on three
different time scales: {\it(i)}~a collision stage involving
penetration, energy loss and generation of ballistic
displacements~($\sim$~10$^{-13}$~s), {\it(ii)}~a thermalization stage,
in which the excess energy is dissipated and the local structure
relaxes~($\sim$~10$^{-12}$~s), {\it(iii)}~a long-term relaxation stage
in which temperature activated processes may occur~($>$~10$^{-10}$~s).
It is postulated that the ions cause local melting near the impact
site, a phenomenon known as "thermal spike" which anneals the surface
region and reduces the stress. 

An alternative model was proposed by Marks \etal \cite{marks_prb96a}.
This suggests that the collision induced stress is stochastically
localized on the surface and not in the subsurface region, arguing
that the growth proceeds directly on the surface and not in the bulk
region. For this, two-dimensional~(2D) MD simulations were done on a
graphite-like substrate with ion energies from 1~eV to 75~eV.  It was
emphasized that the simulations are not an attempt to reproduce the
dynamics of a particular material, it should be viewed as a 2D model
of a hypothetical material using a realistic interatomic potential. It
is really hard to accept any result of this kind of simulations.
However, it was found that the stress calculated in this 2D model
reaches the maximum at 30~eV deposition energy and then gradually
decreases at higher energies.  This transition has not been previously
reproduced by any simulation, but is commonly observed experimentally.

\subsection{Computer simulations}

There is no experimental method for the determination of microscopic
structures in three dimensions.  Efforts to develop simulation
techniques for analyzing atomic scale structures are therefore
continually being made. The research in this area has recently focused
on the construction of realistic model structures for amorphous carbon
using computer simulations by help of two basic methods.  Both, the
Monte Carlo~(MC) and molecular dynamics~(MD) simulations need local
potentials that are usually difficult to derive yet are essential to
simulations. 

The MC method uses a random number generator to discover the phase
space. For the initial structure a random packed matrix~(or continuous
random network) is usually chosen. Then an algorithm which satisfies
the detailed balance is applied. The structure having the lowest
energy is believed to have the best amorphous structure. A~more
efficient algorithm was proposed for continuous disordered systems by
Barkema \etal in 1996 \cite{barkema_prl96}, which is called
activation-relaxation technique~(ART).  Another simulation method
based on MC simulations has also been developed by McGreevy and
Pusztai \cite{McGreevy_MolSim88} for the investigation of
non-crystalline structures.  The Reverse Monte Carlo~(RMC) scheme is
based on the results of diffraction measurements and this is the only
simulation method which is free from the description of the
interaction between atoms.  It was applied for amorphous carbon
\cite{walters_prb98} providing a large scale three-dimensional model.
MC simulations are very successful in finding the minimal energy
structures. However, the dynamics of any system cannot be followed. 

Molecular dynamics simulation is a technique to compute the
equilibrium and transport properties of a classical many-body system.
MD simulations are in many respects very similar to real experiments.
When a MD study is done, a sample is prepared consisting of N
particles and then the Newton equations of motion are solved for this
system. Time averages for such a system almost represents that of a
microcanonical ensemble~(E,N,V), if one assumes that time average is
equivalent to ensemble averages.  The microcanonical ensemble is
inconvenient both for theoretical analysis and also for comparisons of
computed results with experimental data. Canonical and other ensembles
can be derived using extended Hamiltonians.  The reader can find more
details about MC and MD simulations, for example, in the book by Allen
and Tildesley \cite{allen_90}, and in the book by Rapaport
\cite{rapaport_95}.

In the MD simulations two different  interatomic potentials are often
applied.  The first set of potentials are the classical empirical
potentials.  The most empirical potentials fell into two main groups.
Pair potentials like Morse  or Lennard--Jones 6$-$12 interatomic
potentials are inapplicable to covalently bonded systems. The second
group of interactions contain at least one additional term, i.e.
three-body potential is included \cite{tersoff_prl88, brenner_prb90} .
This term could stabilize the covalently bonded structure.  The above
two interactions between carbon atoms are the most commonly applied
classical empirical potentials for large system with several hundred
atoms \cite{tersoff_prl88, kaukonen_prl92, kelires_prb93,
kaukonen_prb00, jaeger_jap00}.

Although simulations based on empirical potentials achieved remarkable
success, extension of the simulation method to quantum mechanics is
important.  A more accurate approach is to derive the interatomic
potential directly from the electronic ground state where the total
energy functional is calculated using density functional theory (DFT)
within the local-density approximation (LDA).  Unfortunately, DFT is
very computer time-consuming, it is therefore unthinkable to perform a
separate self-consistent electronic minimization at every MD step.
The approach called Car-Parrinello method \cite{car_prl85} overcomes
this difficulty and it has been successfully applied for MD
simulations of covalently bonded amorphous systems yielding valuable
structural and electronic information of small systems.  Structural
simulations were carried out using {\it ab initio} or modified
Car-Parrinello DFT method for a$-$C \cite{galli_prl89, marks_prb96b,
mcculloch_prb00}.

Larger systems are necessary to study for understanding complex
dynamics processes such as growth.  An alternative possibility to
Car-Parrinello DFT method is to find tight-binding~(TB) models that
could be applied to larger systems (a few hundred atoms) and could be
less time consuming than DFT \cite{wang_prb89, xu_jpcm92,
sankey_prb89, elstner_prb98}.  In the last decade applications of
different TB potentials were widespread in the MD study of carbon
systems \cite{drabold_prb91, wang_prl93a, wang_prl93b, drabold_prb94,
jungnickel_prb94, porezag_prb95, koehler_prb95, dong_prb98,
uhlmann_prl98, chuchun_prb99}.

The above quantum mechanical potentials were efficiently used to
simulate the bulk of a$-$C structures.  A~common feature of the
simulations was that each had an initial random network which was then
annealed and cooled down near room temperature. Lots of physical
parameters were recovered.  Every model shows a very good agreement in
the short-range order measured in these materials.  Furthermore, other
physical properties were compared like electronic density of states
and vibrational density of states. In 1989, a simulation was made by
Galli and her colleagues \cite{galli_prl89}. The disordered carbon
network was generated by quenching from liquid using the
Car-Parrinello method. The 54-atom-supercell reproduced well, the
short-range order of the atomic structure and the electronic structure
of a$-$C.  In the work of Wang \etal \cite{wang_prl93a} the amorphous
structure was generated by quenching from liquid phase.  The supercell
contained 216 atoms and the ring statistics analysis showed that the
clusters formed graphite-like sheets at 2.20, 2.44 and 2.69 g/cm$^3$
densities as well.  It was also found that at higher densities the
fraction of fourfold coordinated atoms increased.  In addition,
contrary to previous MD quenching \cite{galli_prl89}, twofold
coordinated atoms were found in the amorphous matrix. Their latter
work \cite{wang_prl93b} focused on ta$-$C. The 74\% percentage of
fourfold coordinated atoms was generated in a sample where the initial
density was very high 4 g/cm$^3$. Note that the density of diamond is
close to 3.5 g/cm$^3$. At that density the structure consisted 80\% of
$sp^3$ carbon atoms, and then by decreasing the density, the $sp^3$
content decreased to 74\%. Drabold \etal \cite{drabold_prl94}
criticized the transferability of the potential \cite{xu_jpcm92} used
in the works of Wang \etal \cite{wang_prl93a, wang_prl93b}. However,
Wang and Ho argued with this statement \cite{wang_prl94}. Recently, a
more accurate form of their carbon potential was published
\cite{tang_prb96}. The above preparation methods do not contain the
information of the growth of amorphous carbon.

Atomic scale modeling of ion-beam-induced growth of amorphous carbon
was simulated by Kaukonen and Nieminen \cite{kaukonen_prl92,
kaukonen_prb00} and by J\"ager and Albe \cite{jaeger_jap00}. The work of
Kaukonen and Nieminen \cite{kaukonen_prl92} -- redone later
\cite{kaukonen_prb00} -- report 1$-$150~eV ion beam kinetic energy
window simulations. The empirical Tersoff \cite{tersoff_prl88}
potential described the interatomic interaction between carbon atoms.
It was found in the latter work that the low energy~(10$-$40~eV)
bombardment produces the highest $sp^3$ content in the host matrix,
supporting the model proposed by Marks \etal \cite{marks_prb96a}.
However, the $sp^3$ content was half of the experimental observed
value.  In the work of J\"ager and Albe \cite{jaeger_jap00} a similar
simulation work was carried out with the Tersoff's
\cite{tersoff_prb89} and Brenner's \cite{brenner_prb90} potential.
They have found that the fraction of tetrahedrally coordinated atoms
is too low, even if structures with densities close to diamond are
obtained. However, changing the cutoff distance in Brenner's
potential, the $sp^3$ fraction close to 82\% was found, which is in
good agreement with experiments. This modification has an undesired
side effect, it has increased the fraction of fivefold coordinated
carbon atoms, which was never observed by any quantum mechanical
treatment of interatomic potentials. A general problem of classical
potentials that are fitted to a limited set of materials properties is
their accuracy in describing processes which cover areas lying far
from the fitted points of the potential energy hyperface. In the case
of a$-$C (and t$-$aC) deposition the validity of using classical empirical
potentials can be argued.

It means that quantum mechanical potentials are needed to describe
interatomic potentials. The first pioneer work was done by Uhlmann
\etal \cite{uhlmann_prl98}. In this work the diamond-like carbon~(DLC,
ta$-$C) films were investigated by bombardment with neutral carbon atoms
a 3~g/cm$^3$ amorphous film. The results supported the subplantation
model. It was shown that for E~$>$~30~eV the deposition is a
subplantation process, where the carbon atom species penetrate and
occupy subsurface positions, producing a defective, low density,
$sp^2$ rich surface layer and a dense, $sp^3$-rich layer is formed
below this surface. Chu-Chun Fu \etal \cite{chuchun_prb99} published a
simulation method for the film growth over silicon substrate.  The
silicon surface is assumed to enhance hardness through the formation
of a large proportion of tetrahedral bonds.  They published that a
mixed thin layer in the surface preferring carbon-silicon bonds to
carbon-carbon bonds were formed at impinging energy 7.55~eV of carbon
atoms. This simulation contained only 9 and 16 bombarding carbon
atoms. The above two results were published very recently. The project
which was done in part of this Ph.D. thesis was started long before
these results were achieved and will be summarized in the next
section.

\section{Modeling the growth of amorphous carbon}
\markboth{MD SIMULATION OF a$-$C STRUCTURES}{MODELING THE GROWTH OF
a$-$C}

The aim of this chapter is to describe in detail the low energy
(1$-$5~eV) bombardment of the target surface. If the carbon atom
reaches the surface with such a low kinetic energy the probability of
its penetration under the surface and the probability that it starts a
collision-cascade is very low. It can however sometimes scatter
backwards and can chemically bond on the surface producing the growth
of amorphous carbon. These statements allow us to study the growth of
amorphous carbon in this environment.  We address here the analysis of
the growth of amorphous carbon with quantum mechanical treatment of
the interatomic potential on large scale systems with over 100 atoms.

\subsection{Molecular dynamics calculations}

Molecular dynamics simulations were carried out to study the dynamics
of the growth process.  The tight-binding~(TB) Hamiltonian of Xu,
Wang, Chan, and Ho \cite{xu_jpcm92} was used to calculate the
interatomic potential between carbon atoms. All the parameters and
functions of these methods were fitted to the results of local density
functional calculations.  In that scheme, the system is described by a
Hamiltonian of the form

\beq
 H(\{{\vec{\bf r}_i}\}) = 
 \sum_i {{\vec{\bf p}_i^2}\over{2m_i}} + \sum_n^{occupied} 
 \langle \Psi_n |H_{TB}(\{{\vec{\bf r}_i}\}) |\Psi_n\rangle +
 E_{rep}(\{{\vec{\bf r}_i}\})
 \label{hamilton}
\eeq
where $\{{\vec{\bf r}_i}\}$ denote the positions of the atoms ($i$ =
1, 2, ..., N), and ${\vec{\bf p}_i}$ is the momentum of the $i$th
atom. The first term in Eq.\ (\ref{hamilton}) is the kinetic energy of
the ions. The second term is the electronic band-structure energy
calculated from a parameterized tight-binding Hamiltonian
$H_{TB}(\{{\vec{\bf r}_i}\})$. It consists of the sum of the energy
eigenvalues $E_i$ for the occupied part of the electronic band
structure, which is calculated by the Slater-Koster empirical
tight-binding method \cite{slater_pr54} with the tight-binding
parameters determined by fitting the calculated electronic structure
of carbon. The off-diagonal elements of $H_{TB}$ are described by a
set of orthogonal $sp^3$ two-center hopping parameters,
$V_{ss\sigma}$, $V_{sp\sigma}$, $V_{pp\sigma}$, $V_{pp\pi}$, scaled
with interatomic separation $r$ as a function $s(r)$~(Eq.\ \ref{sr});
and the on-site elements are the atomic orbital energies of the
corresponding atom.  The third term is a short-ranged repulsive
energy. It represents the sum of the ion-ion repulsion and the
correction to the double counting of the electron-electron interaction
in the second term.  In this scheme, the quantum mechanical nature of
the covalent bonding among the carbon atoms is explicitly taken into
account in the electronic structure which is calculated for each of
the atomic configurations in the MD simulations. Details of the
tight-binding model are described in reference \cite{xu_jpcm92}.  The
interatomic potential describes accurately the energetic, vibrational,
and elastic properties of the diamond~(fourfold),
graphite~(threefold), and linear-chain~(twofold) structures in
comparison with self-consistent first-principles density functional
calculations.  The realistic TB potential has already been used for
preparation of amorphous carbon \cite{wang_prl93a}, tetrahedrally
bonded amorphous carbon \cite{wang_prl93b} and successfully applied
for preparation of fullerenes \cite{zhang_prb92, laszlo}.

The orbital basis used is $2s$ and $2p$. The off-diagonal elements for
the tight-binding Hamiltonian are given in terms of a function of the
interatomic distance $s(r)$

\beq
  s(r)=\left(\frac{r_0}{r}\right)^n 
  \exp\left\{n \left[-\left(\frac{r}{r_c}\right)^{n_c} +
  \left(\frac{r_0}{r_c}\right)^{n_c}
  \right]\right\}.
\label{sr}  
\eeq
This function contains four independent parameters, that have been
adjusted with results of {\it ab initio} calculations or experiments.
The diagonal elements, as well as the hopping integrals for the
equilibrium distance for one particular structure have also been
parameterized, and are given in the tables of reference\
\cite{xu_jpcm92}.  The repulsive energy is a function for the sum of
pair interactions, of a similar functional form as $s(r)$, with again
four independent parameters which are also taken from the previous
reference

\beq
  \phi(r) = \phi_0 \left(\frac{d_0}{r}\right)^m 
  \exp\left\{m\left[-\left(\frac{r}{d_c}\right)^{m_c}+
  \left(\frac{d_0}{d_c}\right)^{m_c}
  \right]\right\}.
\eeq
 
To use these interactions in a MD simulation it is necessary to
introduce a cutoff for the interatomic distances. This is particularly
important in cases where periodic boundary conditions are imposed, so
that replicas of the same atom do not interact with each other. 

For integration of Newton's equations of motion the velocity Verlet
algorithm was used with time step of 0.5~fs: 

\ba
 {\bf\vec{r}}_i(t+{{\Delta}}t)
 &=&
 {\bf\vec{r}}_i(t)+{\bf{\vec{v}}}_i(t){{\Delta}}t+
 {{1}\over{2}}{\bf{\vec{a}}}_i(t){{\Delta}}t^{2}+
 {\mathcal O}({{\Delta}t}^3)\\
 {\bf\vec{v}}_i(t+{{\Delta}}t)
 &=&
 {\bf\vec{v}}_i(t)+
 {\frac
 {{\bf\vec{a}}_i(t+{\Delta}t)+{\bf\vec{a}}_i(t)}
 {2}}
 {\Delta}t.
\ea
 
The attractive forces were obtained using the Hellmann-Feynman
theorem. The Hellmann-Feynman force on atom $i$ is

\beq
  {\bf\vec{F}}_i = - \sum_n^{occupied} 
  \langle \Psi_n | 
  \partial H_{TB}(\{{\vec{\bf r}_i}\})/ \partial {\bf\vec{r}}_i | 
  \Psi_n \rangle,
\eeq
where the sum is over occupied levels. The repulsive forces are
directly the gradient of $E_{rep}$. Finally, we would refer to two
articles which describe the above tight-binding molecular dynamics
method in more details: the work of Wang \etal \cite{wang_prb89} with
orthogonal basis, whose scheme is used in this chapter for simulation
of growth process of a$-$C and the work of the Paderborn group
\cite{elstner_prb98, frauenheim_pssb00}, which uses non-orthogonal
basis.

\subsection{First simulations}

The review of our computer simulations will be given in this section.
This project started in 1997 and our aim was to find the best
treatment of the description of a$-$C growth using quantum mechanical
treatment of applied potential.

First, we applied 0~K substrate which is the simplest method to
simulate the energy dissipation process during the collision of
bombarding atoms and the substrate atoms. The quantity {\it critical
distance} was introduced, which was chosen 17\% larger than the first
neighbor bond distance in diamond~(1.54~\AA). We used the following
model.  If a bombarding carbon atom arrived in the neighborhood of a
substrate atom, it became the part of atoms where the kinetic energy
was decreased in every time step with a given number, which was
smaller, but close to one. If the next bombarding atom arrived in the
neighborhood of the substrate atom or the closeness of the atoms whose
kinetic energy decreased in every time step, the similar decrease of
energy was employed.  This model can be considered as the cooling of
the bombarding atom in the region of the impact, decreasing its
kinetic energy in every time step. The zero temperature of the
substrate provides the quickest energy dissipation. The first results
showed that the low energy~(1.29~eV) bombardment gives a more
chainlike structure and the higher energy bombardment close to 10~eV
provides more amorphous network structures \cite{kohary_jncs98,
kugler_fm99}.

To improve the description of energy dissipation in our computer
experiments the task of finding the better temperature control of the
substrate was addressed. There are several opportunities to control
the temperature in MD simulation. The Nos\'e-Hoover thermostat uses
extended Hamiltonian and describes the canonical ensemble in the right
statistical physics form. In the case where the particle number N in
the system tends to infinity, it can be shown that the canonical
ensemble can be realized approximately in equilibrated systems.
However, systems with size N $\approx$ 100 already show very good
results.  Unfortunately, it turned out that the non-equilibrium growth
process cannot be simulated with the above description of constant
substrate temperature \cite{kohary_diploma}. The failure of the use of
Nos\'e-Hoover thermostat does not mean that the constant temperature
cannot be realized in the simulations.

\begin{figure}[t]
\centerline{
\epsfig{file=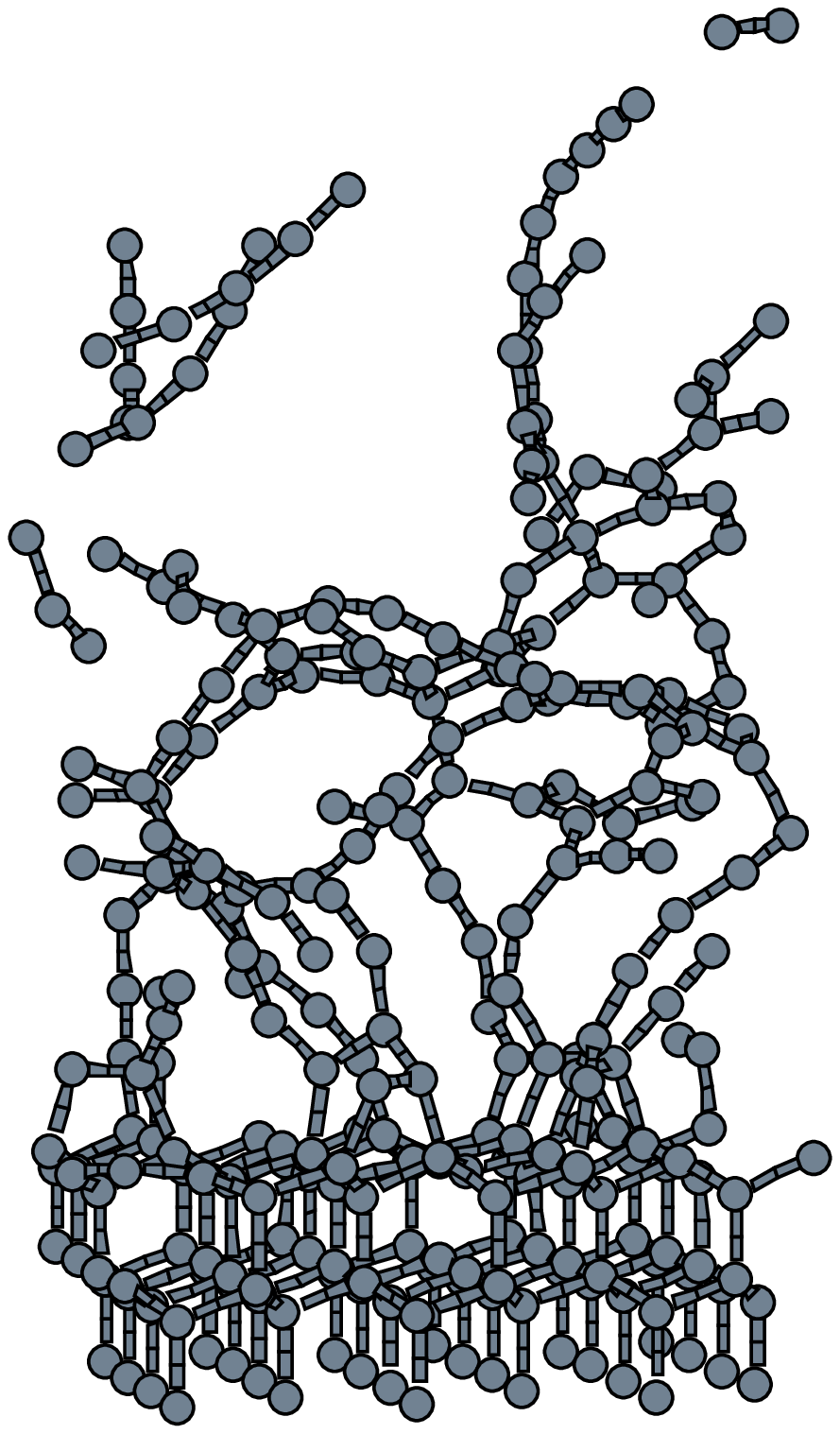,width=7.5truecm}
\epsfig{file=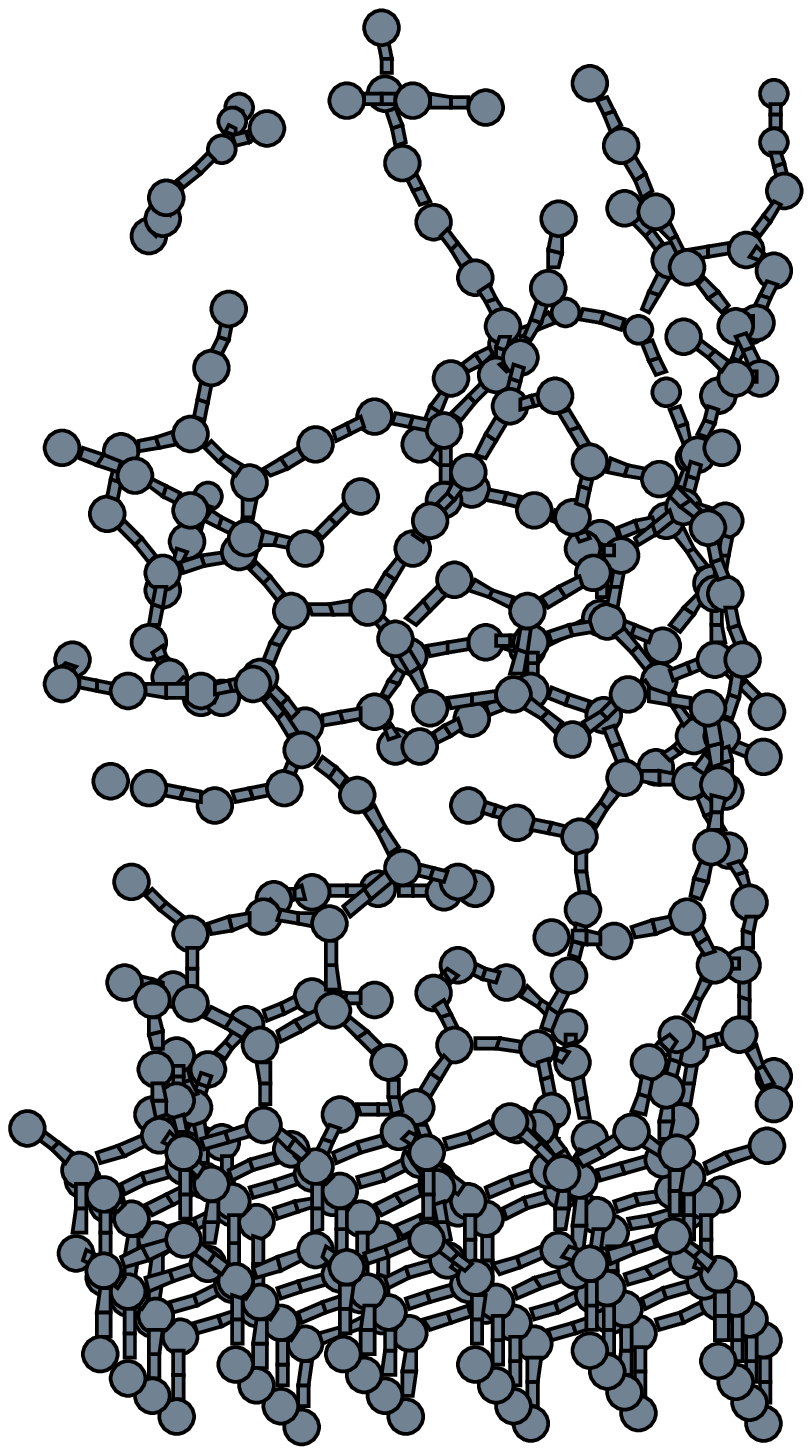,width=7.5truecm}
}
\medskip
\caption{
\small
The final structures of the atom-by-atom deposition. The left model is
chainlike amorphous structure with deposition energy 6.5~eV. It
contains 286 atoms including 120 atoms in the substrate. The right
model shows the structure made after 13~eV deposition. It contains 323
atoms.
} 
\label{fig00_aC} 
\end{figure}

For the next step we decided to use substrates, with fixed atoms in
the bottom layer and giving initially 0~K to the remaining atoms in
the substrate. The results of the simulations provided similar results
as the rigid substrate \cite{kugler_grenoble98} with 6.5 and 13~eV
bombardment.  The snapshots of the final structures are shown on Fig\
\ref{fig00_aC}.  It is supported by the results that lower kinetic
energy bombardment supports more chainlike structures and that larger
energy constructs more amorphous structures.  It was also shown that
the kinetic energy of the atoms in the film decreases near
exponentially after no further bombarding atoms appeared in the film
\cite{kohary_jncs00}. This behavior will be discussed in more detail
later in this chapter.

\subsection{Numerical algorithm and simulation details}
\label{method}
 
In this section a more realistic model of amorphous carbon deposition
will be given \cite{kohary_prb01_aC}. The simulation technique was as
follows. An ideal diamond film consisting of 120 atoms was employed to
model the substrate. The rectangular simulation cell was open along
the positive $z$-axis ([111] direction in our case) and periodic
boundary conditions were used in planar, $x$ and $y$ directions.
Atoms in the bottom substrate layer were fixed at their ideal lattice
positions, whereas the rest of the 96 atoms in the uppermost substrate
layers were allowed to move with full dynamics.   To simulate the
constant substrate temperature the kinetic energy of the moveable
atoms in the substrate were rescaled in every time step. The velocity
Verlet algorithm was used to determine the trajectories of the carbon
atoms.  Time step was chosen to be equal to 0.5~fs.

Before starting the deposition process, substrate was kept at a given
temperature for 0.5~ps in order to make  structural relaxation.  The
neutral bombarding atoms were randomly placed in $x$ and $y$
directions above the substrate as near as possible, but not closer to
any other atom  than the cutoff distance of the potential. The initial
velocities of bombarding atoms were directed to the substrate with
given kinetic energy.   There is no thermal equilibrium so initial
velocities were chosen using the $v=\sqrt{{{2E_{beam}}\over{m}}
(1.2-0.4 \times p)}$ simple relation, where $p$ is a uniformly
distributed random number between 0 and 1.  Directions were determined
by $\theta = 120^\circ+60^\circ \times p$ and $\phi = 360^\circ \times
p$, where $\theta$ and $\phi$ are the polar angles and $p$ is again a
random number.

The frequency of the atomic injection was $f$ = 1/125~fs$^{-1}$ on
average:  in every MD time step a random number was generated and was
compared to a given number (in the present case it was 0.004), to see
if it is greater or lower.  In case of the latter, a new particle is
"created" and it started its motion bombarding the substrate.  It
means that on average in every 0.004$^{-1}$ = 250th MD step, the
random number was smaller than 0.004.  The time step was equal to
0.5~fs, i.e.  250$\times$0.5~fs = 125~fs was the average time period
to put a new atom into the atomic current.  This flux is in the orders
of magnitude larger than the deposition rate usually applied in real
experiments.  We found, however, that even at such a high flux the
overheating effect of the surface (see, e.g., \cite{kaukonen_prl92})
can be negligible and the low-energy evaporation can be modeled.  The
lower substrate temperatures applied in our simulations results in
faster energy dissipation, which compensates for the high deposition
rate.  The initial kinetic energies of the atoms in the beam were
dissipated by interacting with the surface atoms, whereas their
trajectories were followed with full dynamics.  We note here that some
atoms were scattered back and needed to be cut from the system.  This
kind of backscattering can be found in the experiments as well.

\subsection{Energy dissipation}

\begin{figure}
\centerline{
\epsfig{file=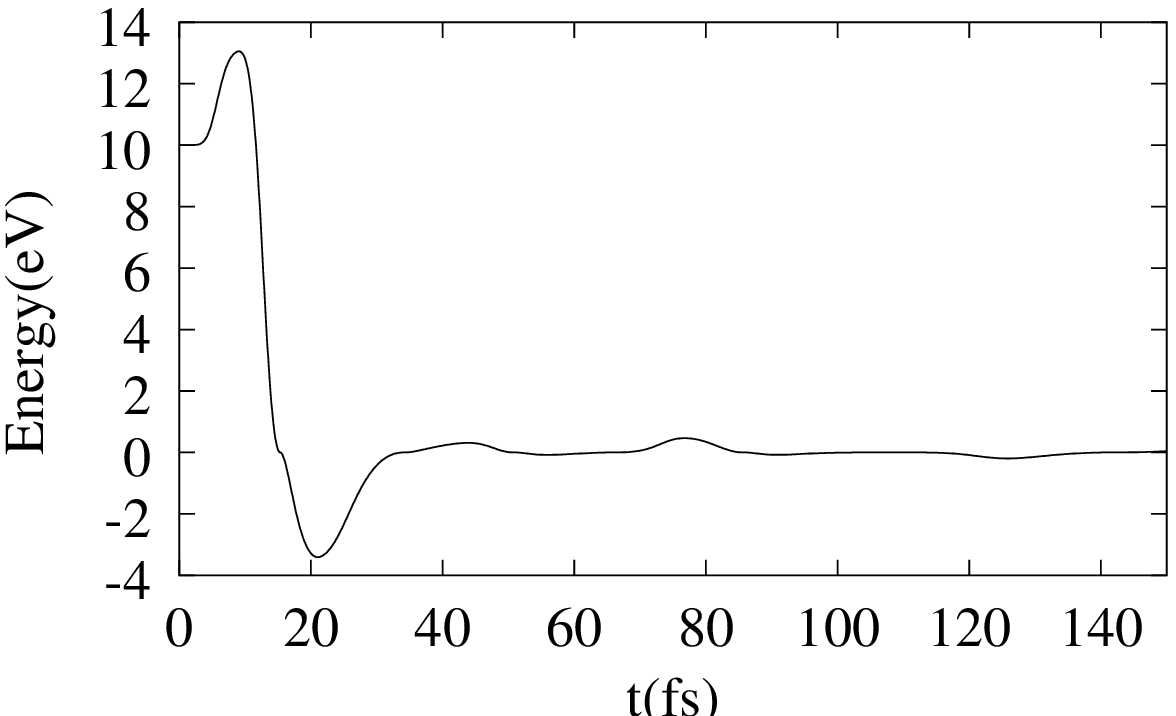,width=7.3truecm}
\hspace{0.5truecm}
\epsfig{file=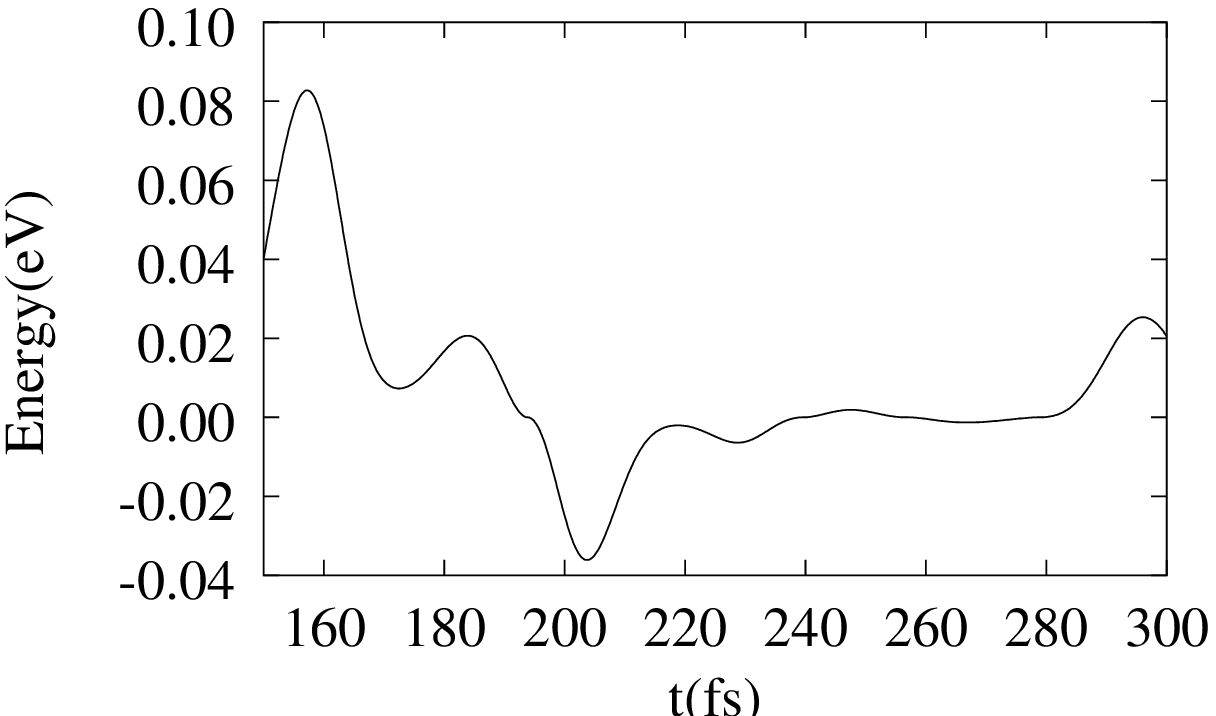,width=7.3truecm}
}
\medskip
\caption{
\small
The $z$-component kinetic energy of the bombarding atom.
In the figure on the left the time interval 0$-$150~fs, while
in the figure on the right the interval 150$-$300~fs can be seen. 
} 
\label{fig0_aC} 
\end{figure}

In the experiments the magnitude of the atom flux in the atom-by-atom
deposition techniques is 10$^{14}$~cm$^{-2}$s$^{-1}$. It means, that
on average one atom bombards the approximately 10$\times$10~\AA$^2$ \
substrate surface in one second. Computers cannot perform simulations
in the above time scale.  The typical time step chosen is around 1~fs
and the calculations over several picoseconds last several
weeks~(months) for systems having 100$-$200 atoms and if the
interatomic potential is described by quantum mechanical treatment.
Due to the problems mentioned above, all the details of the growth
cannot be realized by the help of computer simulations.  However, the
short time processes, like energy loss and local thermalization can be
simulated, as these processes are in picosecond range.  It is an
approximation that the growth process can be simulated with the above
time step, when the frequency of bombarding is close to
1/125~fs$^{-1}$.  Kaukonen and Nieminen \cite{kaukonen_prl92} reported
that an overheating effect can sometimes be observed. To decide
weather it plays an important role in our quantum mechanical treatment
of the problem, we performed computer simulations with only one
bombarding atom over the 80 atoms substrate surface with 100~K.  The
time step $\Delta$t equal to 0.1~fs was chosen. A carbon atom was
placed 2.75 \AA \ over the substrate.  The initial kinetic energy of
the bombarding carbon atom was chosen to 10~eV and has only
$z$-directional velocity.  We performed simulation placing the carbon
atom in high symmetrical points over the surface and one general
place.  The Fig.\ \ref{fig0_aC} shows the $z$ component kinetic energy
changes for the bombarding atom. It can be seen that the carbon atom
bounced back after 15~fs and loses its initial kinetic energy after
40~fs.  Then oscillations of kinetic energy around 0~eV with 0.1~eV
deviations can be found, as it is clear from the Fig.\ \ref{fig0_aC}.
This shows that the 1/125~fs$^{-1}$ flux is acceptable in our computer
simulations.

\subsection{Structural properties and discussion}

\begin{figure}[htbp]
\centerline{\epsfig{file=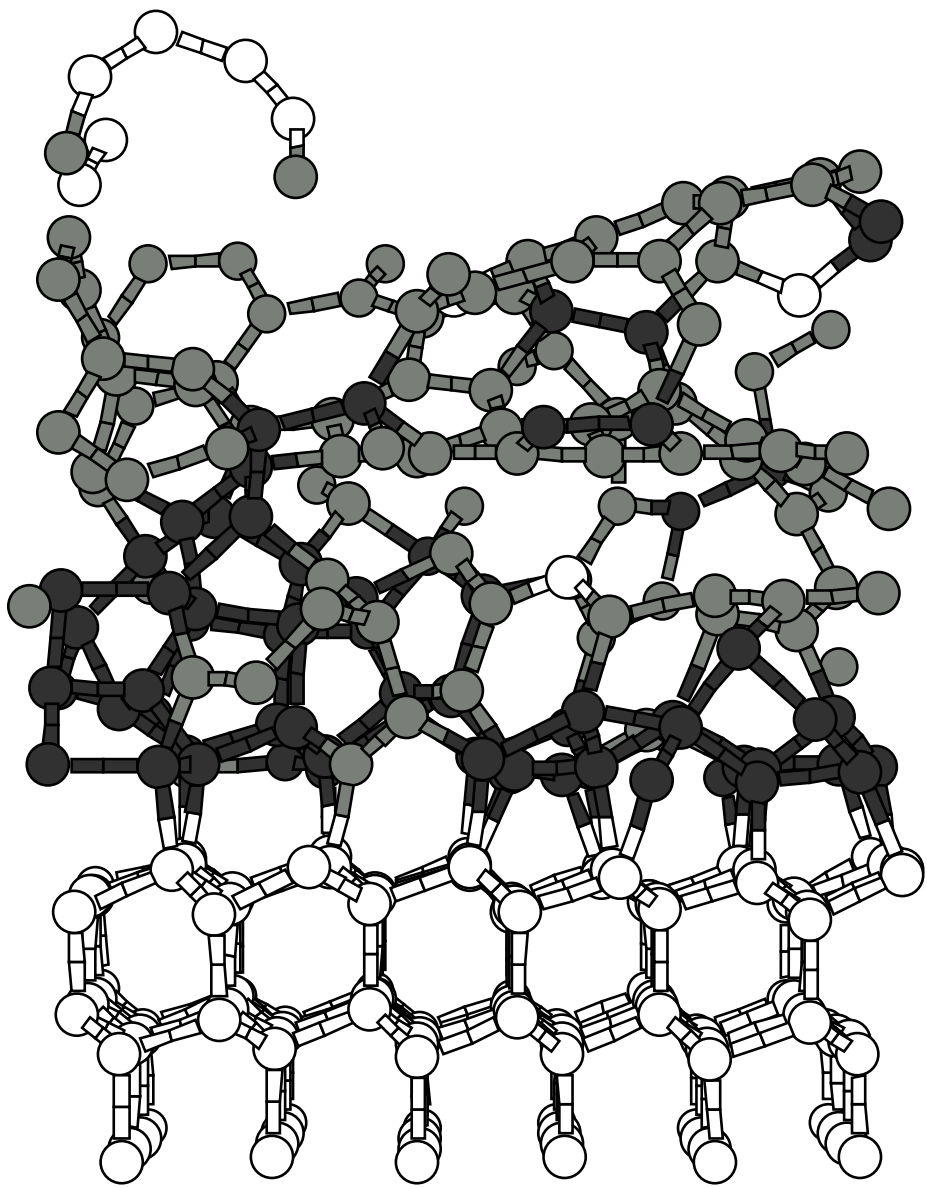,height=17.0truecm}}
\medskip
\caption{
\small
A snapshot of the e5T100L model is shown after growth and relaxation.
The substrate (open circles at the bottom) at T$_{sub}$ = 100~K
remained similar to the crystal lattice during growth. Black and grey
atoms are fourfold and threefold coordinated, respectively. The rest
of the open circles correspond to twofold and one onefold coordinated
atoms. 
For a color version see:
http://www.phy.bme.hu/$\sim$kohary/aCfigures.html
} 
\label{fig1_aC} 
\end{figure}

\begin{figure}
\centerline{\epsfig{file=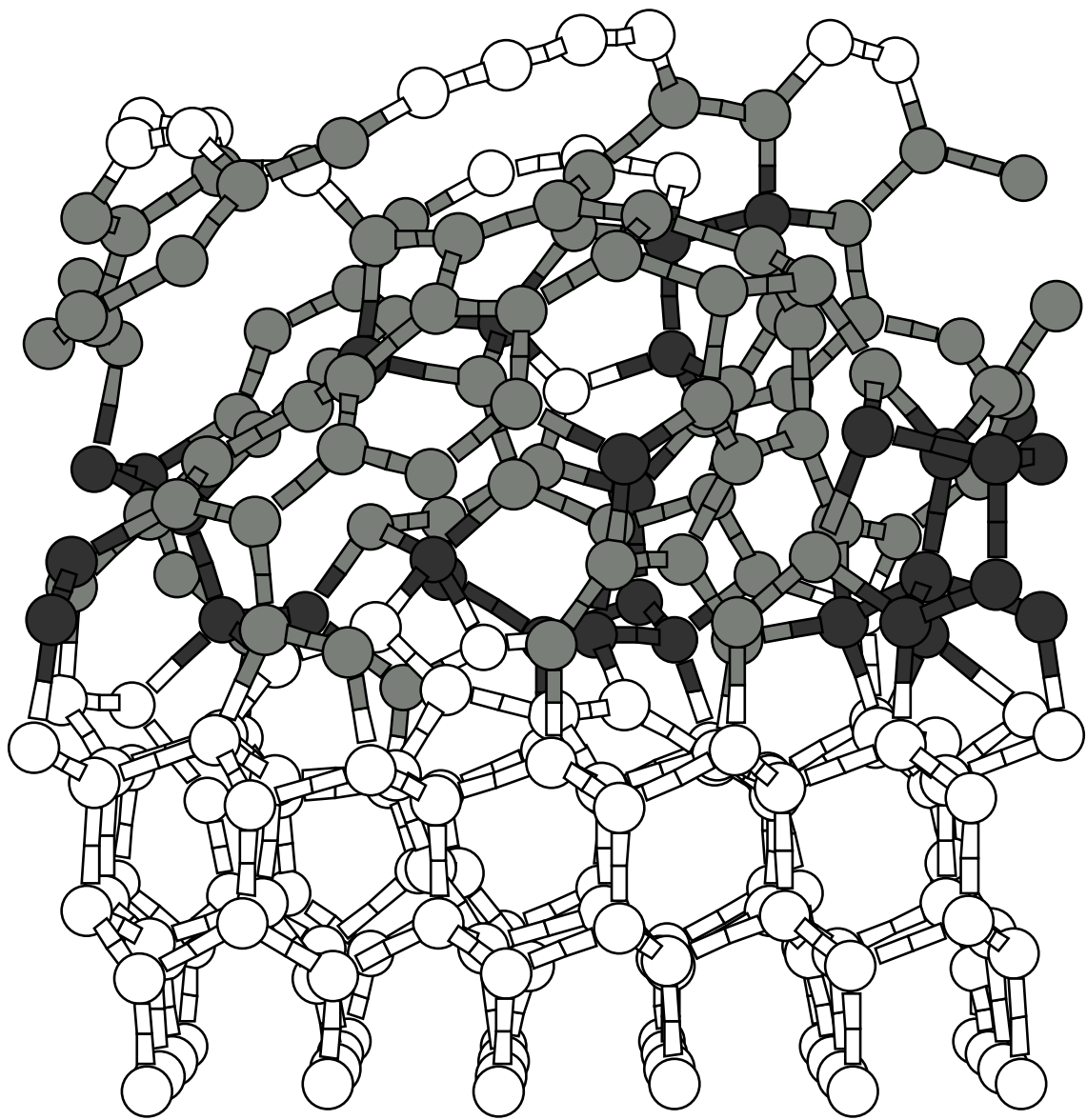,height=17.0truecm}}
\medskip
\caption{
\small
A snapshot of the e5T300 is shown after growth and relaxation.  The
substrate at T$_{sub}$ = 300~K lost even the topological structure of
a perfect crystal during growth. An atom (open circle) in the middle
of the amorphous part in figure originally belonged to the substrate.
For a color version see:
http://www.phy.bme.hu/$\sim$kohary/aCfigures.html
}
\label{fig2_aC} 
\end{figure}

Six different structures were made by our method~(see section
\ref{method}):  models constructed with 25~ps and 40~ps~(denoted by L)
injection times, and with average kinetic energy E$_{beam}$ = 1~eV and
5~eV at T$_{sub}$ = 100~K substrate temperature (e1T100, e5T100,
e1T100L and e5T100L) and two other models by 25~ps injection,
E$_{beam}$ = 1 and 5~eV energy with substrate temperature of 300~K
(e1T300, e5T300).  In all cases the structures were relaxed for 5~ps
after the deposition, i.e. a bombarding atom no longer appeared in the
system.  The final structures of larger models (e1T100L and e5T100L)
made by different bombarding  energies consist of almost the same
number of atoms~(177 and 172), with a thickness of 12.65~\AA \ and
10.98~\AA.  Models e1T100, e5T100, e1T300 and e5T300 have 126$-$129
atoms.  Typical central processing unit~(CPU) times of simulations are
about 50$-$70 days on DEC-Alpha workstations.  Snapshots of two films
(e5T100L and e5T300) are shown in Fig.\ \ref{fig1_aC} and\
\ref{fig2_aC} after growth and relaxation.  The substrate at T$_{sub}$
= 100~K remained similar to crystal lattice, while atoms left their
position and  the topological order became broken during growth at
T$_{sub}$ = 300~K.  An atom (open circle) in the middle of the
amorphous part in Fig.\ \ref{fig2_aC} originally belonged to  the
substrate.

\subsubsection{Time development of relaxations}

\begin{figure}
\centerline{\epsfig{file=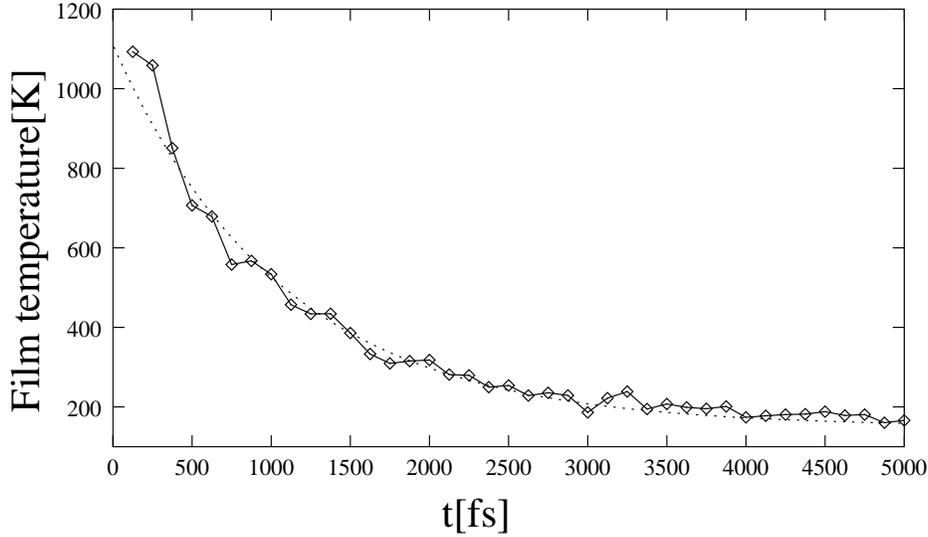,height=7.0truecm}}
\medskip
\caption{
\small
The temperature versus time during relaxation of e1T100L model is shown.  The 
fit is described by  $T_{film}$ = ($\exp({{t[fs]}/{1069.03}} + 6.8) + 150$)~K,
which has an incorrect asymptotic behavior.
}
\label{fig3_aC} 
\end{figure}

After the growth process the films relaxed with full  dynamics for
5~ps.  During this period the substrate was kept at constant
temperature while the deposited networks were cooling down.  The
temperature versus time relation of this non-equilibrium process is
exponential function in the interval~[0:5000] fs as shown in Fig.\
\ref{fig3_aC} for e1T100L model.

The best fit was done according to $T_{film}(t) = c + \exp(a \cdot t +
b)$~K.  The constant $c$~=~150~K was chosen rather than 100~K, which
has a false asymptotic behavior, but it gave a better fit.  Fitting
parameters are $a$ = -0.935~ps$^{-1}$ and $b$ = 6.8671 and
$c$~=~150~K. This means that the relaxation time $\tau = -{{1}/{a}}$=
1069.03~fs~$\approx$~1.07~ps.  

The time dependence of bond length and bond angle deviations were also
investigated.  In crystals, bond lengths and bond angles only have
thermal disorder components as a result of the thermal fluctuations.
Non-crystalline materials have an additional broadening contribution
to the bond length and bond angle distribution resulting from static
disorder.  First we calculated the standard deviations $\sigma(t_k)$
at discreet times $t_k$ in the following way:

\begin{equation}
{\sigma(t_k)}
= {\sqrt{ {{1}\over{N}} \sum_j^N \left( d_j(t_k) -
{{\sum_{i=1}^N d_i(t_k)}\over{N}}
\right)^2 }}
\end{equation}
where $N$ is the number of bonds and $d_i(t_k)$ is the $i$th bond
distance.  Values between 0.095 \AA \ $ < \sigma_{b} < $ 0.11 \AA \
were obtained for bond length deviations. (The index $b$ is for bond
length.) Similar calculations for bond angle deviations provided
$\sigma_{\Theta}(t_k)$ within an interval of
11.7$^\circ$-12.8$^\circ$.   The contribution to standard deviation of
thermal fluctuation was also studied as a function of time using the
following term:

\begin{equation}
{\sigma^{th}(t_k)}
= {\sqrt{ {{1}\over{N}} \sum_i^N \left( d_i(t_k) - \langle d_i
\rangle \right)^2 }}
\end{equation}
where time average $\langle d_i \rangle$ of bond $d_i$ is used instead
of $d_i$(T~=~0~K).

\begin{figure}[t]
\centerline{
\epsfig{file=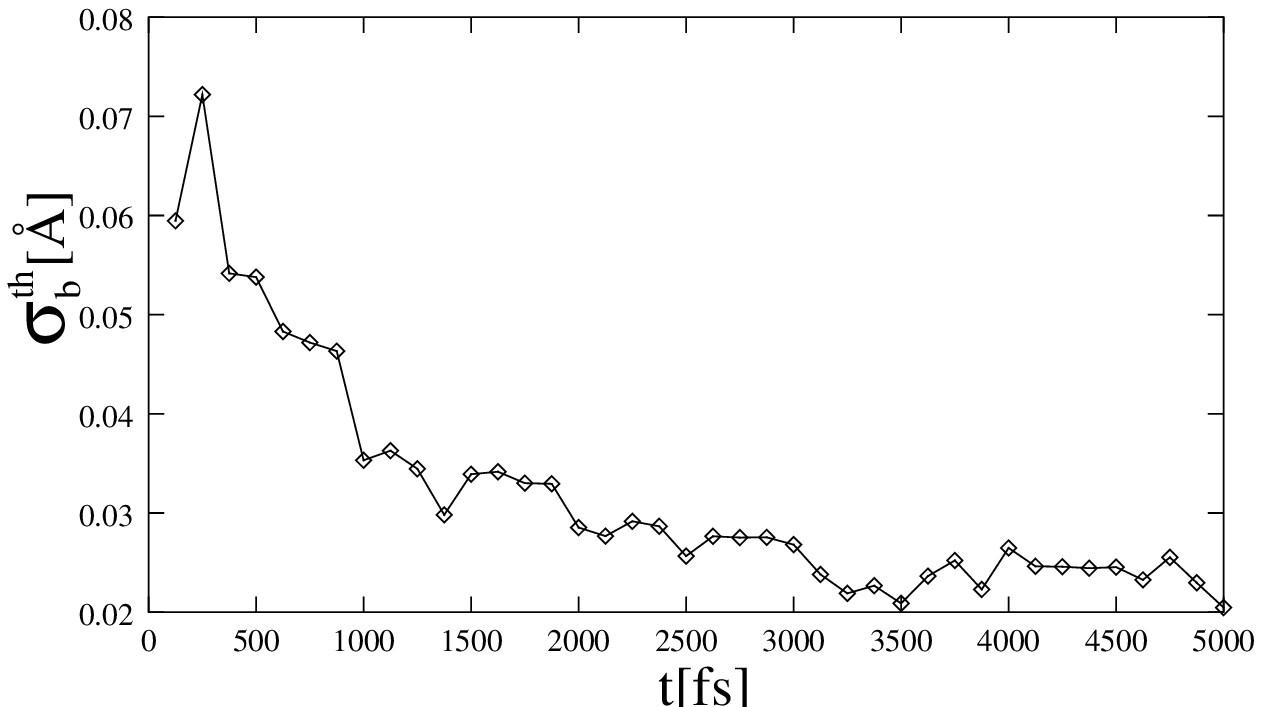,width=7.5truecm}
\epsfig{file=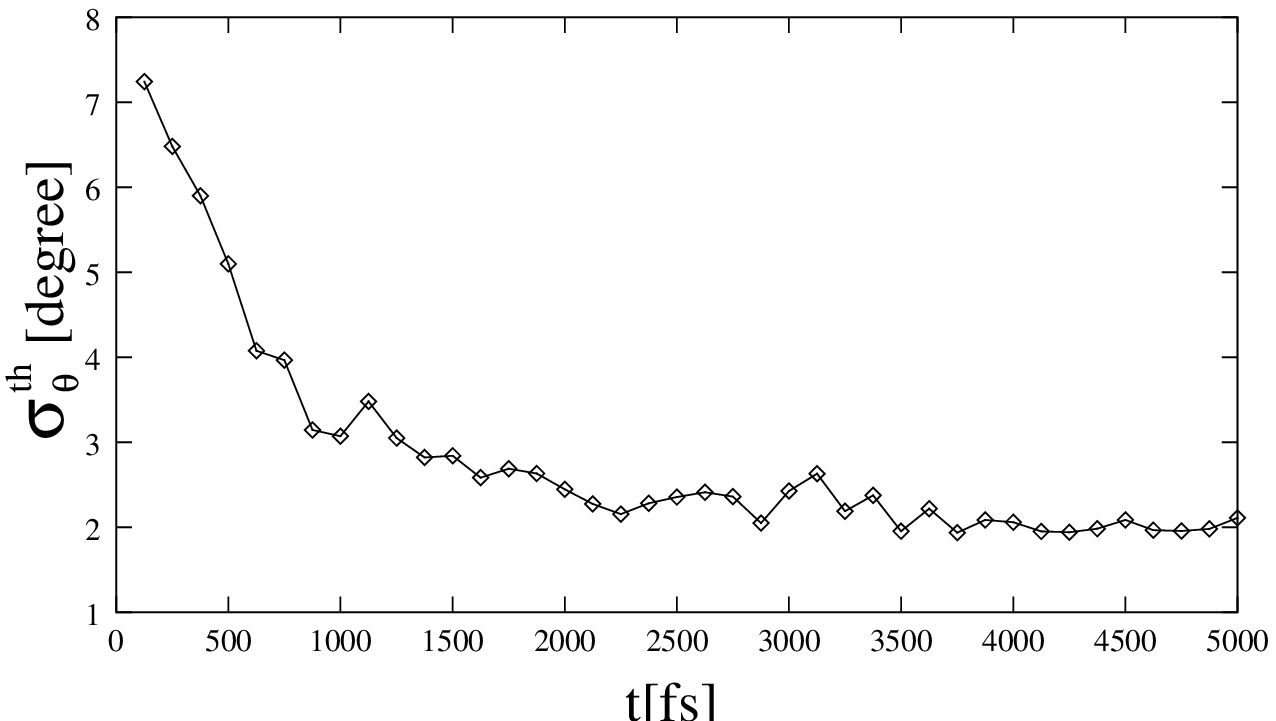,width=7.5truecm}
}
\medskip
\caption{
\small
As shown, an expected $\sigma_{b}^{th}$ and $\sigma_{\Theta}^{th}$
versus time functions have an exponential decay.  At the beginning of
relaxation when the temperature of the film is about 1000~K the
thermal part of the fluctuation is important while at the end it plays
only a minor role.  
}
\label{fig4_aC} 
\end{figure}

\begin{figure}[b]
\centerline{
\epsfig{file=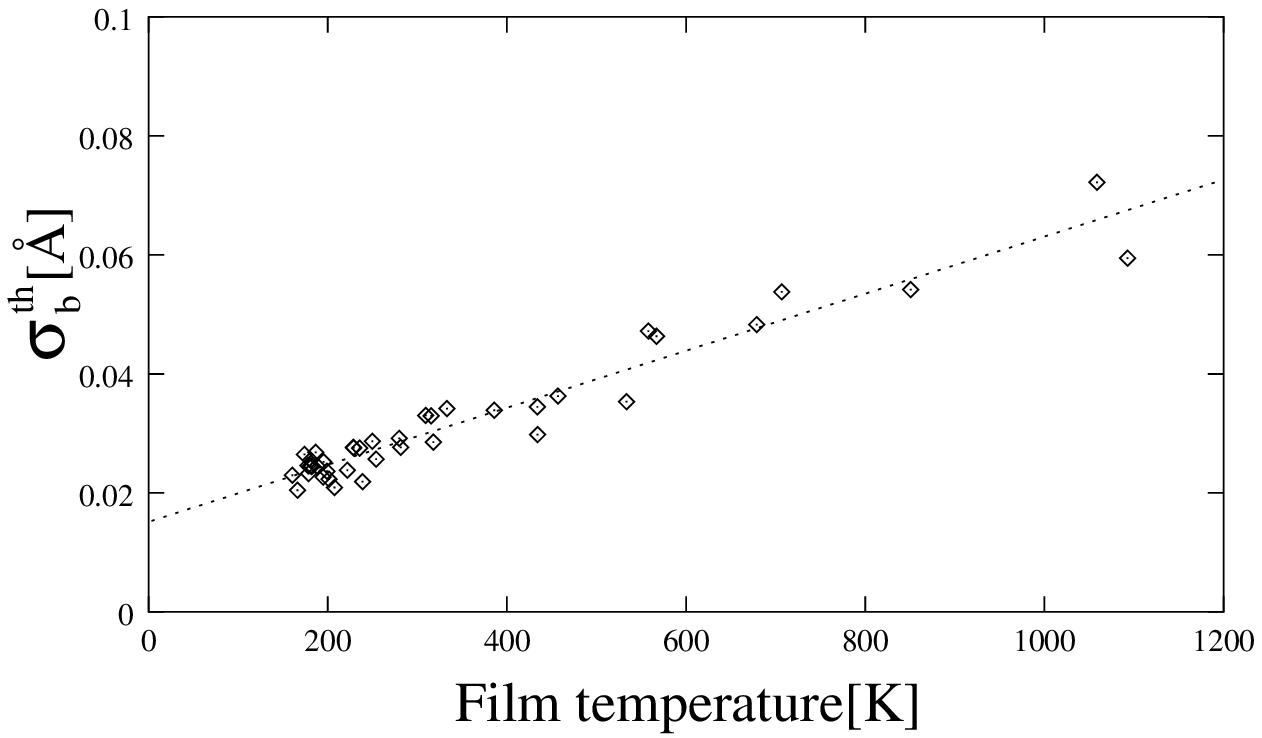,width=7.5truecm}
\epsfig{file=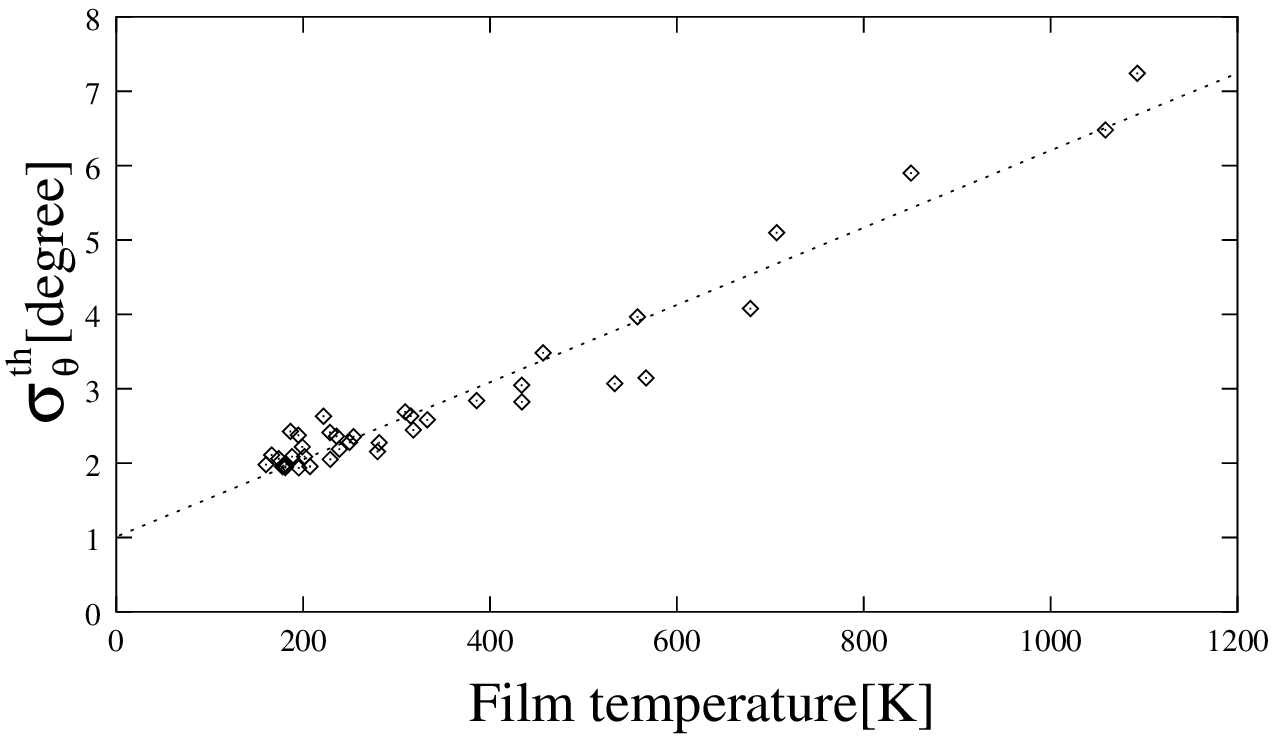,width=7.5truecm}
}
\medskip
\caption{
\small
The $\sigma_{b}^{th}$ and $\sigma_{\Theta}^{th}$ in function of
temperature show linear relationships.  Straight lines do not intersect
origin because usually $\langle d_i \rangle \neq d_i(T=0) $.  
}
\label{fig5_aC} 
\end{figure}

As expected $\sigma_{b}^{th}$ and $\sigma_{\Theta}^{th} $ versus time
functions have an exponential decay as shown in Fig.\ \ref{fig4_aC}.
At the beginning of relaxation, when the temperature of the film is
about 1000~K, the thermal part of the fluctuation is important. At the
end it plays only a minor role.  The $\sigma_{b}^{th}$ and
$\sigma_{\Theta}^{th}$ values as a function of temperature show linear
relationships, as displayed in Fig.\ \ref{fig5_aC}.  Straight lines do
not intersect origin because usually $\langle d_i \rangle \neq
d_i$(T~=~0~K).

\subsubsection{Density}

For density calculations, two different volumes were defined: we will
refer to them as cells for the total sample and for the bulk.  The
bottom of the cells were 0.77 \AA \ lower than the bottom carbon atom
in the amorphous network. This is almost at half bond distance of
bonds between the substrate atoms and the bottom carbon atoms in the
film.  The top of the cells were determined differently. For the total
sample the top of the cell was the highest $z$-coordinate that occurs
for atoms in the network. For the bulk, this $z$ coordinate was
decreased by 3~\AA.  The $x$ and $y$ size of the cell was determined
according to two-dimensional periodic boundary conditions.  Table\
\ref{table1_aC} contains the densities of different models.  Each row
is divided into further two rows. The upper rows constantly refer
to the bulk and the lower rows refer to the total sample.  For bulk
models the densities are always larger than for the total sample.  At
T$_{sub}$ = 100~K bulk densities are between 2.7 and 3.0 g/cm$^3$
except for model e1T100L where the density is equal to 2.4 g/cm$^3$.  The
reason for lower density is that on the top of the network a void
appeared, thus drastically decreasing the local density.  At T$_{sub}$
= 300~K the structures are less dense (2.0$-$2.3~g/cm$^3$).

\begin{table}
\caption{
It contains the number of atoms~(No.), the percentage of atoms with
different coordination numbers (Z), the average coordination number
$\langle Z \rangle$, the thickness z$_d$, and the density $\rho$ of
different models. The two rows belong to bulk~(top) and to
total~(lower) sample, respectively. For bulk models the densities are
always larger than for the total sample.  
\label{table1_aC}
}
\vspace*{0.5truecm}
\centerline{
\begin{tabular}{|c|c|c|c|c|c|c|c|}
\hline
Model& No. & Z=2 & Z=3 & Z=4 & $\langle Z \rangle$ & z$_d$(\AA) & $\rho(g/cm^3)$\\
\hline
\hline
e1T100 &   113 &  0.9 & 59.3 & 39.8 &  3.4 &  6.5 &  3.0 \\
 &   129 &  7.0 & 56.6 & 34.9 &  3.2 &  9.5 &  2.3 \\
\hline
e5T100 &   124 &  4.8 & 58.1 & 37.1 &  3.3 &  7.7 &  2.7 \\
 &   129 &  7.0 & 55.8 & 35.7 &  3.3 & 10.7 &  2.0 \\
\hline
e1T300 &   121 &  5.8 & 71.9 & 21.5 &  3.1 &  9.8 &  2.0 \\
 &   130 &  7.7 & 70.8 & 20.0 &  3.1 & 12.8 &  1.6 \\
\hline
e5T300 &    92 &  0.0 & 68.5 & 31.5 &  3.3 &  6.3 &  2.3 \\
 &   126 & 11.1 & 64.3 & 24.6 &  3.1 &  9.3 &  2.1 \\
\hline
e1T100L &   160 &  5.0 & 65.0 & 30.0 &  3.2 & 11.0 &  2.4 \\
 &   177 &  9.0 & 62.1 & 27.1 &  3.1 & 14.0 &  2.1 \\
\hline
e5T100L &   151 &  2.0 & 55.6 & 42.4 &  3.4 &  9.4 &  2.7 \\
 &   172 &  4.7 & 56.4 & 38.4 &  3.3 & 12.4 &  2.3 \\
\hline
\end{tabular}
}
\end{table}

\begin{figure}
\centerline{
\epsfig{file=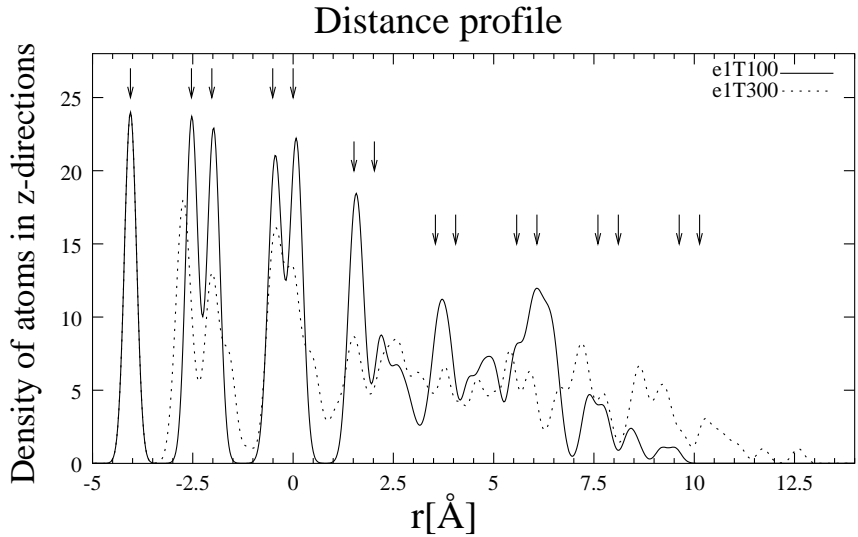,height=7.0truecm}
}
\medskip
\caption{
\small
Density profiles of two films, e1T100 (solid line) and e1T300
(dotted line), are displayed perpendicular to the substrate surface.
Arrows represent the layer positions in a perfect diamond crystal.
The difference in temperature between the two substrates causes a
difference between memory effects. 
}
\label{fig6_aC} 
\end{figure}

In Fig.\ \ref{fig6_aC} density profiles of two networks (e1T100 and
e1T300) are displayed perpendicular to the substrate surface.  The
arrows represent the layer positions in a perfect diamond crystal.
The difference in temperature between two substrates causes a
difference between memory effects.  The structure on the surface at
100~K has a more pronouncing layering effect than the other network on
substrate at room temperature.  In the first 3$-$4~\AA \ thick layers
over the [111] surface the $sp^3$ content is high due to the memory
effect.  In the rest of the bulk the $sp^2$ content is dominant.

\subsubsection{Radial distribution function}

\begin{figure}
\centerline{
\epsfig{file=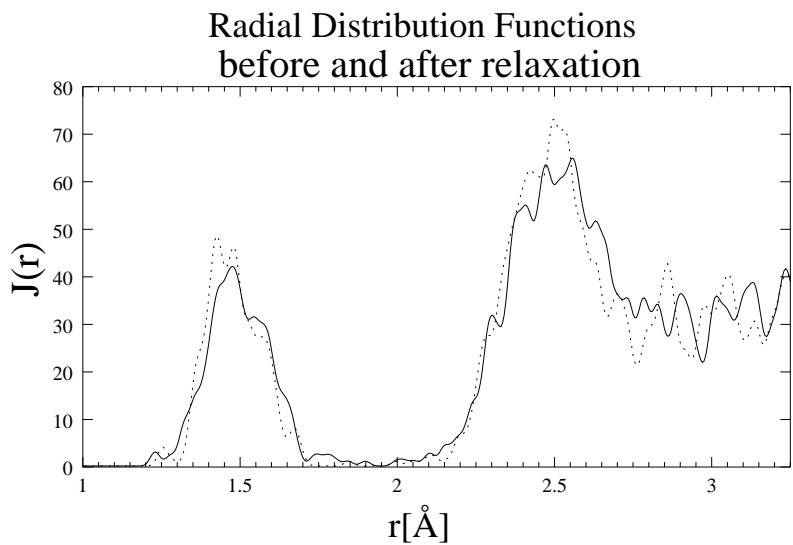,height=7.0truecm}
}
\medskip
\caption{
\small
Two radial distribution functions of model e1T100L are displayed here.
The solid line shows the first and second neighbors contributions to
RDF before relaxation, while the dashed line shows these
contributions after relaxation.  First and second neighbor peaks in
the final structure are a bit narrower.
}
\label{fig7_aC} 
\end{figure}

\begin{figure}
\centerline{
\epsfig{file=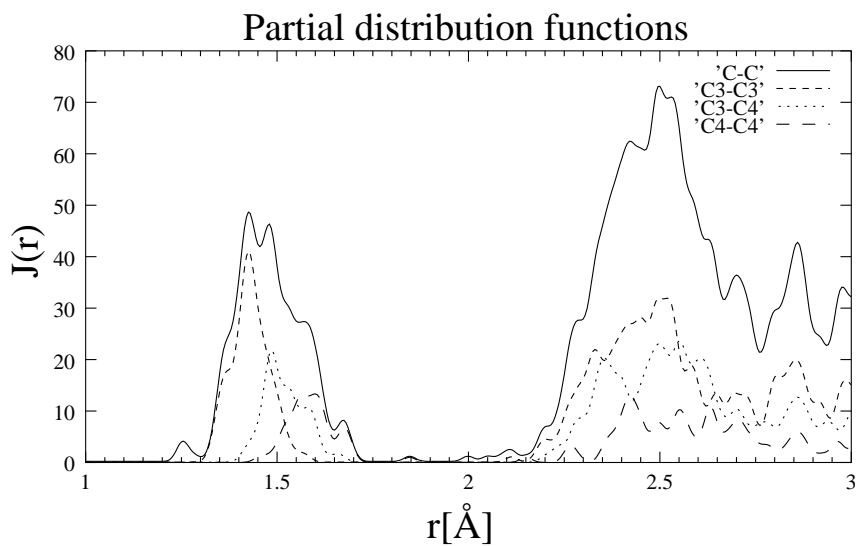,height=7.0truecm}
}
\medskip
\caption{
\small
The total~(C$-$C) and partial~(C3$-$C3, C3$-$C4 and C4$-$C4) distribution are
shown in e1T100L model.  The '3' and '4' denote the coordination number
three and four.  The average bond distances are 1.43 \AA, 1.52 \AA,
and 1.59 \AA \ for C3$-$C3, C3$-$C4 and C4$-$C4, respectively.
}
\label{fig8_aC} 
\end{figure}

The two radial distribution functions of model e1T100L are displayed
in Fig.\ \ref{fig7_aC}. The solid line shows the first and second
neighbor contributions to RDF before relaxation,  while the dashed
line belongs to these contributions after relaxation.  Peaks in the
final structure seem to be a bit narrower and in the interval of
1.6$-$2.2 \AA \ there are less bond lengths after relaxation.
Numerical analysis provides the same conclusion. In Fig.\
\ref{fig8_aC} partial distribution functions $J_{i,j}$ of e1T100L can
be seen, where $i$ and $j$ denote the coordination numbers of the
connected atoms.  Table\ \ref{table2_aC} contains the average bond
distances which are 1.43 \AA, 1.52 \AA, and 1.59 \AA, for $J_{3,3}$,
$J_{3,4}$ and $J_{4,4}$, respectively.

\begin{table}
\caption{
Average bond distance statistics
\label{table2_aC}
}
\vspace*{0.5truecm}
\centerline{
\begin{tabular}{|c|c|c|c|c|c|c|c|c|}
\hline
Model& \multicolumn{8}{c|}{C$-$C distance(\AA)}\\
\hline
& C$-$C & $\sigma_{C-C}$ & C3$-$C3 & $\sigma_{C3-C3}$ & C3$-$C4 & $\sigma_{C3-C4}$& C4$-$C4 & $\sigma_{C4-C4}$ \\
\hline
\hline
e1T100 &     1.51 &     0.09 &     1.41 &     0.04 &     1.53 &     0.05 &     1.60 &     0.07 \\
 &     1.50 &     0.10 &     1.41 &     0.04 &     1.53 &     0.05 &     1.60 &     0.07 \\
\hline
e5T100 &     1.50 &     0.09 &     1.44 &     0.05 &     1.53 &     0.05 &     1.58 &     0.06 \\
 &     1.50 &     0.09 &     1.44 &     0.05 &     1.53 &     0.05 &     1.58 &     0.06 \\
\hline
e1T300 &     1.48 &     0.09 &     1.43 &     0.05 &     1.54 &     0.06 &     1.58 &     0.09 \\
 &     1.48 &     0.09 &     1.43 &     0.05 &     1.55 &     0.06 &     1.58 &     0.09 \\
\hline
e5T300 &     1.50 &     0.09 &     1.43 &     0.05 &     1.54 &     0.06 &     1.58 &     0.06 \\
 &     1.48 &     0.09 &     1.43 &     0.05 &     1.54 &     0.06 &     1.58 &     0.06 \\
\hline
e1T100L &     1.49 &     0.09 &     1.43 &     0.05 &     1.52 &     0.05 &     1.59 &     0.07 \\
 &     1.49 &     0.09 &     1.43 &     0.05 &     1.52 &     0.05 &     1.59 &     0.07 \\
\hline
e5T100L &     1.52 &     0.09 &     1.44 &     0.06 &     1.54 &     0.06 &     1.59 &     0.06 \\
 &     1.51 &     0.09 &     1.44 &     0.05 &     1.54 &     0.06 &     1.59 &     0.06 \\
\hline
\end{tabular}
}
\end{table}

\subsubsection{Coordination number}

For bond counting we used a value of 1.85~\AA \  as an upper limit of
bond length inside the first coordination shell.  Table\
\ref{table1_aC} shows the percentage of different coordinations.
There is no fivefold atom in the structures. More than half of the
atoms have the $sp^2$ bonding configuration in the models as shown in
Table\ \ref{table1_aC}.  The average coordination numbers in the bulk
were slightly over the graphite coordination.  The dominating part of
fourfold coordinated atoms is near the substrate and the contribution
of twofold coordinated atoms is due to the atoms at the top of the
amorphous network (see Fig.\ \ref{fig1_aC} and\ \ref{fig2_aC}).  The
e5T100L model contains a onefold coordinated carbon atom at the end of
a chain. When we compare the average coordination numbers there are no
drastic differences among the simulated models.  A~MD simulation of
Kaukonen and Nieminen \cite{kaukonen_prl92}, using the classical
empirical potential of Tersoff, shows the same portion of $sp^2$ but
the $sp^1$:$sp^2$:$sp^3$ ratio is slightly different at T$_{sub}$ =
300~K.  At the E$_{beam}$ = 1~eV beam energy 5.8:71.9:21.5 is the
ratio in the e1T300 model, while it is 19.0:73.2:6.7 in their case.
Our sample has a higher portion of diamond-like atoms and less twofold
coordinated carbon atoms at the same condition.

\subsubsection{Bond angle distribution}

\begin{figure}
\centerline{
\epsfig{file=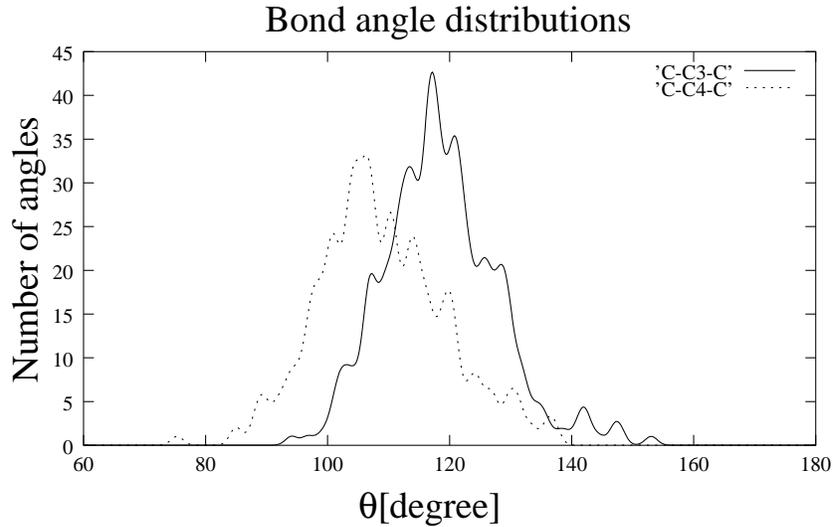,height=7.0truecm}
}
\medskip
\caption{
\small
The bond angle distribution of model e1T100L, where the central atoms
are threefold (C$-$C3$-$C) and fourfold coordinated (C$-$C4$-$C). 
}
\label{fig9_aC} 
\end{figure}

The angles can be analyzed in detail according to the coordination
numbers of the neighbors and the center (e.g.  C3$-$C4$-$C3,
C4$-$C4$-$C4, etc.) From diffraction data this causes more
difficulties. In Fig.\ \ref{fig9_aC} we display the bond angle
distribution of model e1T100L, where the central atoms  are threefold
(C$-$C3$-$C) and fourfold coordinated (C$-$C4$-$C).  In graphite-like
crystal and in diamond lattice these angles are 120$^\circ$ and
109.47$^\circ$, respectively.

The average values of partial bond angle distributions and deviations
are gathered in Table\ \ref{table3_aC} and Table\ \ref{table4_aC}.
Atoms with $sp^3$ local arrangements have nearly the same average
values in the C3$-$C4$-$C3, C3$-$C4$-$C4, and C4$-$C4$-$C4 cases, which are around
the tetrahedral angle.  Clear differences are observed for $sp^2$
configurations. C4$-$C3$-$C4 angles have much less average values than the
others. The latter average bond angles are close to 120$^\circ$.  It
means that threefold coordinated central atoms with at least two C3
first neighbors are always near planar structures even if the third
one is C4.  We have not investigated  dihedral angle distribution
because it is not so characteristic at a$-$C than at amorphous silicon
or germanium.

\begin{table}
\caption{
The average values of partial bond angle distributions and deviations.
\label{table3_aC}
}
\vspace*{0.5truecm}
\centerline{
\begin{tabular}{|c||c|c||c|c|c|c|c|c|}
\hline
Model&\multicolumn{8}{|c|}{C$-$C3$-$C}\\
\hline
& 
  CC$_3$C & $\sigma$ & 
  C$_3$C$_3$C$_3$ & $\sigma$ & 
  C$_3$C$_3$C$_4$ & $\sigma$ &
  C$_4$C$_3$C$_4$ & $\sigma$  \\
\hline
\hline
e1T100 &    118.1 &      9.8 &    118.3 &      9.6 &    118.3 &     10.2 &    113.5 &      7.3 \\
 &    118.1 &      9.5 &    118.3 &      9.3 &    118.3 &     10.1 &    113.5 &      7.3 \\
\hline
e5T100 &    118.0 &     10.5 &    117.7 &     11.1 &    119.9 &      9.9 &    110.8 &      8.0 \\
 &    118.0 &     10.5 &    117.7 &     11.1 &    119.9 &      9.9 &    110.8 &      8.0 \\
\hline
e1T300 &    118.3 &      9.5 &    119.4 &      9.4 &    116.4 &      9.0 &    113.9 &     11.1 \\
 &    118.3 &      9.6 &    119.3 &      9.6 &    116.4 &      9.0 &    113.9 &     11.1 \\
\hline
e5T300 &    118.1 &     10.0 &    119.5 &      9.3 &    118.0 &     10.4 &    109.9 &      7.3 \\
 &    118.2 &      9.4 &    119.3 &      8.9 &    117.8 &     10.2 &    110.0 &      7.1 \\
\hline
e1T100L &    118.6 &      9.4 &    119.1 &      9.2 &    118.1 &      9.9 &    114.9 &      9.9 \\
 &    118.6 &      9.4 &    119.0 &      9.2 &    118.1 &      9.9 &    114.9 &      9.9 \\
\hline
e5T100L &    117.9 &     10.4 &    120.2 &     10.3 &    115.9 &      9.1 &    112.7 &     14.7 \\
 &    117.8 &     10.2 &    119.5 &     10.2 &    115.7 &      9.1 &    112.7 &     14.7 \\
\hline
\multicolumn{9}{c}{}\\
\multicolumn{9}{c}{}\\
\hline
Model&\multicolumn{8}{|c|}{C$-$C4$-$C}\\
\hline
& CC$_4$C& $\sigma$ &
  C$_3$C$_4$C$_3$ & $\sigma$ & 
  C$_3$C$_4$C$_4$ & $\sigma$ &
  C$_4$C$_4$C$_4$ & $\sigma$ \\
\hline
\hline
e1T100 &    108.9 &     11.1 &    107.1 &      8.8 &    110.2 &     11.2 &    108.1 &     11.1\\
 &    108.9 &     11.1 &    107.1 &      8.8 &    110.2 &     11.2 &    108.1 &     11.1\\
\hline
e5T100 &    109.1 &      9.7 &    110.6 &      9.8 &    110.3 &     10.7 &    108.1 &      8.8\\
 &    109.1 &      9.7 &    110.6 &      9.8 &    110.3 &     10.7 &    108.1 &      8.8\\
\hline
e1T300 &    109.2 &      9.7 &    109.2 &     11.4 &    108.3 &      9.2 &    111.1 &      9.1\\
 &    109.2 &      9.7 &    109.2 &     11.4 &    108.3 &      9.2 &    111.1 &      9.1\\
\hline
e5T300 &    109.0 &     10.0 &    108.8 &      9.5 &    109.3 &     10.6 &    108.9 &      9.4\\
 &    109.1 &      9.8 &    109.0 &      9.3 &    109.1 &     10.4 &    108.8 &      9.3\\
\hline
e1T100L &    109.0 &     10.6 &    107.8 &     10.3 &    110.3 &     10.9 &    107.9 &     10.6\\
 &    109.0 &     10.6 &    107.8 &     10.3 &    110.3 &     10.9 &    107.9 &     10.6\\
\hline
e5T100L &    109.1 &      9.8 &    105.9 &     11.5 &    111.3 &      9.9 &    108.3 &      9.3\\
 &    109.1 &      9.8 &    106.2 &     11.7 &    111.3 &      9.9 &    108.3 &      9.3\\
\hline
\end{tabular}
}
\end{table}

\begin{table}
\caption{
The average values of partial bond angle distributions and deviations.
\label{table4_aC}
}
\vspace*{1.0truecm}
\centerline{
\begin{tabular}{|c||c|c||c|c|c|c|}
\hline
Model&\multicolumn{2}{|c||}{C$-$C$-$C}&\multicolumn{4}{|c|}{Ring statistics}\\
\hline
& C$-$C$-$C & $\sigma$& 4 & 5 & 6 & 7 \\ 
\hline
\hline
e1T100 &    112.8 &     11.5 &&&&\\
 &    113.5 &     12.2 & 1 &    21 &    40 &    37 \\
\hline
e5T100 &    113.3 &     11.4 &&&&\\
 &    113.6 &     12.1 &0 &    26 &    43 &    42\\
\hline
e1T300 &    115.3 &     11.3 &&&&\\
 &    115.6 &     11.7 &     0 &    25 &    21 &    17\\
\hline
e5T300 &    113.8 &     10.9 &&&&\\
 &    115.2 &     12.2 &     0 &    18 &    32 &    20 \\
\hline
e1T100L &    114.1 &     11.2 &&&&\\
 &    114.6 &     12.1 &     1 &    31 &    50 &    40 \\
\hline
e5T100L &    112.6 &     10.9 &&&&\\
 &    113.0 &     11.2 &     0 &    44 &    54 &    58\\
\hline
\end{tabular}
}
\end{table}

\subsubsection{Ring statistics}

In our analysis the definition of the ring is a closed path, which
starts from a given atom walking only on the first neighbor bonds.
Every atom is visited only once.  The size of the ring is the number
of atoms forming the closed path. The ring statistics for our models
are displayed in Table\ \ref{table4_aC}.  A fourfold ring was found in
the sample made with $E_{beam}$ = 1~eV at T$_{sub}$ = 100 K substrate
temperature but there was no threefold ring in our simulation. In the
graphitic and diamond network, sixfold, eightfold, etc. rings are
present.  In our models even-numbered as well as odd-membered rings
can be found. It seems to be clear from the Table\ \ref{table4_aC}
that lower beam energy results in less 4$-$7 membered rings than at a
higher bombarding energy.

\subsection{Summary}

In summary, low energy molecular dynamics simulations of atomic beam
growth on a diamond [111] surface were carried out using two different
average carbon beam energies E$_{beam}$ = 1 and 5~eV and two different
substrate temperatures T$_{sub}$ = 100 and 300~K during 30 ps.  To
obtain larger structures the simulations were prolonged by 15 ps at
T$_{sub}$ = 100~K.  Six networks were prepared by atom deposition with
periodic boundary conditions in two dimensions and the most important
structural parameters were compared. Our aim was to construct large
scale amorphous carbon models of over 100 atoms by quantum-mechanical
treatment.  The atomic interactions were described by a well-tested
tight-binding potential.  The structures were grown by atom-by-atom
deposition onto substrates. The growing process was described as in
real experiments without any artificial model of energy dissipation
\cite{kaukonen_prl92, kohary_jncs00}, however, the deposition rate was
much higher than is usually applied in experiments. Other studies on
the growth process use similar rates. On the basis of our time
development investigation we expect that much lower deposition rates do
not cause a significant difference in our models. The influence of our
lower substrate temperature is that the energy dissipation is faster
and this slightly compensates for the high deposition rate.
Unfortunately, it is clear that real experimental rates cannot be
implemented in computer simulation in the near future, even if less
time-consuming classical empirical potentials are employed.

\chapter{Atomistic simulation of the bombardment process during the
BEN phase of diamond CVD}
\markboth{\cxhy \ BOMBARDMENT ON AMORPHOUS CARBON}{\cxhy \ BOMBARDMENT ON AMORPHOUS CARBON} 
\label{CxHy}

In this chapter the detailed examination of \cxhy \ projectile
bombardments onto hydrogenated amorphous carbon~(a$-$C:H) is studied by
means of computer simulations. This is the first work addressing
these kind of processes with a reasonable quantum mechanical
treatment.  We investigated the conditions of hydrocarbon ion
bombardments in an amorphous carbon layer. This involves the
penetration depth  study of different ion species by different
bombarding energies. Special attention was paid to the critical energy
needed for penetration and to the structural rearrangements in the
amorphous carbon layer caused by the bombardment. We present here the
first simulation with realistic conditions for bias enhanced
nucleation microwave plasma assisted chemical vapor deposition of
diamond used by K\'atai \etal \cite{katai_jap99_CxHy,
katai_drm00_CxHy, katai_dis_CxHy}.

\section{Introduction}

\subsection{The bias enhanced nucleation CVD growth of diamond}

Chemical vapor deposition~(CVD) techniques are frequently used to
prepare different (hydrogenated) amorphous carbon films, as well as
diamond layers. Bias enhanced nucleation~(BEN) is used to achieve a
higher nucleation density in the latter case.  It can be assumed, that
during the deposition, the different projectile ions and their kinetic
energy with which they arrive at the surface, play an important role.
The identification of the most abundant ionic species and the
measurement of their energy distributions under typical microwave
plasma enhanced CVD diamond nucleation conditions are still an
important issue to be addressed if one attempts to validate any model
of the nucleation process. K\'atai \etal reported a detailed mass
selective energy analysis of the incoming ions during BEN nucleation
of chemical vapor deposition of diamond \cite{katai_jap99_CxHy,
katai_drm00_CxHy, katai_dis_CxHy}.  They measured the different ion
energies and fluxes as a function of bias voltage at 25~mbar and
800~$^\circ$C.  The total flux of hydrogen species~(about
0.17$\times$10$^{16}$~cm$^{-2}$s$^{-1}$) was considerably less than
for hydrocarbons~(around 1.1$\times$10$^{16}$~cm$^{-2}$s$^{-1}$). At
zero bias voltage ions have energies between 5 and 10~eV and the
average ion energy increases for most species with increasing bias
voltage in a linear way in the low bias regime. The overall ion flux,
which is integrated over the energy distribution and summed for all
ion species, increases with ascendant bias voltage almost
exponentially.  However, the fluxes of several species show a maximum
versus bias voltage between 0 and 200~V. It was shown that about 85\%
of the total hydrocarbon flux is determined by C$^+$, CH$^+$,
CH$_2^+$, CH$_3^+$, C$_2$H$_2^+$, C$_2$H$_3^+$ species, with average
kinetic energies between 50 and 70~eV. The full width at half of the
maximum energy distributions were between 20 and 30~eV. It was
manifested that the lower limit for nucleation is around 100~V bias
voltage, and the optimum is around 200~V.  It was observed that among
C$_1$H$_y$ (y = 1, 2, 3) projectiles, the flux versus energy of
CH$_3^+$ distribution is broader and shifted to lower energies with a
total flux only marginally lower than those of CH$_{0-2}^+$. The
C$_2$H$_2^+$ had the highest total flux of all hydrocarbon species.
Finally, the above mentioned experiment supported the subplantation
model proposed by Lifshitz \etal \cite{lifshitz_prl89_CxHy,
lifshitz_prb90_CxHy}.

\subsection{Proposed model for the formation of diamond nuclei in the BEN
process}

A speculative model to explain the formation of the heteroepitaxial
alignment of diamond nuclei with the underlying substrate in the BEN
process was proposed on the basis of results from {\it in situ} plasma
analytical measurements, the results from surface analytical
measurements and from literary data \cite{katai_dis_CxHy}.  Here we
will give just a short summary of this model.

In the initial phase of BEN a polycrystalline $\beta$-SiC 3$-$4~nm
thick layer is formed. It is epitaxially oriented with respect to the
underlying silicon substrate. This silicon-carbon layer is not always
observed in the experiments. However, from the point of view of the
proposed model, its presence or absence is of no importance.  In the
next step of a typical BEN process, an amorphous carbon layer forms on
the surface of this silicon-carbon layer. The thickness of the
amorphous carbon layer is controlled by the balance between
simultaneous growth and the etching of the hydrocarbon and the
hydrogen species.  The thickness of this layer was found to be around
5$-$8~\AA.  

The penetration depth of ions below the surface during the BEN process
depends on their energy. Uhlmann {\it et al.} \cite{uhlmann_diss_CxHy,
uhlmann_prl98_CxHy} executed molecular dynamics simulations to study
the penetration of carbon atoms in diamond like amorphous carbon
films, which are frequently referred to as tetrahedrally bonded amorphous
carbon~(ta$-$C).  It was shown for carbon atoms having kinetic energies
bigger than 30~eV that the deposition is a subplantation process, where the
carbon atom species penetrate and occupy subsurface positions. The
penetration depth was found to be energy dependent.  The results of
these computer simulations permit the assumption that bombarding
hydrocarbon ion species having a larger kinetic energy than 30$-$40~eV
in BEN, penetrate 4$-$8~\AA \ below the amorphous carbon surface. 
Consequently, hydrocarbon species penetrate into the vicinity of the
epitaxially oriented silicon-carbon layer (or the silicon surface),
but they do not damage its structure, as the amorphous carbon
layer is sufficiently thin to give protection.  However, a large
fraction of hydrocarbon ions appears close to the interface of
silicon-carbon and the a$-$C layer, increasing the local density in that
region of the amorphous carbon layer. 

Following this, the formation of diamond nuclei starts. The proposed
model for the BEN process did not give a qualitative explanation of
this mechanism, but two main points were emphasized. The diamond
nucleation occurs close to the interface of SiC layer and a$-$C layer,
and the silicon-carbon layer is protected from ion damage, thus it can
provide orientation information. After the nuclei is formed, the
kinetic energies for almost all of the ions are below the displacement
energy of diamond~(80~eV) and they do not damage the nuclei.

In this chapter, we concentrate on the incubation period of nucleation. We
study the penetration characteristics of hydrocarbon ion species by
means of computer simulations. For this, molecular dynamics
simulations were performed by a reasonable quantum mechanical treatment.
First, a realistic surface model of hydrogenated amorphous
carbon~(a$-$C:H) film is prepared. Then the bombardments of different
hydrocarbon ions is studied by different energies. This involves the
penetration depth study of different ion species by different
bombarding energies.  Special attention was paid to the critical
energy needed for penetration and to the possible description of the
formation of diamond nuclei.  Finally, we present here the first
simulation with realistic conditions for BEN as used in experiments.

\section{Surface model of hydrogenated amorphous carbon}

\subsection{Generation of the surface model}

The surface for our molecular dynamics~(MD) simulations was generated
from a hydrogenated amorphous carbon bulk supercell with an original
density of 2.2~g/cm$^3$. This structure was prepared by long time
annealing of randomly distributed 128 carbon and 69 hydrogen atoms
\cite{frauenheim_prb93_CxHy, frauenheim_prb94_CxHy, jungnickel_prb94_CxHy}.  The
final arrangement of this model has 1.6\%, 53.1\%, 2.3\% and 41.4\%
fraction of $sp^1$, $sp^2$, $sp^{2+x}$ and $sp^3$ bonded carbon
{atoms, respectively.%
\renewcommand {\thefootnote} {\fnsymbol{footnote}}%
\footnote[2]{The introduction of $sp^{2+x}$
hybrid denotes a carbon atom with a coordination equal to three, but the
geometry of the bond having a similar structure to the $sp^3$ hybrid with
one "dangling bond" in the fourth direction.}}  
The average first neighbor coordination number is 2.88 in the cubic
cell with an edge equal to 10.66~\AA.  For bond counting we used the
value of 1.85~\AA \  as an upper limit of bond length inside the first
coordination shell.  We followed previous works \cite{dong_prb98_CxHy,
uhlmann_prl98_CxHy} to prepare a surface for \cxhy \ ion bombardments.
First, this structure was doubled in one direction, which we refer to
as the $z$-axis. Then to model the surface, we broke the periodic
continuation along the $z$ direction to transform the periodically
extended cube into an infinite slab with two free surfaces.  One
labeled as the "top" surface and the other as the "bottom" surface.
We then cut the sample along the $z$-axis, 15~\AA \ below the top
layer, and the atoms below this were removed from the sample.  In
other words, we had a new bottom surface. Henceforth, it shall be
referred to as the bottom surface.  The cut of periodic continuation
leads to dangling bonds on the surfaces.  We chose the top surface as
the target for bombardments. For the determination of the low energy
structure of this film we undertook the following.  The broken
$\sigma$ bonds at the bottom were saturated by hydrogen atoms to
retain the electronic structure there.  To achieve local rearrangement
on the top surface~(similar to the surface reconstruction in
crystals), the sample was treated under an annealing process at 2000~K
over about 1~ps, with time step $\Delta$t equal to 41~a.u. ($\approx$
1.0~fs).  This equilibrated structure was then subsequently cooled
down to 30~K within about 1~ps.  Then the sample relaxed with
conjugent gradient method.  Finally, this low energy structure was
annealed at 1000~K for about 1~ps to prepare the substrate for \cxhy \
projectile bombardment. Our purpose was to do simulations with realistic
conditions for BEN experiments where the temperature is around
800~$^\circ$C.

\begin{figure}
\centerline{
\epsfig{file=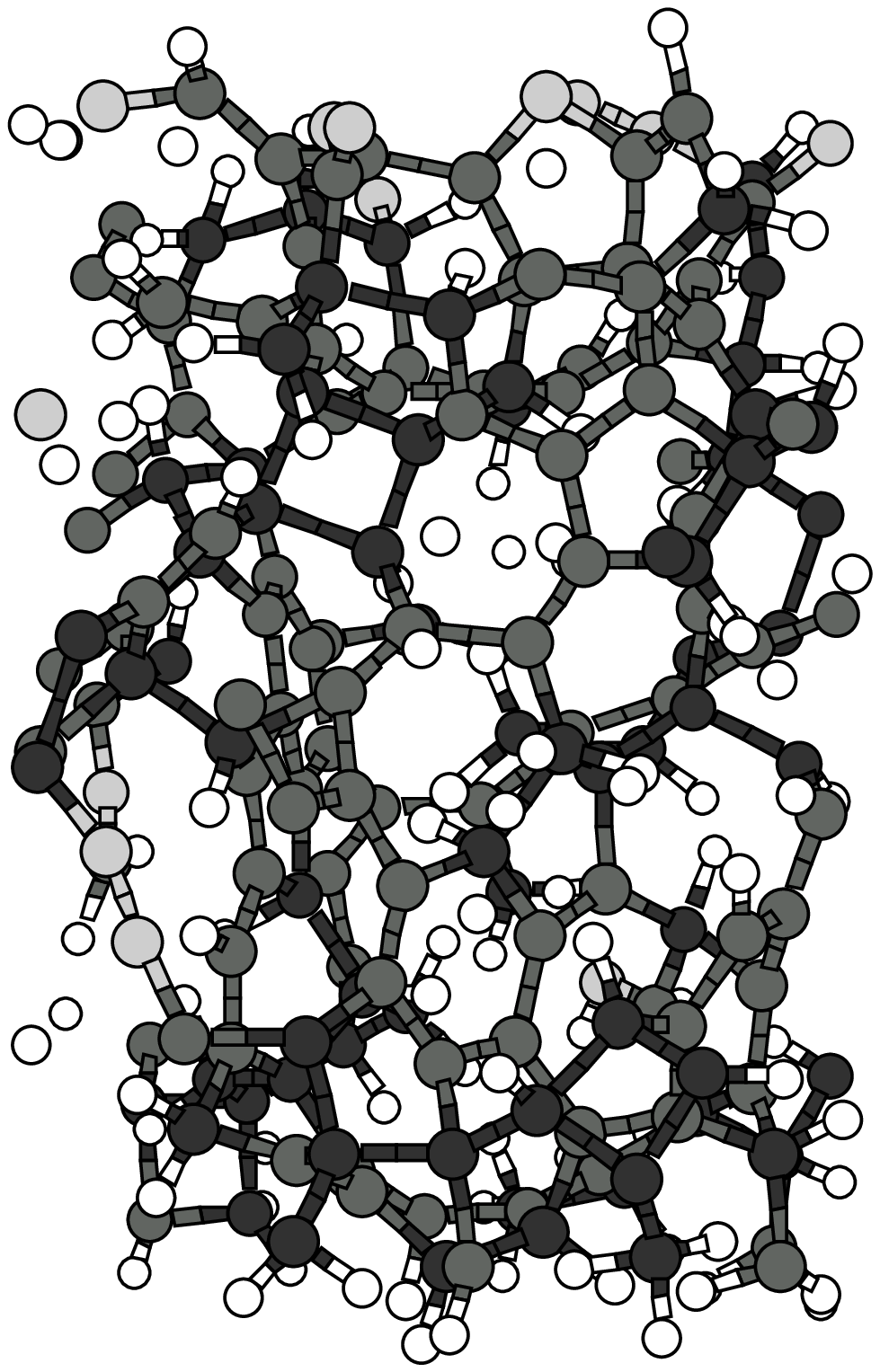,height=7.0truecm}
\hspace*{1.0truecm}
\epsfig{file=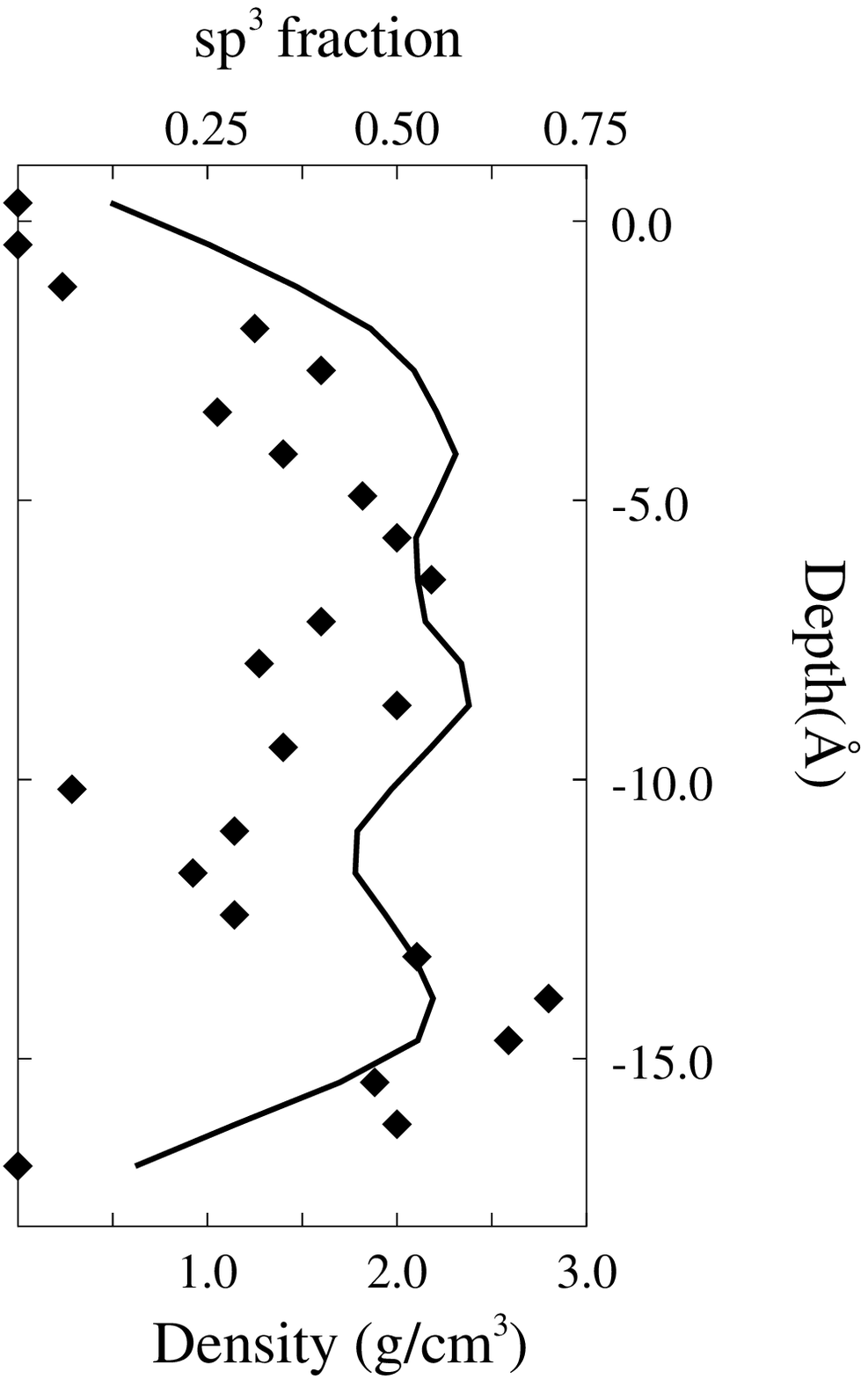,height=7.0truecm}
}
\medskip
\caption{
\small
The substrate structure: $sp^3$ fraction, and mass density profile.
The black atoms show carbon atoms with a coordination equal to
four~(Z~= 4).  Grey atoms have Z = 3, while light grey atoms have
lower coordinations. The white circles show hydrogen atoms.  The color
version of the structure can be found at
http://www.physik.uni$-$marburg.de/$\sim$kohary/CxHybomb.html. The solid line in
the figure on the right shows the density of the film and the filled
diamonds show the $sp^3$ fraction.
}
\label{fig1_CxHy}
\end{figure}

\subsection{Properties of the model}

The final structure of our surface a$-$C:H film contained 182 carbon and
107 hydrogen atoms, 16 of the latter replace the broken $\sigma$
bonds at the bottom. We chose a 11.1~\AA \ thick moveable part
from the top, which contained 190 atoms. It is sufficient to model the
bombardment experiments because the deepest penetration length is estimated
to be around 10~\AA \ for the projectiles having kinetic energy lower
than 100~eV.  The rest of the slab at the bottom, with a thickness
4~\AA, were fixed in all simulations.  The density of the sample was
1.98~g/cm$^3$.  The bulk density was equal to $\rho_{bulk}$ =
2.05~g/cm$^3$. We will refer to the bulk, as the structure 3~\AA \ below
the highest $z$ coordinated atom and we will refer to the surface, as the
structure above the bulk.  We used periodic boundary conditions in
plane, with the original cell size equal to 10.66~\AA.  The bonding
arrangement contained 6.6\%, 48.4\%, 5.5\% and 38.5\% fraction of
$sp^1$, $sp^2$, $sp^{2+x}$ and $sp^3$ bonded carbon atoms,
respectively.  The mass profile in Fig.\ \ref{fig1_CxHy} shows that
there is not an important deviation from the average density, the
minimum is about 1.8~g/cm$^3$ at 11~\AA \ and the maximum is about
2.4~g/cm$^3$ at 8~\AA \ below the surface.  

\begin{figure}
\centerline{\epsfig{file=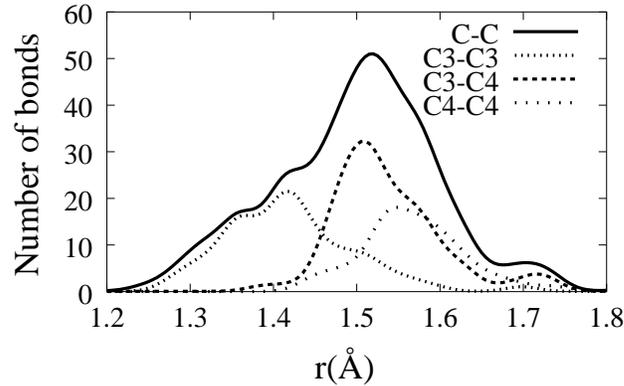,height=5.0truecm}}
\medskip
\caption{
\small
Total~(C$-$C) and partial~(C$i$$-$C$j$) bond length distribution for the
substrate film.  (The $i$ and $j$ denote the coordination of the carbon
atoms.) 
}
\label{fig2_CxHy}
\end{figure}

In Fig.\ \ref{fig2_CxHy} the total and partial bond length
distributions can be seen. The partial distributions are centered at
1.42, 1.54, 1.57~\AA, for C3$-$C3, C3$-$C4 and C4$-$C4 respectively.
(C$i$$-$C$j$ means the distribution of distances between carbon atoms
with coordination number $i$ and $j$). The bond length distributions
for the first neighbor distance show wide fluctuations around the
crystalline graphite~(1.42~\AA) and diamond~(1.54~\AA) bond lengths.

\section{\cxhy \ projectile bombardments. The simulation details }

In our work we aimed to study the ion bombardment in an $sp^2$-rich
a$-$C:H structure by means of computer simulations
\cite{kohary_tobepub_CxHy}.  We were interested {\it(a)}~in the
determination of average penetration depths, {\it(b)}~in the
examination into the possible decay of the projectiles, and {\it(c)}~in
the structural change of the target substrate.  We studied the effects
of the bombardment by different bombarding energies and atomic
contents of projectiles.  The experimental results showed that the
most important ions during the bias enhanced nucleation~(BEN) are
C$^+$, CH$^+$, CH$_2^+$, CH$_3^+$, C$_2$H$^+$, C$_2$H$_2^+$ and
C$_2$H$_3^+$ \cite{katai_jap99_CxHy}. In fact, C$_2$H$_2^+$ has the
highest total flux among hydrocarbon ions, with two or three times
larger flux compared to any hydrocarbon and carbon ions. 

We report here, the incubation period of the bias enhanced microwave
plasma assisted CVD~(BEN MW$-$CVD) phase of diamond nucleation growth.
We studied the ion bombardment with two relevant projectiles:
acetylene~(\ctht) and methyl~(\cht). The interaction between the atoms
was described by the density functional tight-binding~(DFTB) method
\cite{elstner_prb98_CxHy, frauenheim_pssb00_CxHy}, which was effective
to describe the ground state geometries and physical properties of
different carbon systems \cite{koehler_prb95_CxHy,
frauenheim_jncs95_CxHy} and was also applied in the previous carbon
bombardment simulations \cite{uhlmann_diss_CxHy, uhlmann_prl98_CxHy}.
In the work of Uhlmann \etal  the effect of neutral carbon atoms on
a$-$C substrate was studied \cite{uhlmann_prl98_CxHy}.  However, in
those simulations the conditions were different to ours.  They
simulated the carbon atom bombardment which resulted in creating a
diamond like amorphous carbon~(DLC or ta$-$C), which has a high
percentage of $sp^3$ carbon atoms~(even near to 100\%). We were
however interested in finding a realistic simulation for BEN MW$-$CVD
nucleation of diamond and checking the validity of the model for the
description of heteroepitaxial alignment for diamond nuclei in the BEN
process, as proposed by K\'atai \etal \cite{katai_dis_CxHy}.

\begin{figure}
\centerline{
(a)
\epsfig{file=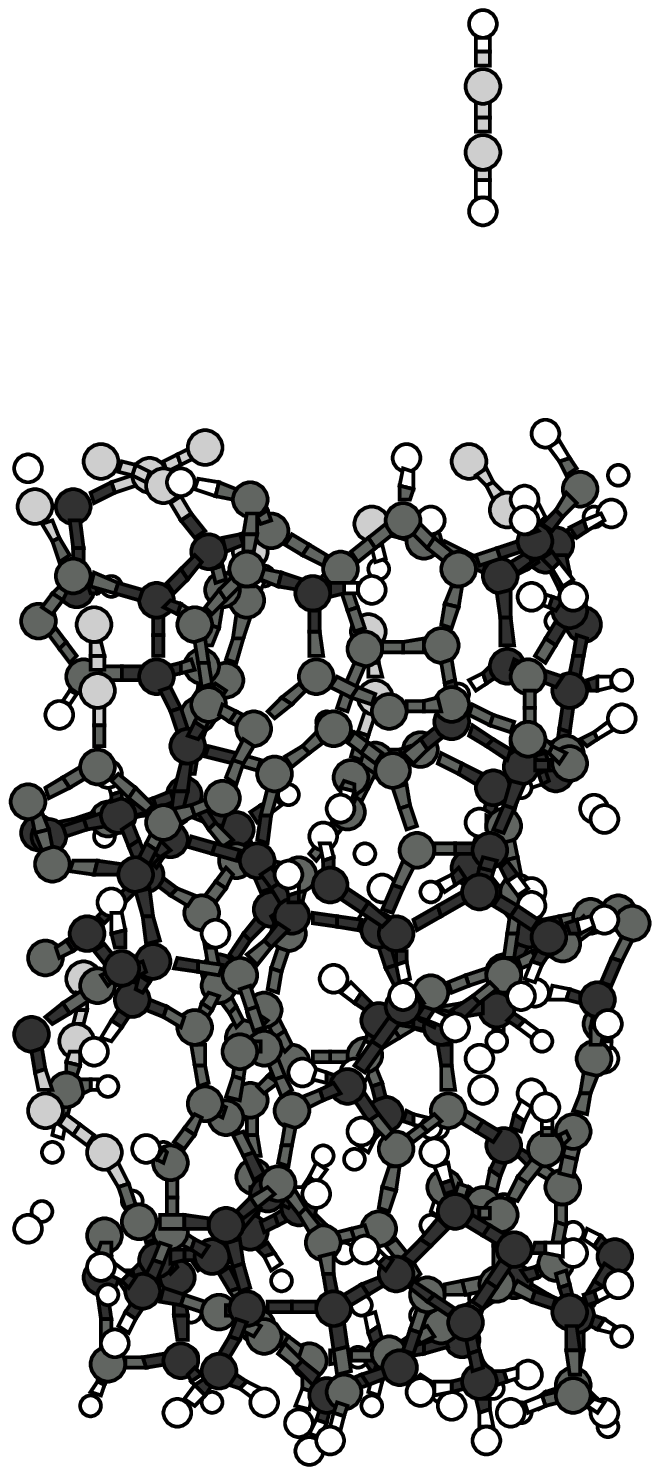,height=9.0truecm}
\hspace*{1.0truecm}
(b)
\epsfig{file=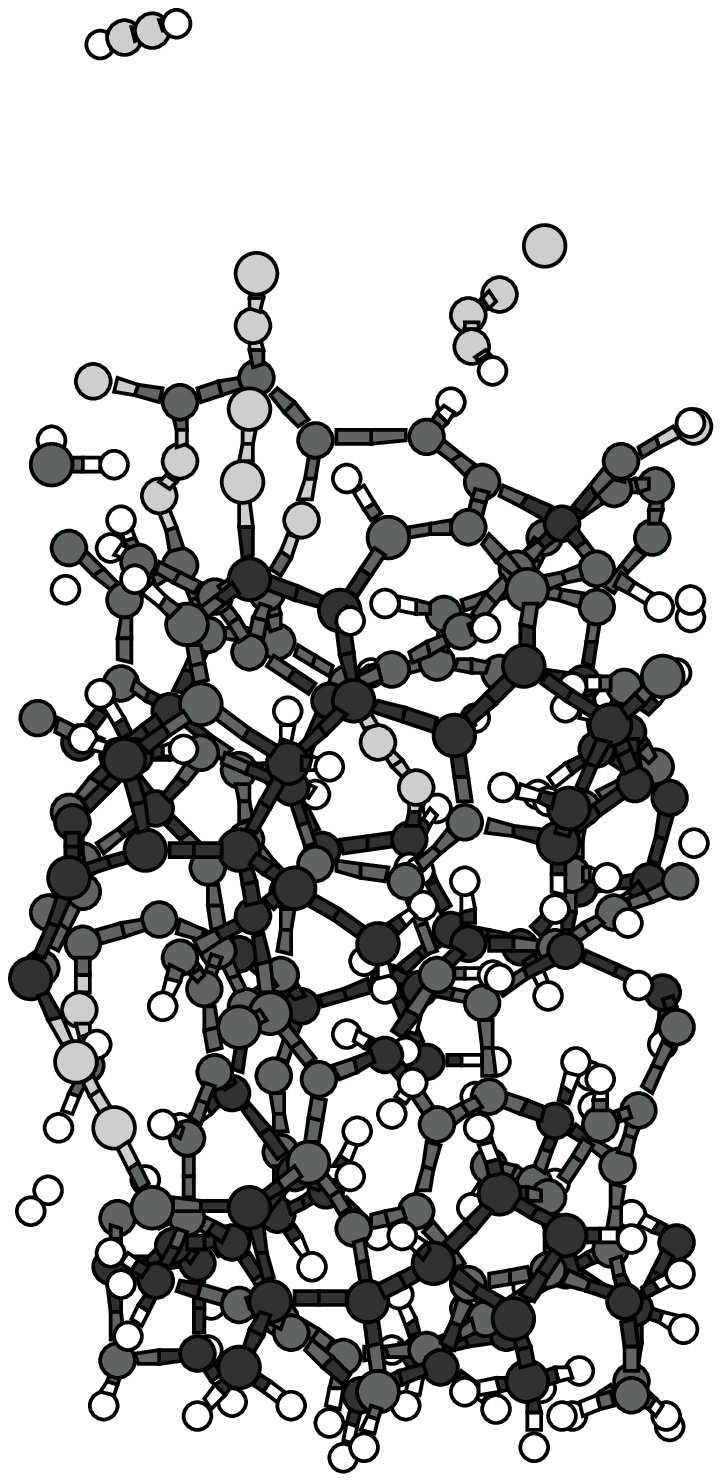,height=9.0truecm}
}
\vspace*{1.0truecm}
\centerline{
(c)
\epsfig{file=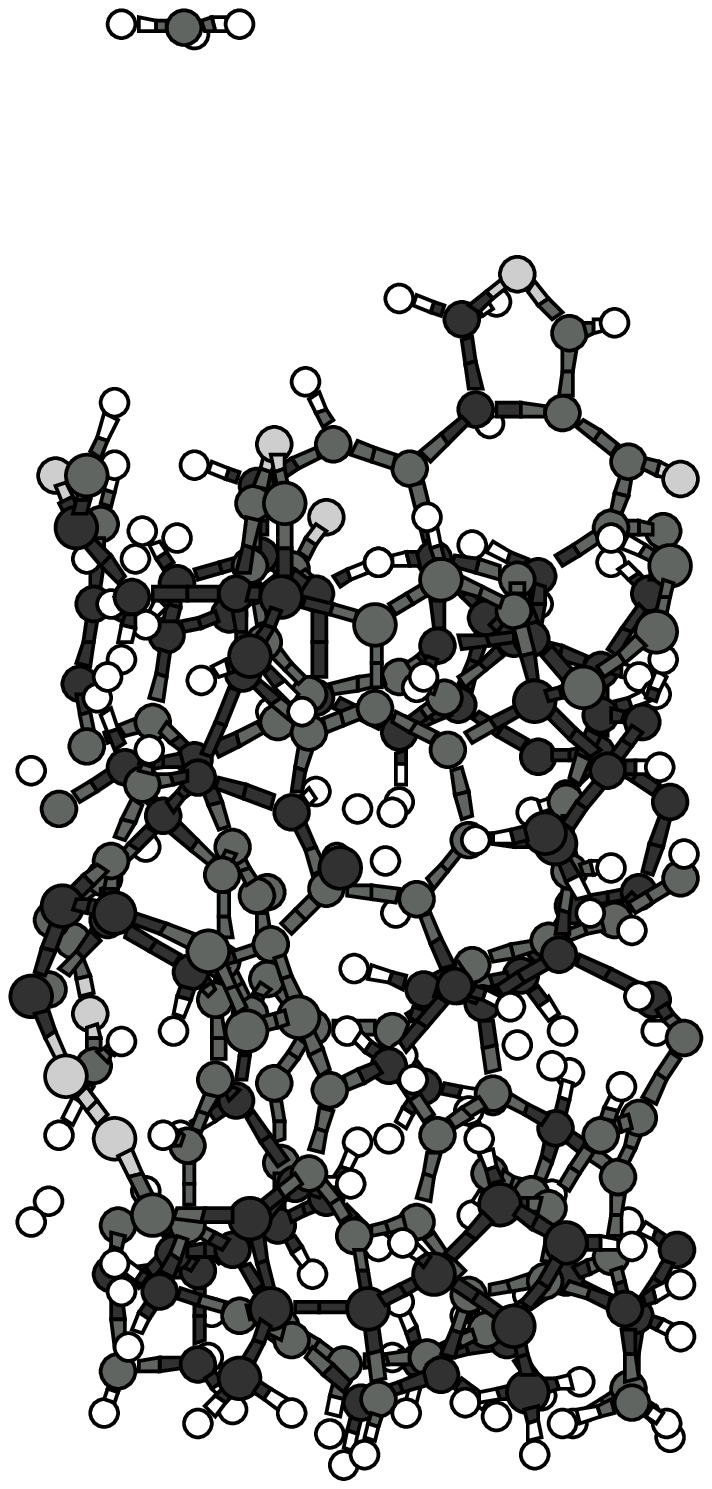,height=9.0truecm}
\hspace*{1.0truecm}
(d)
\epsfig{file=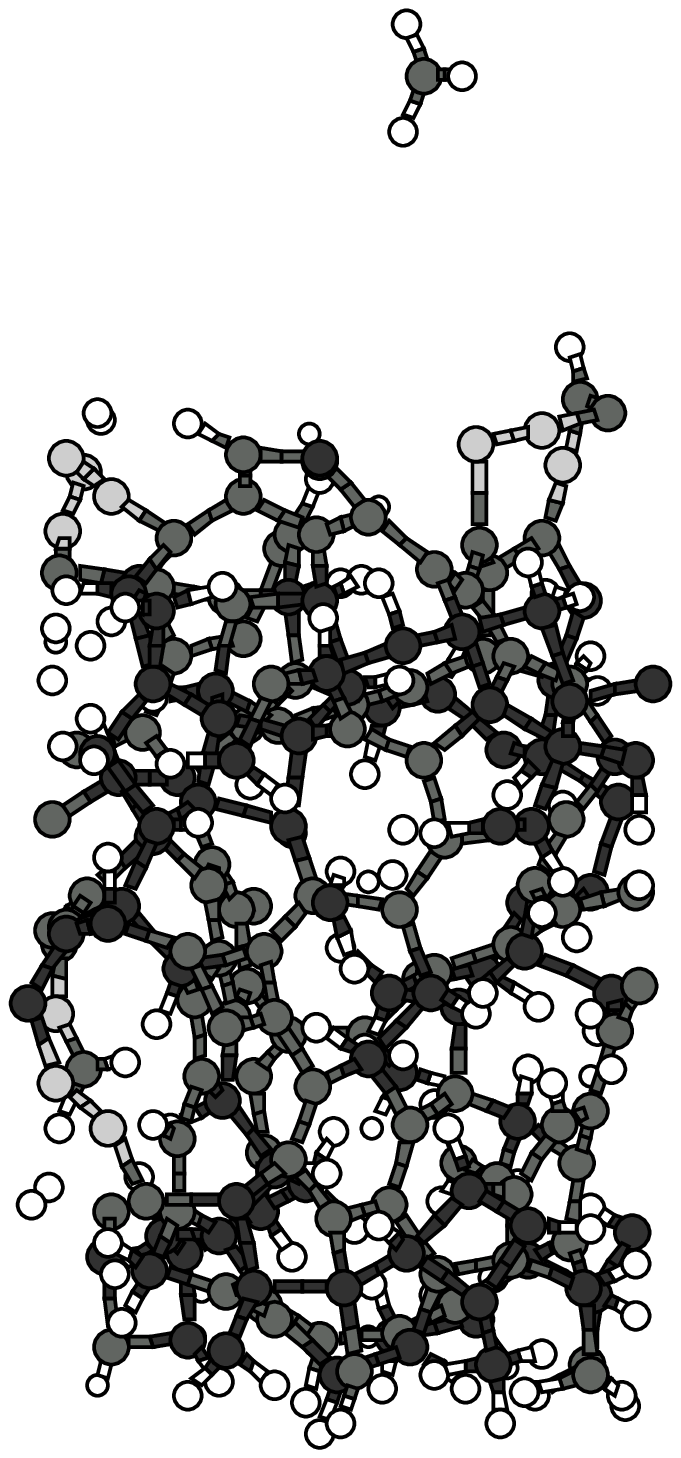,height=9.0truecm}
}
\medskip
\caption{
\small
The four different projectile orientations. The names are 
(a) POINT, (b) LINE, (c) PLANE, (d) EDGE. We used the same
coordination scheme as in Fig.\ \ref{fig1_CxHy}.
For a color version see:
http://www.physik.uni$-$marburg.de/$\sim$kohary/CxHybomb.html
}
\label{fig3_CxHy}
\end{figure}

\begin{figure}
\centerline{\epsfig{file=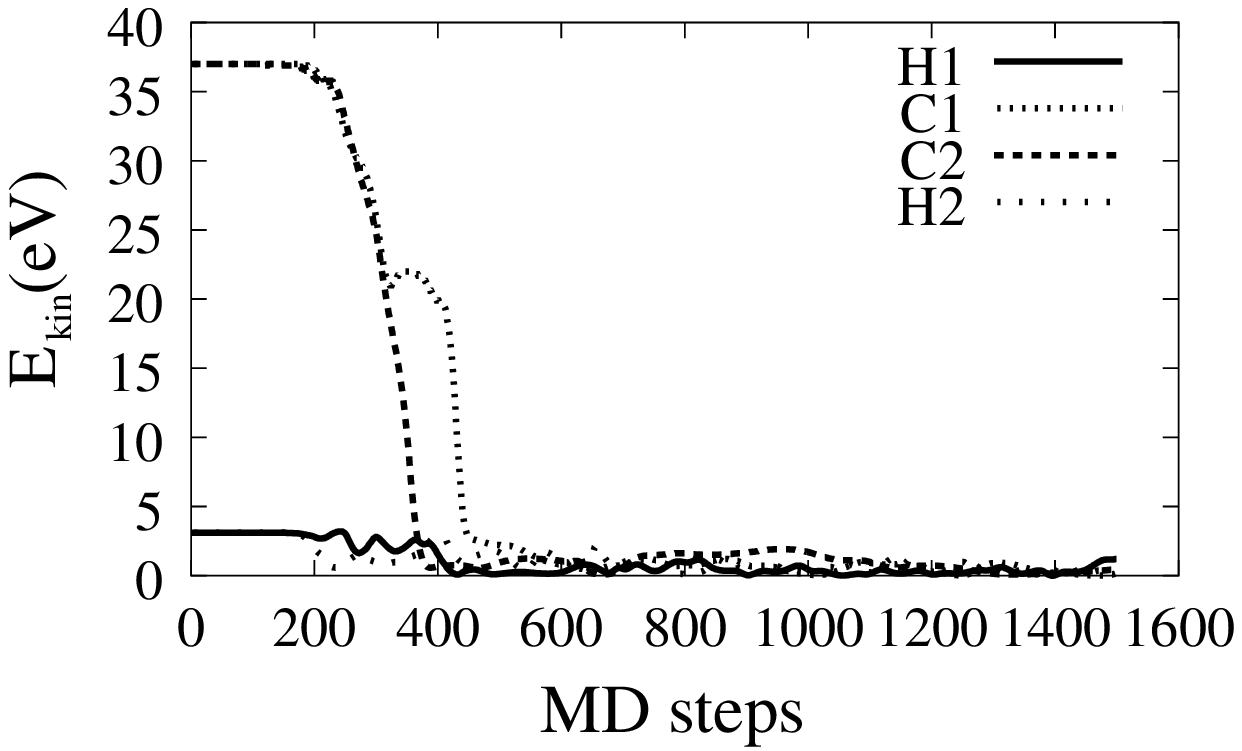,width=7.5truecm}
\epsfig{file=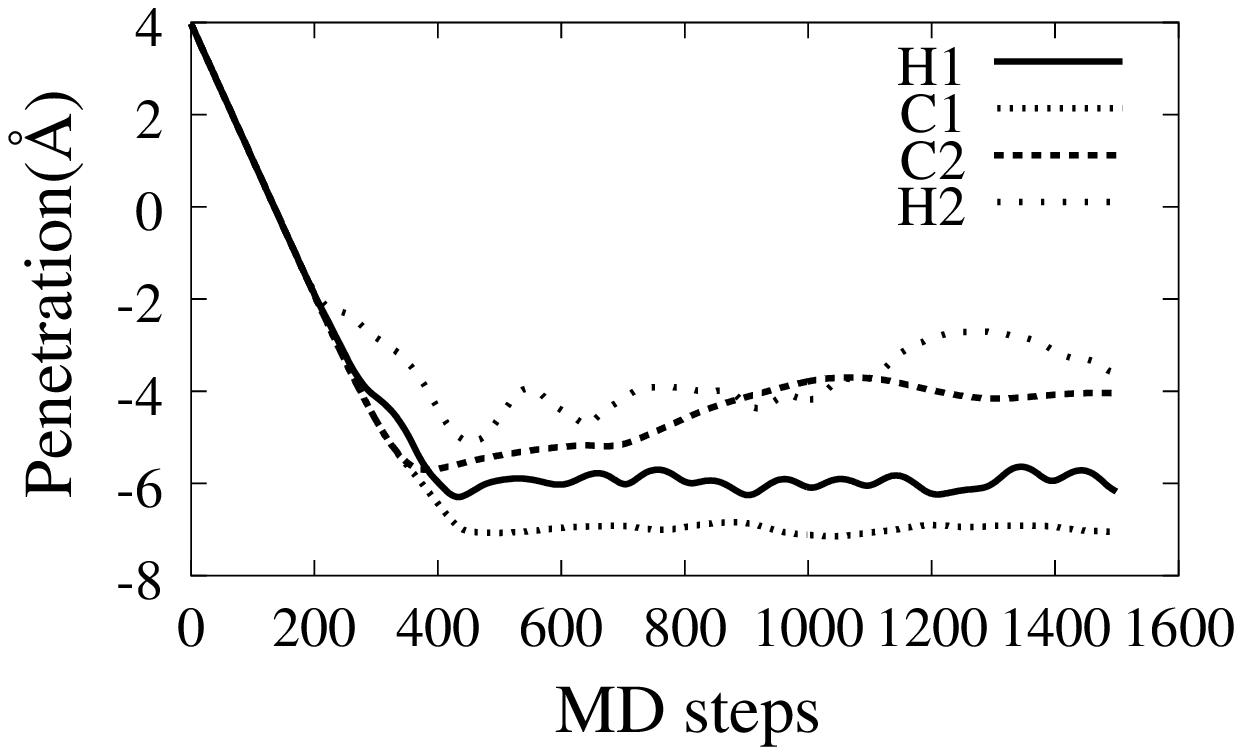,width=7.5truecm}}
\medskip
\caption{
\small
In the figure on the left, the kinetic energy versus time for the LINE
projectile starting with 80~eV kinetic energy is shown. The carbon
atoms have initially approximately 37~eV kinetic energy, while
hydrogens have much lower energy.  One MD step is equal to 5~a.u. in
this case.  In the figure on the right, the penetration of the atoms
of the projectile is shown. The surface of the film is at 0~\AA.
}
\label{fig4_CxHy}
\end{figure}

The simulation details are the following. In the first {\it collision
stage}, which describes the penetration and the energy loss of the
species, a projectile with given kinetic energy was directed towards
the substrate.  Random numbers were used for determination of the
coordinates of the mass center and for the local configurations of the
species.  The lowest atom of the projectile was placed consistently
4~\AA \ over the substrate, where the interatomic potential almost
vanishes.  Simulations with three different kinetic energies were
done~(20, 40 and 80~eV) for \ctht \ and with 20 and 40~eV for \cht \
to study the energy dependences of ion bombardment. The choice of
these two projectiles and the kinetic energies were governed by
experimental results \cite{katai_jap99_CxHy, katai_drm00_CxHy,
katai_dis_CxHy}. The C$_2$H$_2^+$ projectile has the largest flux
among hydrocarbon species.  It can be predicted from previous single
carbon projectile bombardments that 30$-$40~eV kinetic energy will be
the critical energy for penetration of different ion species under the
surface.  We have chosen 20, 40 and 80~eV kinetic energies according
to the above assumptions. The 20~eV energy bombardment describes the
effects of low energy projectiles, whose energy is much lower than the
energy needed for nucleation processes. The nucleation starts at 100~V
bias voltage.  The \ctht \ projectiles have approximately 40~eV and
\cht \ ion species have around 35~eV kinetic energy in this case.
At~200~V bias, at the optimal voltage for nucleation, the \cht \
projectiles still have kinetic energy around 40~eV, but the \ctht \
ions species, which possess the highest flux amongst all hydrocarbon
ions, have kinetic energy close to 80~eV \cite{katai_jap99_CxHy}.
This energy is the displacement energy in diamond as well.  From
C$_1$H$_y^+$ projectiles \cht \ was chosen because its flux is just
slightly lower than those of C$_1$H$_{0-2}^+$. Furthermore,
C$_1$H$_y^+$ projectiles with low hydrogen content (y~$\le$~2) will
warrant similar effects to single carbon bombardment, which was
simulated earlier, but at different conditions
\cite{uhlmann_prl98_CxHy}.  The collision stage was simulated for
7500~a.u.  ($\approx$ 182~fs).  During this period the kinetic energy
of the substrate was rescaled according to 1000~K by a velocity
rescaling method which simulates the constant temperature in the
sample. For the different kinetic energy ion implantations, different
time steps were used: $\Delta$t equal to 15, 10 and 5 a.u., for
E$_{kin}$ = 20, 40 and 80~eV, respectively.  We studied four
independent orientations of the bombarding species as it is shown in
Fig. \ref{fig3_CxHy}.  The first orientation of \ctht \ is called
POINT, as the top view shows the projectile as a point.  The second
orientation of \ctht \ is called LINE.  In the case of \cht \
projectile, two different orientations were called PLANE and EDGE,
according their orientation towards the surface at the start of the
bombardments.  The reason for choosing these orientations is because
they are the extreme cases. In experiments projectile orientations are
a mixture of these extremes.  In Fig.\ \ref{fig4_CxHy} the time
dependence of the kinetic energy of the POINT projectile starting with
80~eV can be seen.  The carbon atoms have approximately 37~eV kinetic
energy, which decreases to 1$-$5~eV after 400 MD steps, when the
species arrives to the maximal penetration depth in the film.  The
time step for this case was 5~a.u.~($\approx$ 0.125~fs). The hydrogen
atoms already have low kinetic energy at the start~(t = 0).

In the second {\it thermalization stage}, the local structure relaxes
and the ion energy dissipates into the lattice. The projectile atoms
become part of the substrate, their velocities were also rescaled
according to the substrate temperature~(1000~K) for approximately
0.5~ps~($\Delta$t $\approx$ 0.5~fs).  The trajectories of the next
projectile were started over this final substrate.  In summary, ten
different projectiles were started over the initial substrate for all
different orientations and energies.  We note here that some
projectile atoms were scattered back and some hydrogen atoms were
jostled out from the film as well. In Fig.\ \ref{fig3_CxHy} we
demonstrate projectiles over the final (i.e. after ten bombarding
events) structures for EDGE40, PLANE40,
LINE80 and POINT80.  The central processing unit~(CPU) time for the
simulations was 12$-$15 weeks on HP ALPHA workstations.

\section{Results}

\subsection{Dissociation of the projectiles}

One of the main interests of our computer experiments was the time
evolution study of the species.  Monitoring the trajectories of the
projectiles showed that their dissociations are the function of their
kinetic energy~(see Table\ \ref{table1_CxHy}).  For acetylene the
carbon-carbon bond was almost never broken at low energies~(20,
40~eV). That bond was dissociated only twice for LINE orientation at
40~eV.  At large bombarding energies~(80~eV), the carbon-carbon bond
was broken nine times for LINE orientation and only five times for
POINT orientation. Yet, in the experiments the \ctht \ projectiles
have random orientations when they arrive at the substrate surface.
This implies the fact that at high energies~(around 80~eV) the
carbon-carbon bond in \ctht \ is broken between 50 and 90\%.  The
number of broken hydrogen-carbon bonds in the projectiles increased
with the bombarding energy.  One can estimate from Table\
\ref{table1_CxHy} that more than 60\% of these kinds of bonds are
broken in the experiments. It was also evident from the simulations,
that after the dissociation, the hydrogen atoms like to form H$_2$
molecules, particularly in the case of \cht.  The final configurations
of the projectile carbon atoms in the film can be seen in Table\
\ref{table2_CxHy}. They mostly have $sp^{1+x}$ and $sp^2$ hybrids.
For methyl species, usually one hydrogen atom stays in a bond with the
carbon atom. Also, for acetylene, for lower energy bombardment~(20 and
40~eV) the hydrogen atoms stay bonded with the projectile carbon atom.
We conclude from the data that when all the bonds are broken in the
projectile during the collision phase, i.e. when the carbon atoms
become naked, they have a lower coordination number in contrast to
projectile carbon atoms, when not all the bonds are broken.

\begin{table}
\caption{
The number of broken bonds during the implantation of the projectiles.
H1~(C1) is the deepest hydrogen~(carbon) atom in C$_2$H$_2$
species~(POINT orientation). H1, H2, H3 are the three hydrogen atoms
in the CH$_3$ projectile.
\label{table1_CxHy}
}
\vspace*{0.5truecm}
\centerline{
\begin{tabular}{|c|c||c|c|c|}
\hline
C$_2$H$_2$ & E(eV) & H1$-$C1 & C1$-$C2 & C2$-$H2 \cr
\hline
\hline
     & 20 & 3 & 0 & 3 \cr
\cline{2-5}     
LINE & 40 & 6 & 2 & 8 \cr
\cline{2-5}
     & 80 & 8 & 9 & 8 \cr
\hline
\hline
      & 20 & 1 & 0 & 3 \cr
\cline{2-5}     
POINT & 40 & 6 & 0 & 8 \cr
\cline{2-5}
      & 80 & 9 & 5 & 10 \cr     
\hline      
\multicolumn{5}{c}{}\\
\hline
CH$_3$ & E(eV) & C$-$H1 & C$-$H2 & C$-$H3 \cr
\hline
\hline
     & 20 & 1 & 7 & 5 \cr
\cline{2-5}     
EDGE & 40 & 8 & 9 & 9 \cr
\hline
\hline
      & 20 & 7 & 3 & 5 \cr
\cline{2-5}     
PLANE & 40 & 7 & 7 & 8 \cr
\hline      
\end{tabular}
}
\end{table}

\begin{table}
\caption{
The final orientations of the projectile carbon atoms in the films. 
C1~(C2) is the deeper~(higher) carbon atom in C$_2$H$_2$
species~(POINT orientation).
\label{table2_CxHy}
}
\vspace*{0.5truecm}
\centerline{
\begin{tabular}{|c||c|c|c|c|c|c||c|c|}
\hline
 & \multicolumn{6}{|c||}{LINE} & \multicolumn{2}{|c|}{PLANE} \cr
\hline
& \multicolumn{2}{|c|}{20~eV} & \multicolumn{2}{|c|}{40~eV} &
        \multicolumn{2}{|c||}{80~eV} & 20~eV & 40~eV \cr
\hline
       & C1 & C2 & C1 & C2 & C1 & C2 & 
	 C & C \cr
\hline
\hline
$sp^0$     & 2 & 2 & 1 & 2 & 3 & 3 & 3 & 1\cr 
\hline                                      
$sp^1$     & 2 & 1 & 1 &   & 2 & 3 &   & 1\cr 
\hline                                      
$sp^{1+x}$ &   & 1 & 3 & 1 & 2 & 2 & 1 & 3\cr 
\hline                                      
$sp^2$     & 3 & 2 & 3 & 4 & 3 & 2 & 1 & 3\cr 
\hline                                      
$sp^{2+x}$ &   &   & 1 & 1 &   &   & 1 &  \cr 
\hline                                      
$sp^3$     & 3 & 4 & 1 & 2 &   &   & 4 & 2 \cr 
\hline
\multicolumn{9}{c}{}\\
\hline
 & \multicolumn{6}{|c||}{POINT} & \multicolumn{2}{|c|}{EDGE}\cr
\hline
& \multicolumn{2}{|c|}{20~eV} & \multicolumn{2}{|c|}{40~eV} &
        \multicolumn{2}{|c||}{80~eV} & 20~eV & 40~eV \cr
\hline
       & C1 & C2 & C1 & C2 & C1 & C2 & 
	 C & C \cr
\hline
\hline
$sp^0$     & 1 & 1 & 2 & 3 & 1 & 1 &   & 2\cr
\hline
$sp^1$     & 2 & 2 & 2 & 2 & 1 &   &   & 2\cr
\hline
$sp^{1+x}$ & 4 & 3 & 3 & 2 & 2 & 5 & 5 & 1\cr
\hline
$sp^2$     & 3 & 4 & 3 & 3 & 3 & 2 & 2 & 3\cr
\hline 
$sp^{2+x}$ &   &   &   &   &   & 1 & 2 & 1\cr
\hline
$sp^3$     &   &   &   &   & 2 & 1 & 1 & 1 \cr
\hline
\end{tabular}
}
\end{table}

\subsection{Penetration of the projectiles}

It is important to study the penetration of the projectile atoms to
retrieve information about processes during nucleations.  In our
computer simulations 10~\AA \ was found to be the deepest penetration
length in the sample under the surface. This demonstrates that our
choice of 11~\AA \ as the moveable part of the film was reasonable.
The average of the maximum penetration depths are orientation, energy
and projectile dependent.  Table\ \ref{table3_CxHy} shows that higher
energy provides a deeper penetration depth. There is a significant
difference between the implantation depth for POINT and LINE
projectiles, if the average final penetration depth is considered.
This distance hardly changes for the orientation LINE increasing the
subplantation energy from 20 to 40~eV.  However, in the same
conditions, there are significant changes for POINT orientation.  We
declare here, that orientation has a decisive role in the penetration
behavior of the projectiles, however, in experiments the species
arrive with random orientations to the surface. 

We now turn to the question which was raised by the proposed model for
the description of diamond nuclei growth in the BEN process
\cite{katai_dis_CxHy}.  In that model it was assumed that the
hydrocarbon ion species having kinetic energies bigger than
30$-$40~eV, penetrate under the amorphous carbon surface in the
vicinity of the silicon-carbide layer.  In the experiments, the
thickness of the amorphous carbon layer was found to be around
5$-$8~\AA, depending on the balance between simultaneous growth and
the etching of hydrocarbon and hydrogen species.  From Table\
\ref{table3_CxHy} it can be asserted that for low energy bombardments
(with 20~eV), the penetration of the projectile ions take place in
surface regions. It seems reasonable to affirm that penetration in
deeper regions starts at around 30$-$40~eV.  For bigger kinetic
energies~(80~eV), the penetration depths of the projectiles are close
to 4$-$8~\AA, i.e. in the vicinity of the silicon-carbon substrate in the
experiments.

\begin{table}
\caption{
The average maximal penetration depths for the projectile carbon
atoms.  The average final penetration depths can be seen in the
brackets. For the latter case only projectile atoms under the original
surface were considered.
\label{table3_CxHy}
}
\vspace*{0.5truecm}
\centerline{
\begin{tabular}{|c|c|c|c|c|}
\hline
C$_2$H$_2$ & E(eV) & C1(\AA) & C2(\AA) & Mass Center(\AA)\\
\hline
     & 20 &  -2.40 ( -1.31) &  -2.64 ( -1.48) &  -2.38 ( -1.41) \\
\cline{2-5}                                                                                           
 LINE& 40 &  -2.83 ( -2.17) &  -2.73 ( -1.98) &  -2.50 ( -2.05) \\
\cline{2-5}                                                                                           
     & 80 &  -5.83 ( -5.52) &  -4.71 ( -4.96) &  -4.83 ( -3.98) \\
\hline                                                                                                
\hline                                                                                                
     & 20 &  -2.71 ( -1.77) &  -2.81 ( -2.31) &  -2.61 ( -2.02) \\
\cline{2-5}                                                                                           
POINT& 40 &  -4.76 ( -4.80) &  -4.89 ( -4.61) &  -4.51 ( -4.32) \\
\cline{2-5}                                                                                           
     & 80 &  -6.24 ( -6.09) &  -5.80 ( -4.59) &  -5.59 ( -4.73) \\
\hline
\end{tabular}
}
\vspace*{0.5truecm}
\centerline{
\begin{tabular}{|c|c|c|c|}
\hline
CH$_3$ & E(eV) & C(\AA) & Mass Center(\AA)\\
\hline
     & 20 &  -2.84 ( -2.07) &  -2.74 ( -2.13) \\ 
\cline{2-4}                                                                                          
EDGE & 40 &  -3.93 ( -2.93) &  -3.30 ( -2.02) \\
\hline                                                                                               
\hline                                                                                               
     & 20 &  -2.74 ( -1.83) &  -2.45 ( -1.59) \\ 
\cline{2-4}                                                                                          
PLANE& 40 &  -3.43 ( -3.19) &  -2.84 ( -2.17) \\
\hline
\end{tabular}
}
\end{table}

\subsection{Structural rearrangement}

The bombardment of the projectile atoms cause structural
rearrangements in the substrate and contribute to the formation of
diamond nuclei. In the model describing the bias enhanced microwave
plasma assisted CVD of diamond nuclei, the formation of diamond nuclei
was not explained because it was beyond of scope of that model
\cite{katai_dis_CxHy}. It is assumed in the model that the presence of
silicon-carbon or silicon substrate contribute to the formation of
diamond nuclei, as it has orientation information. It is beyond the
work of our computer simulation to report on this process here, due to
the fact that we did not have a silicon substrate. In this section we
will give a possible explanation of the initial process into the
formation of diamond nuclei, which involves the occurrence of
increasing fraction of $sp^3$ atoms in deeper regions after the
bombardments.  

Taking into consideration the changes in the total film, our results
show that the thickness of films have grown by 1$-$3~\AA \ in $z$
direction after ten subsequent bombardments for all different cases.
The $sp^3$ content in the films increased with 1$-$5\%, while the
$sp^2$ content decreased by the same magnitude. There are no
significant changes in the total density of the films and on average
partial first neighbor distances.

\begin{figure}
\centerline{
\epsfig{file=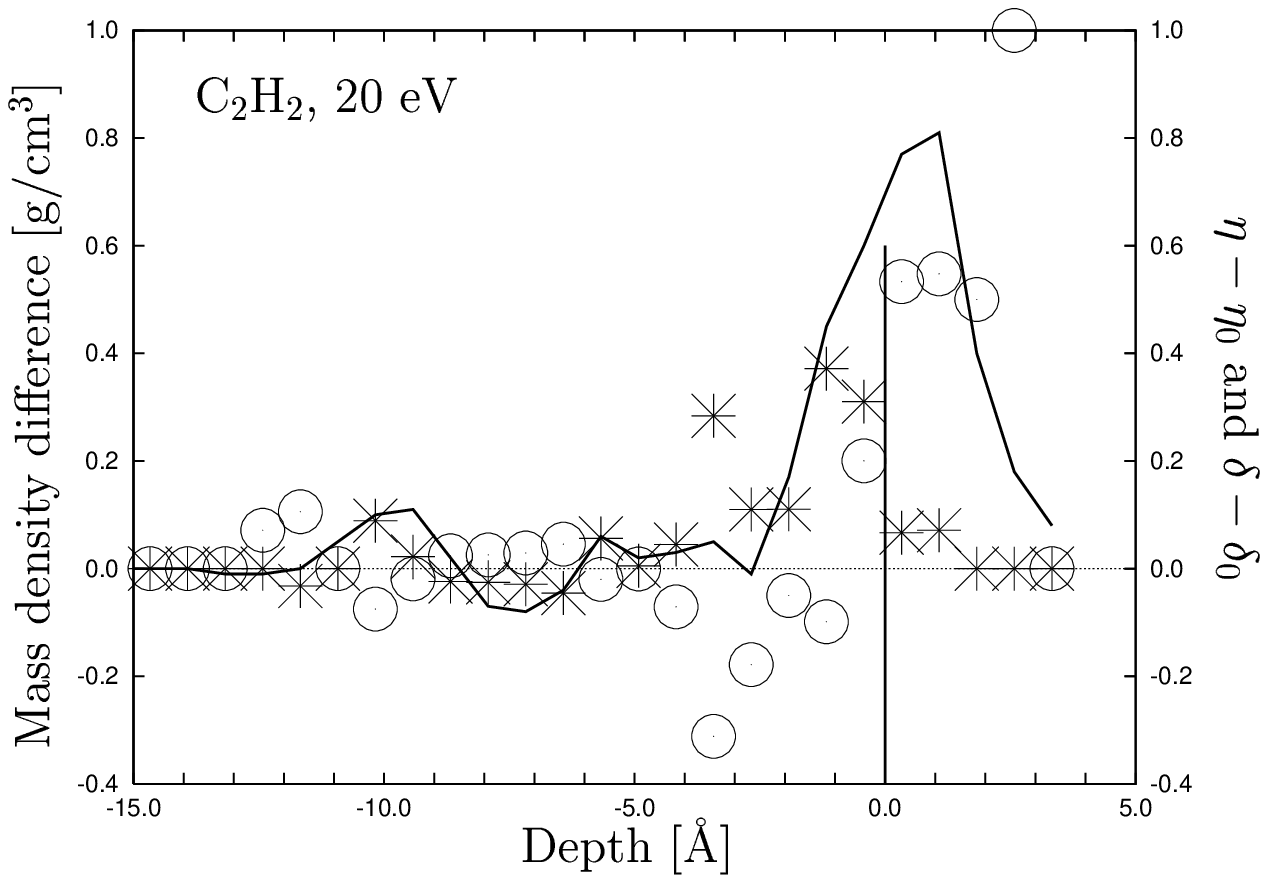,width=7.5truecm}
\epsfig{file=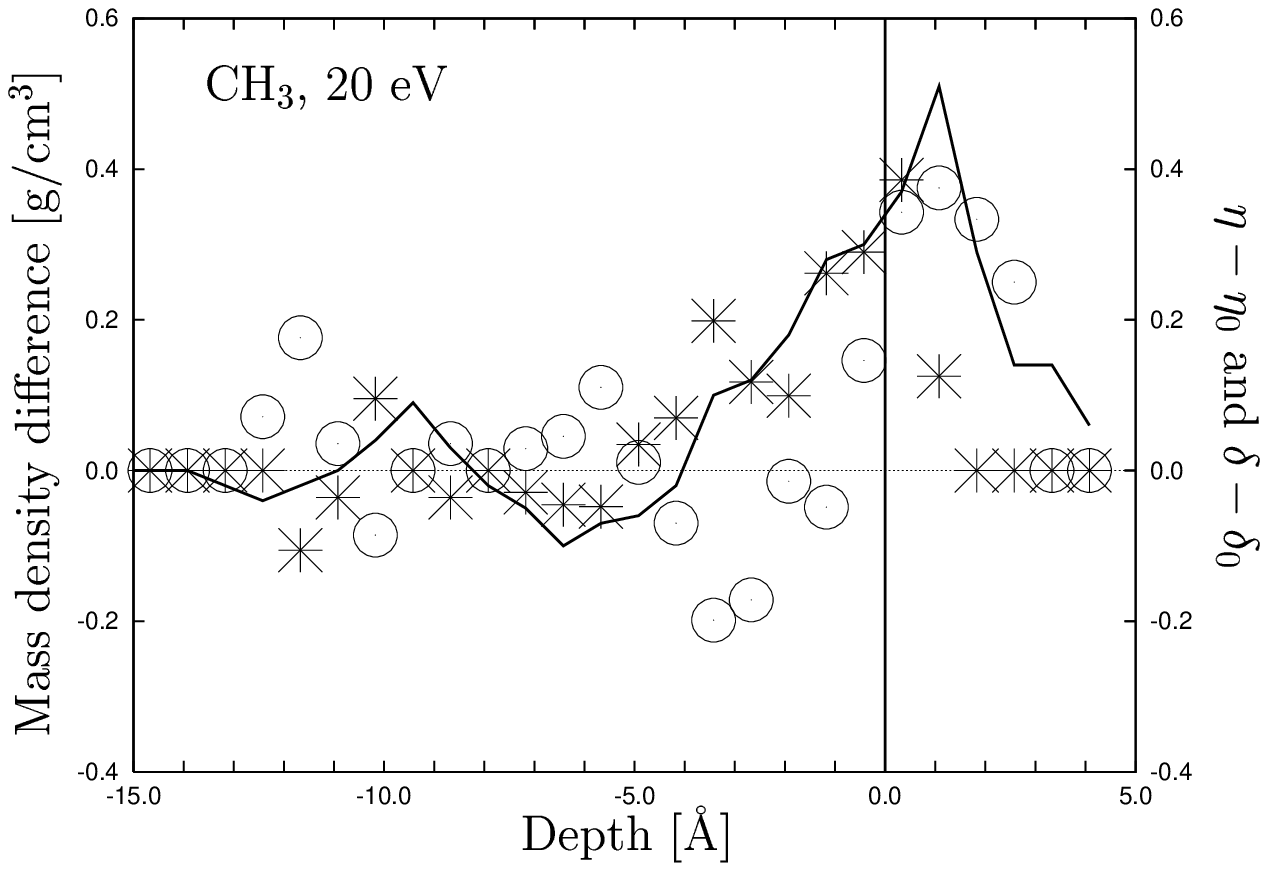,width=7.5truecm}
}
\centerline{
\epsfig{file=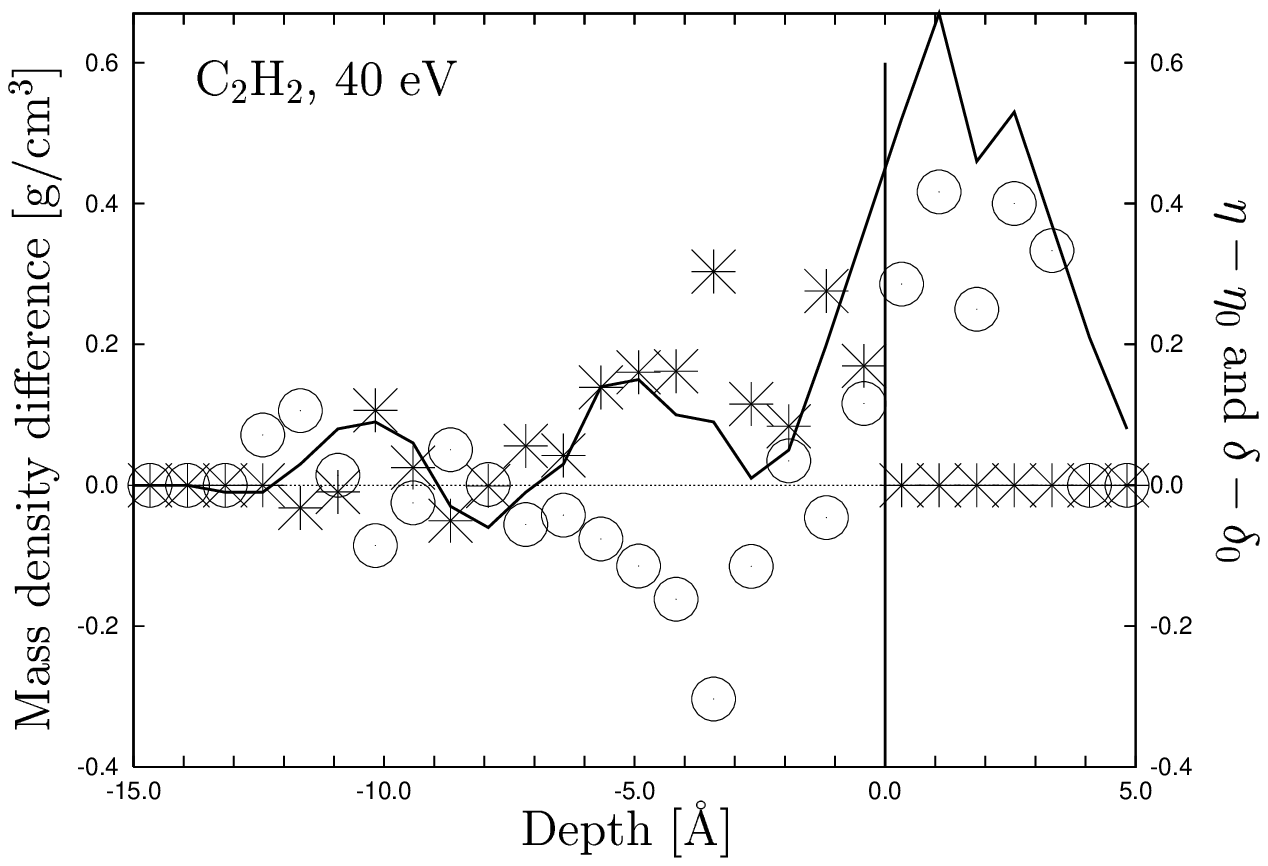,width=7.5truecm}
\epsfig{file=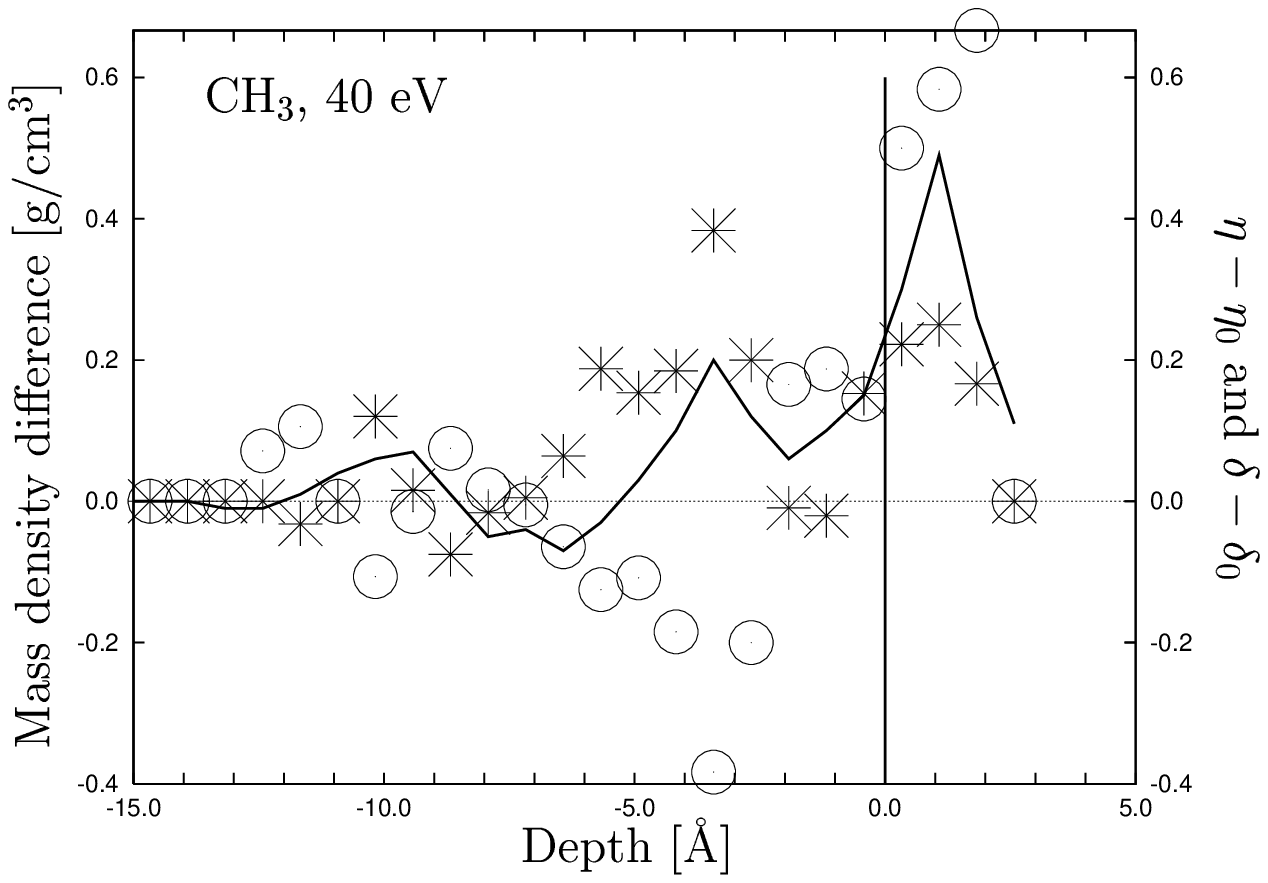,width=7.5truecm}
}
\centerline{
\epsfig{file=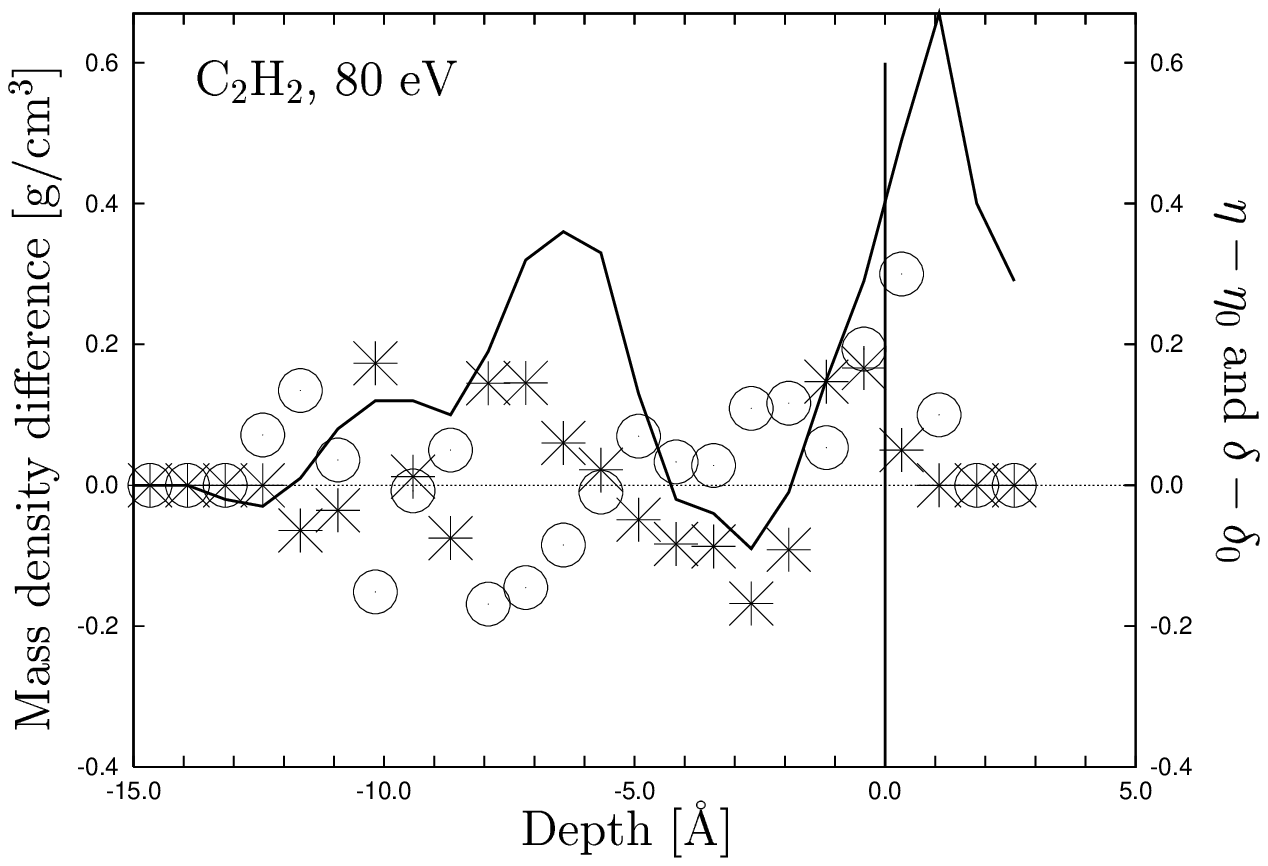,width=7.5truecm}
}
\medskip
\caption{
\small
The depth profiles for amorphous carbon films after ten bombardment
events.  The solid line show the mass density difference profile
between the final and the initial films. The circles and stars denote
the $sp^2$ and $sp^3$ fraction differences, respectively
($\eta-\eta_0$ and $\delta - \delta_0$).  The vertical straight lines
show the original top of the initial film.
}
\label{fig5_CxHy}
\end{figure}

Here, we concentrate on the changes in the different depths of the
film. It was reported in a similar study by Uhlmann \etal
\cite{uhlmann_prl98_CxHy} that the carbon
bombardments form an $sp^2$ rich defective surface layer, whose
thickness scales with ion energy. Below that layer an $sp^3$ rich
layer was found. This was one of the main results in their study,
which supports the subplantation model in the physical vapor
deposition method of diamond like amorphous carbon films.
Nevertheless, the conditions in the BEN process are different.  The
temperature is around 1000~K compared to the typically low temperature
(T $<$ 150~K) usually applied during the preparation of diamond-like
carbon.  Furthermore, the plasma contains several different
hydrocarbon species in BEN, contrary to diamond-like carbon growth,
where single monoenergetic particles (mostly single carbon ions) are
used.  However, our computer simulations of bombardments with \ctht \
and \cht \ projectiles show similar conclusions to the work of Uhlmann
\etal \cite{uhlmann_prl98_CxHy}.  In Fig.\
\ref{fig5_CxHy} the mass density difference and the discrepancy in the
$sp^2$, $sp^3$ fraction for the final and initial films can be seen
for all cases.  Structural rearrangements even for low bombarding
energy with 20~eV can be seen. The main contribution is close to the
surface in the region between -4.0~\AA \ and -2.0~\AA, measured from
the top of the initial substrate. The $sp^3$ content increases with
20$-$30\% and the $sp^2$ content decreases by the same magnitude after
ten bombardment events.
For 40~eV the changes are significant at the deeper region between
-6.0~\AA \ and -3.0~\AA. The change in the $sp^2$ and $sp^3$ content
is broader compared to the change at 20~eV.  At 80~eV, for \ctht
\ projectile bombardment, the penetration regime moves deeper and
can be found between -10.0~\AA \ and -5.0~\AA.  In all cases the
structural rearrangement occurs together with the mass density
changes.  The effect is the largest at 80~eV. We note here that the
mass density increase over the top of the initial surface comes from
the slow growth of the top surface. Therefore, changes in Fig.\
\ref{fig5_CxHy} over 0~\AA \ are irrelevant.

It is known from the experiments of K\'atai \etal
\cite{katai_jap99_CxHy} that \cht \ projectiles have 40~eV kinetic
energy at 200~V bias, whereas for, \ctht \ it is 80~eV.  Nevertheless,
the kinetic energies of the carbon atoms in the projectiles are the
same. It is around 37~eV for \cht \ ion species with 40~eV kinetic
energy and approximately the same for \ctht \ projectile. Yet the
latter one has two carbon atoms in one projectile.  If one compares
the effects of these two projectiles at the optimal bias
voltage~(200~V) for the nucleation process, considerable discrepancies
are noticeable.  The penetration of \ctht \ projectiles occurs in
deeper regions.  In addition, the effects of caused structural
rearrangements are different. The changes are in deeper regions for
\ctht \ ion species in contrast to \cht \ projectiles. We can state
here that \ctht \ projectiles which has the largest flux among the
hydrocarbon species, strongly contributes to the local density
increase and structural rearrangement in the 4$-$8~\AA \ region under
the surface. The increase of the content of $sp^3$ hybrids at the
optimal 200~V bias voltage for diamond nucleation in the region
4$-$8~\AA \ under the surface -- which is the vicinity of silicon
substrate and has orientation information in the experiments -- can
cause the possible formation of diamond nuclei in a "percolated way".
This suspicion should be verified by computer simulation and it is our
main goal in the future.

\section{Summary}

The structural rearrangements in a$-$C:H films were investigated under
bombardments with \cxhy \ projectiles. Special attention was paid to
the study of different projectiles at different bombarding energies.
Larger kinetic energies produce deeper subplantations and lead to a
higher probability of the dissociation of the projectiles. For
acetylene the carbon-carbon bond is hardly ever broken at low
energies~(20, 40~eV) and it is broken in 50$-$90\% of cases for larger
energies~(80~eV). The computer experiments showed that around 60\% of
projectile hydrogen-carbon bonds are broken during the subplantation.
Usually, one of the hydrogen-carbon bond are not broken for \cht \ and
the other two hydrogens form H$_2$ molecules.  At the typical 200~V
bias voltage, which is optimal for diamond nuclei formation in BEN,
the \ctht \ projectiles have 80~eV and \cht \ ion species 40~eV
kinetic energy.  There are significant differences in their effects on
the structural rearrangements in the films.  For \ctht \ projectiles,
the penetration is deeper and these species cause changes in deeper
regions, than for \cht \ projectiles.  Whereas, the \cht \ projectiles do
not contribute to the deep changes and cause structural rearrangements
only in surface regions.  The structural changes in the film showed
that larger kinetic energy ionic species under the average deposition
depth generate a thicker $sp^3$ rich film whilst the surface mostly
consists of atoms with $sp^2$ hybrids.  The border between these two
regimes descend with larger deposition energies. The results support
the subplantation model. It will definitely be a fascinating task in
the future to do simulations over a silicon substrate and to study the
effect of this crystalline layer on the formation of diamond nuclei in
the amorphous film.


\chapter[Growth of amorphous silicon]{Growth of amorphous silicon.
Low energy molecular dynamics simulation of atomic bombardment}
\markboth{GROWTH OF a$-$Si}{GROWTH OF a$-$Si}
\label{Si}

In this chapter the growth of amorphous silicon thin films was studied
using the tight-binding molecular dynamics technique. The atomic
deposition program code, which had been developed for studying the
growth of amorphous carbon structures and which was reported in
chapter\ \ref{aC}, was applied.

\section{Introduction}

Amorphous silicon~(a$-$Si) has been the subject of numerous
experimental and theoretical works in recent decades
\cite{Morigaki99}. Several diffraction measurements were employed on
a$-$Si samples in order to retrieve information about the atomic
arrangement \cite{Moss70, Barna77, Fortner89, Kugler89, Cicco90,
Kugler93}, i.e.\ the radial distribution function~(RDF), which
contains integrated information about the amorphous structures.  The
results of these experiments suggest that the covalently bonded
amorphous silicon are not completely disordered. The bonds between
atoms and the coordination numbers are very similar to the crystalline
phase.  First and second neighbor peaks are broadened, but the
positions are the same as in the crystalline structure, while the
third peak disappears in the measured RDF. The absence of the third
peak confirms that there is no characteristic dihedral angle.

In 1985, a continuous random network model of amorphous
silicon~(WWW~model) was constructed by Wooten {\it et al.}
\cite{Wooten85} who carried out Monte Carlo~(MC) simulations using the
classical empirical potential of Keating \cite{Keating66}. This
defect-free network includes fivefold and sevenfold rings in addition
to the sixfold rings of the diamond structure.  Since then, several
computer generated models have been constructed using classical
empirical potentials \cite{Keriles88, Luedtke89, Ishimaru97} or
applying different quantum mechanical methods \cite{Stich91, Toth94,
Hensel96, Herrero00}, but still the WWW model is considered to be the
best three-dimensional atomic scale representation of the amorphous
silicon network.  Dangling bonds or even floating bonds (if they
exist) are believed to be very rare in this condensed phase of silicon
atoms and are not included in the WWW model.  Recently, an accurate
X$-$ray diffraction measurement \cite{Laaziri99} was published in the
wide Q range of 0.03$-$55 \AA$^{-1}$.  The most important conclusion
of this measurement, apart from confirming earlier results, is that
the coordination number is less than four in implanted a$-$Si sample,
which is controversial with the WWW model of amorphous silicon.

An alternative to MC simulations is molecular dynamics~(MD).  On the
basis of the Car--Parrinello density functional theory~(DFT) method,
MD structural simulation was carried out for a$-$Si having 54 atoms
\cite{Stich91}.  Larger systems are needed for investigating complex
dynamics processes, such as, for example those involved in growth.  A
frequently applied method is to find convenient tight-binding~(TB)
models, which could be employed over a hundred atoms and could be less
time consuming than DFT.  In our simulations the TB Hamiltonian of
Kwon {\it et al.} \cite{Kwon94} was applied to describe the
interaction between silicon atoms. This group developed an excellent
TB potential for carbon systems and we successfully applied it to the
amorphous carbon growth \cite{Kohary01}. All the parameters and
functions of the interatomic potential for silicon were fitted to the
results of local density functional calculations. This TB potential
reproduces energies of different cluster structures, elastic
constants, the formation energies of vacancies and interstitials in
crystalline silicon.  According to the authors' article, the only
disadvantage of this TB model is that the bond lengths inside small
clusters are a bit longer than those derived from {\it ab initio}
calculations or from the experiments.

\section{Simulation details}

Rapid cooling of the liquid phase is frequently applied to construct
amorphous and glassy structures \cite{Stich91, Hensel96, Ishimaru97,
Yang98}.  The system is cooled down to room temperature at a rate of
10$^{11}$ -- 10$^{16}$~K/s.  Another possible approach for
constructing an amorphous structure \cite{kugler_tobepub_Si} is
motivated by the fact that amorphous silicon is usually grown by
different vapor deposition techniques in laboratories.  We have
developed a MD computer code~(see section\ \ref{method}) to simulate
the real preparation procedure of an amorphous structure, which is
grown by atom-by-atom deposition on a substrate.  In our recent work
\cite{Kohary01}, the growth of amorphous carbon films were simulated
by this method.  A brief summary of our simulation technique is
described here~(for details, see reference \cite{Kohary01} or section\
\ref{method}).  A~rectangular diamond lattice cell containing 120
silicon atoms was employed to mimic the substrate.  There were 16
fixed atoms at the bottom. The remaining atoms could move with full
dynamics.  The simulation cell was open along the [111] direction
(positive $z$-axis) and periodic boundary conditions were applied in
planar $x$, $y$ directions.  The kinetic energy of the atoms in the
substrate were rescaled at every time step~(${\Delta}t$ = 0.5~fs) in
order to keep the substrate at a constant temperature.  First, the
substrate was kept at a given temperature for 0.5~ps to enable
structural relaxation. In the deposition process the frequency of the
atomic injection was on average 1/125~fs$^{-1}$.  This flux is orders
of magnitude higher than the deposition rate commonly applied in
experiments.  However, the low substrate temperatures~(100 and 300~K)
during simulations cause quick energy dissipation and this may
compensate the high deposition rate.

In our first set of simulations the initial kinetic energies of the
atoms in the above case were dissipated by the interaction with
surface atoms. To speed up the energy dissipation, we have developed a
different simulation method as well. The velocities of the atoms in
the atomic beam were modified by the following procedure. If a
bombarding silicon atom arrived closer than the "{\it critical
distance}" to any substrate atom, it became part of the "group" of
substrate atoms. In other words, their total kinetic energy was
rescaled in every time step to a given (substrate) temperature.  If
the next bombarding atom arrived in the neighborhood of any of the
atoms in this group, the similar procedure was employed.  The chosen
critical distance was 15\% larger than the bond length in crystalline
silicon~(2.36~\AA).  This is a simple model to speed up the inelastic
collision between bombarding and target atoms and a plausible
interpretation of the phenomena.  Nine different structures have been
constructed by the methods mentioned above:

({\it i}) Six models constructed with 25~ps injection time, and with
average kinetic energy E$_{beam}$ = 1~eV and 5~eV, at T$_{sub}$ =
100~K substrate temperature (e1T100R5, e5T100R5, e1T100R10, e5T100R10,
e1T100R20, and e5T100R20). R5, R10, and R20 indicate different
relaxation times~(5, 10, and 20~ps, respectively) for both structures
after 25~ps injection.

({\it ii}) Three models (e1T300, e5T300, e10T300) were constructed at
T$_{sub}$ = 300~K substrate temperature using our model for quick
energy dissipation.  Average kinetic energies of bombarding atoms were
E$_{beam}$ = 1~eV, 5~eV, and 10~eV, respectively. The simulation time
was 5~ps for these three models. Due to the preparation method the
temperature of the films remained 300~K, therefore no relaxation was
applied.

({\it iii}) In order to retrieve information on the difference between
rapid cooling \cite{Stich91, Hensel96, Ishimaru97} and atom-by-atom
deposition on a substrate (melt-quenching and vapor-quenching), we
prepared an additional, tenth model~(e1T100Q; Q is for quenching) in
the following way. The temperature of the film e1T100R20 was increased
to 3500~K. Considering this, as an initial state, the trajectories of
the silicon atoms were followed by full dynamics for 7.5~ps. The
substrate temperature was kept at 100~K again, which leads to the
cooling of the film above the substrate. After 7.5~ps, the temperature
of the film decreased into the region at around 200~K.  This technique
can be considered as the computer simulation of splat cooling, where
small droplets of melt are brought into contact with the chill-block.

The main goal of our work was to simulate amorphous silicon
structures.

\section{Structural properties}

\begin{figure}
\centerline{
\epsfig{file=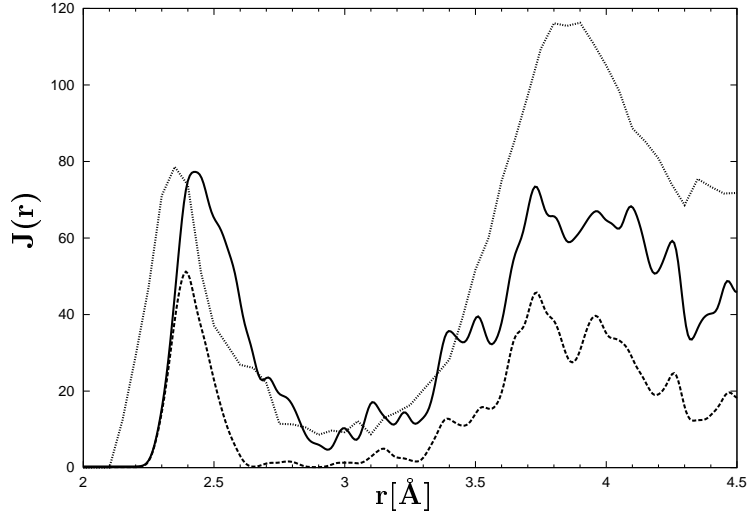,height=7.0truecm}
}
\medskip
\caption{
\small
The first and second neighbor peaks in the radial distribution
functions for e1T100Q model~(solid line). The dotted line shows the
experimental silicon RDF, reported in \cite{Kugler93}. It is clear
that there is not a determined distance, which can be considered as
the border for the first neighbor shell, and the first and second
neighbor peaks overlap. In addition, it can be seen that the model
potential applied in these simulations overestimates the first
neighbor distances.  The dashed line represents the J$_{4,4}$ partial
radial distribution function, which belongs to distances between Z = 4
coordinated silicon atoms, taking the first neighbor shell to 2.6~\AA.
}
\label{fig1_Si}
\end{figure}

\begin{figure}
\centerline{
\epsfig{file=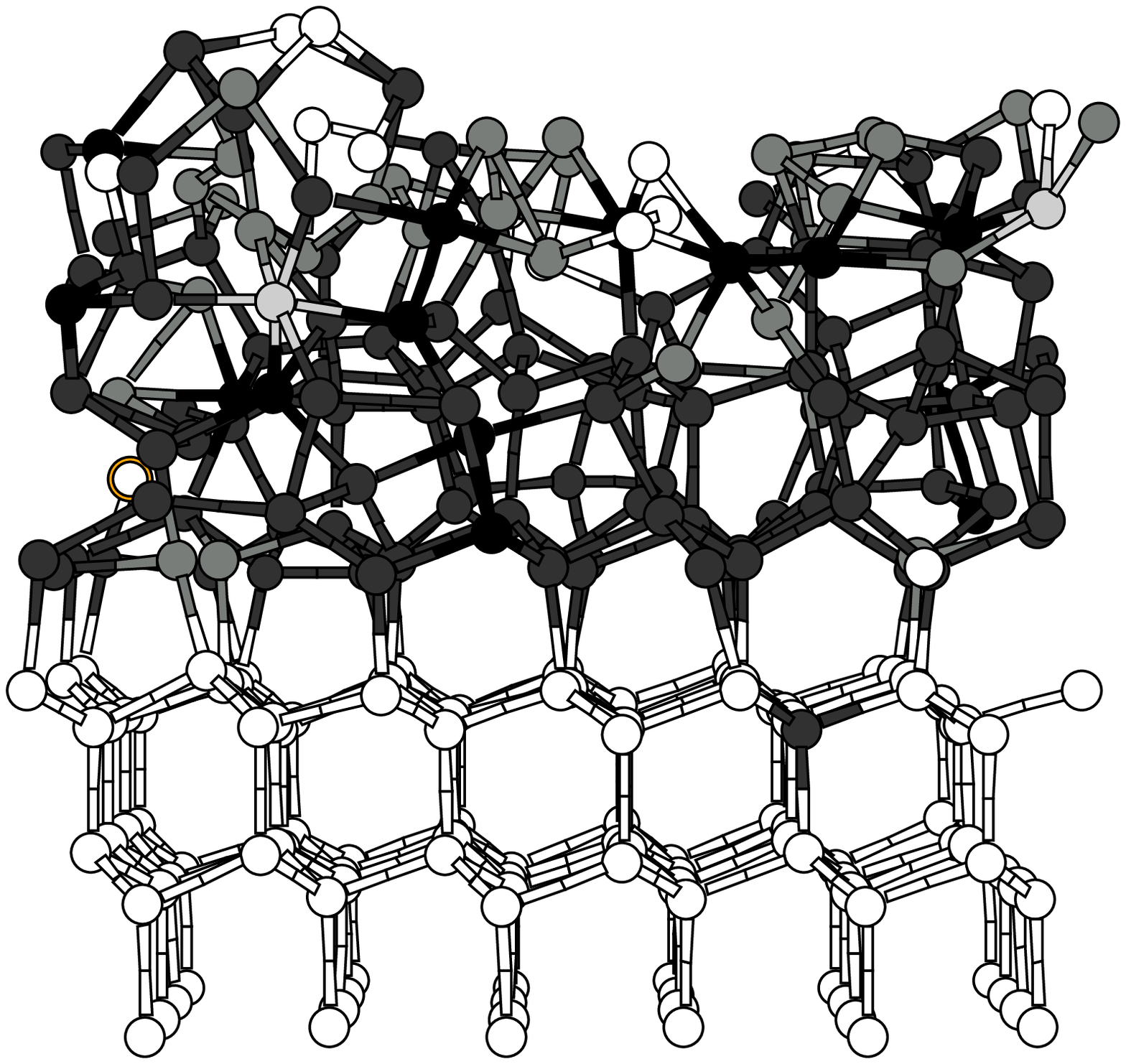,height=7.0truecm}
\epsfig{file=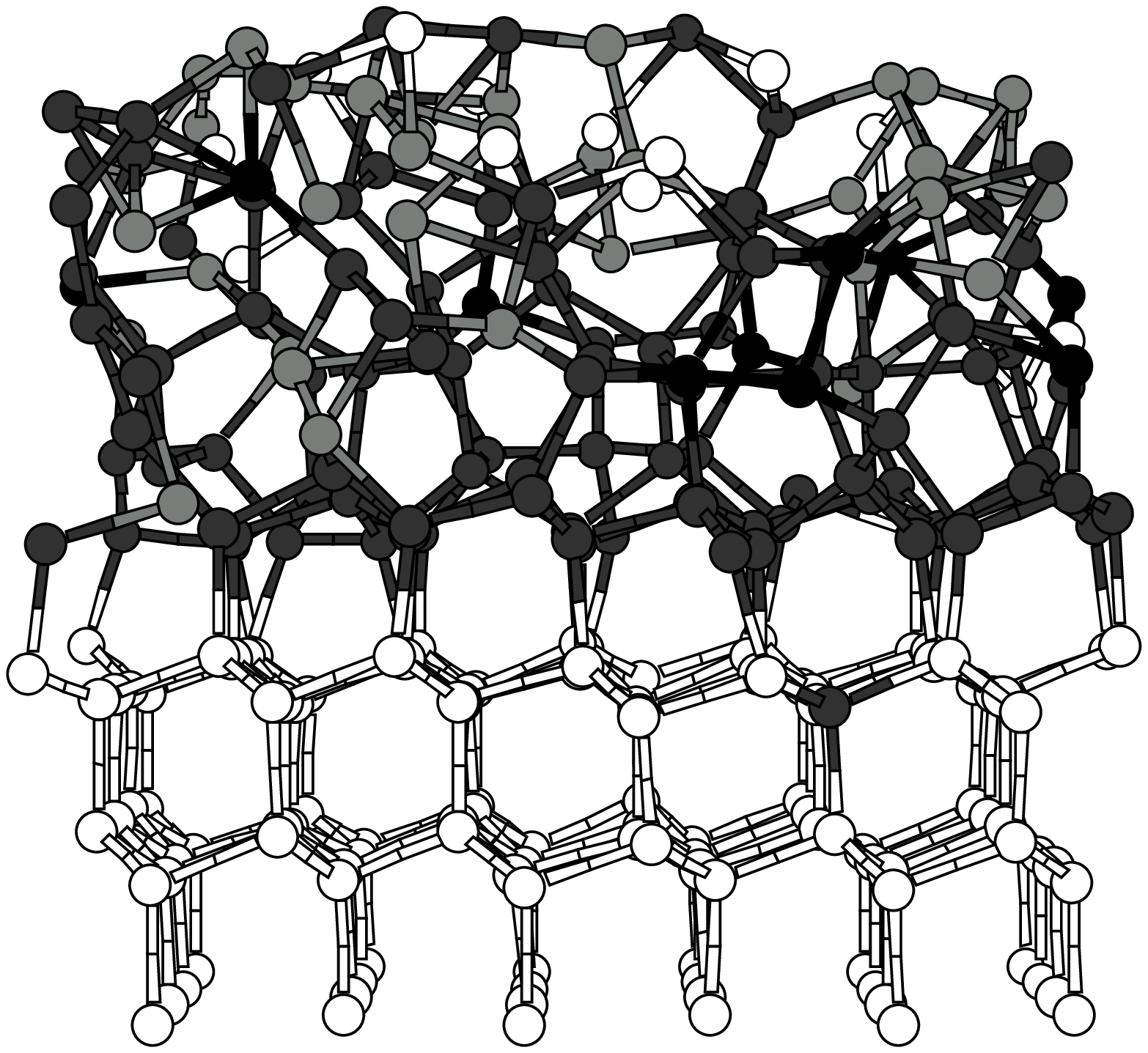,height=7.0truecm}
}
\medskip
\caption{
\small
Snapshots of e1T100R20 (on the left side) and e1T100Q (on the right
side) models are shown after growth and relaxation.  The substrates
(open circles at the bottom) with T$_{sub}$ = 100~K remained similar
to the crystal lattice during growth. Light grey, grey and black atoms are
threefold, fourfold and fivefold-coordinated, respectively.  The rest
of the open circles correspond to twofold- and one one-fold
coordinated atoms.  For a color version see:
http://www.physik.uni$-$marburg.de/$\sim$kohary/aSifigures.html
}
\label{fig2_Si}
\end{figure}

The final structures of different models consist almost of the same
number of atoms (between 162 and 177), with a thickness of about
13~\AA~(see Table\ \ref{table1_Si}). In Fig.\ \ref{fig1_Si} the first
and second neighbor peaks are shown in e1T100Q model.  Contrary to
carbon systems~(see section \ref{aC}, fig.\ \ref{fig7_aC}), there is
not a clear, determined distance, which can be considered as a border
for the first neighbor shell in case for a$-$Si.  We have chosen 2.6,
2.7, 2.82 and 3.0~\AA \ as a limit for the first neighbor shell to
calculate the statistics of the models. Snapshots of films e1T100R20
and e1T100Q are shown in Fig.\ \ref{fig2_Si}, taking the first
neighbor shell to 2.6~\AA.  The substrates with T$_{sub}$ = 100~K
remained similar to crystal lattice.

\begin{table}
\caption{
The tables below contain the number of silicon atoms~(No.) in the
films, in addition to the percentage of atoms with different
coordination numbers (Z), the average coordination number $\langle Z
\rangle$, the thickness z$_{d}$ and the density $\rho$ of the films.
Two rows belong to bulk~(top) and to total~(lower) sample,
respectively. The numbers in the top table~(Table~A) were calculated
by taking the first neighbor shell to 2.6~\AA, and in the bottom
table~(Table~B) by taking 2.82~\AA.
\label{table1_Si}
}
\vspace*{0.5truecm}
\centerline{
\footnotesize
\begin{tabular}{|c||c||c|c|c|c|c||c||c|c|}
\hline
Table~A~(2.6~\AA)& No. & Z=2 & Z=3 & Z=4 & Z=5 & Z=6 & $\langle Z \rangle$ & z$_{d}$(\AA) & $\rho(g/cm^3)$\\
\hline
\hline
e1T100R5 &   140 &  7.1 & 10.7 & 70.0 &  7.1 &  0.0 &  3.7 & 10.1 &  2.5 \\
 &   166 &  9.6 & 15.7 & 62.7 &  8.6 &  0.0 &  3.6 & 13.1 &  2.2 \\
\hline
e1T100R10 &   134 &  3.0 & 14.9 & 67.9 & 12.7 &  0.7 &  3.9 &  9.8 &  2.4 \\
 &   166 &  3.6 & 18.7 & 62.7 & 14.9 &  0.7 &  3.8 & 12.8 &  2.2 \\
\hline
e1T100R20 &   134 &  3.0 & 12.7 & 71.6 & 10.4 &  1.5 &  3.9 &  9.8 &  2.4 \\
 &   166 &  6.6 & 18.7 & 63.3 & 11.2 &  1.5 &  3.8 & 12.8 &  2.2 \\
\hline
e5T100R5 &   151 &  7.9 & 23.8 & 51.7 & 10.6 &  0.0 &  3.5 & 10.6 &  2.3 \\
 &   162 &  9.9 & 24.1 & 50.0 & 10.6 &  0.0 &  3.5 & 13.6 &  1.9 \\
\hline
e5T100R10 &   152 &  5.9 & 23.0 & 50.0 & 20.4 &  0.7 &  3.9 & 10.7 &  2.3 \\
 &   162 &  6.2 & 24.1 & 49.4 & 21.1 &  0.7 &  3.8 & 13.7 &  1.9 \\
\hline
e5T100R20 &   152 &  4.6 & 18.4 & 51.3 & 23.0 &  2.0 &  4.0 & 10.7 &  2.3 \\
 &   162 &  4.3 & 19.8 & 51.2 & 23.7 &  2.0 &  4.0 & 13.7 &  1.9 \\
\hline
\hline
e1T300 &   151 &  7.9 & 16.6 & 53.6 & 15.2 &  2.6 &  3.8 & 10.8 &  2.4 \\
 &   177 & 11.3 & 18.6 & 48.0 & 15.9 &  2.6 &  3.6 & 13.8 &  2.2 \\
\hline
e5T300 &   135 &  6.7 & 19.3 & 47.4 & 20.0 &  3.0 &  3.9 &  9.4 &  2.8 \\
 &   171 & 12.3 & 22.8 & 40.9 & 20.7 &  3.0 &  3.6 & 12.4 &  2.5 \\
\hline
e10T300 &   142 &  7.0 & 19.7 & 52.1 & 19.7 &  0.7 &  3.9 &  9.7 &  2.7 \\
 &   175 & 11.4 & 24.0 & 45.7 & 20.4 &  0.7 &  3.7 & 12.7 &  2.4 \\
\hline
\hline
e1T100Q &   122 &  1.6 & 13.9 & 74.6 &  8.2 &  0.0 &  3.9 &  9.1 &  2.4 \\
 &   166 &  4.2 & 22.3 & 63.3 &  8.2 &  0.0 &  3.6 & 12.1 &  2.3 \\
\hline
\end{tabular}
}
\vspace*{0.5truecm}
\centerline{
\footnotesize
\begin{tabular}{|c||c||c|c|c|c|c||c||c|c|}
\hline
Table~B~(2.82~\AA)&No. & Z=2 & Z=3 & Z=4 & Z=5 & Z=6 & $\langle Z \rangle$ & z$_{d}$(\AA) & $\rho(g/cm^3)$\\
\hline
\hline
e1T100R5 &   140 &  0.0 &  0.7 & 55.0 & 31.4 & 11.4 &  4.6 & 10.1 &  2.5 \\
 &   166 &  0.0 &  3.0 & 51.2 & 39.3 & 13.6 &  4.6 & 13.1 &  2.2 \\
\hline
e1T100R10 &   134 &  0.0 &  1.5 & 53.0 & 29.9 & 11.9 &  4.6 &  9.8 &  2.4 \\
 &   166 &  0.0 &  3.6 & 47.0 & 39.6 & 16.4 &  4.7 & 12.8 &  2.2 \\
\hline
e1T100R20 &   134 &  0.0 &  1.5 & 53.7 & 28.4 & 15.7 &  4.6 &  9.8 &  2.4 \\
 &   166 &  0.0 &  3.6 & 50.0 & 35.1 & 20.1 &  4.6 & 12.8 &  2.2 \\
\hline
e5T100R5 &   151 &  0.0 &  6.0 & 31.8 & 43.7 & 15.9 &  4.8 & 10.6 &  2.3 \\
 &   162 &  0.0 &  5.6 & 32.7 & 47.0 & 16.6 &  4.8 & 13.6 &  1.9 \\
\hline
e5T100R10 &   152 &  0.0 &  2.6 & 34.9 & 40.1 & 15.8 &  4.9 & 10.7 &  2.3 \\
 &   162 &  0.0 &  3.1 & 34.6 & 42.8 & 17.1 &  4.9 & 13.7 &  1.9 \\
\hline
e5T100R20 &   152 &  0.0 &  2.6 & 35.5 & 43.4 & 15.8 &  4.8 & 10.7 &  2.3 \\
 &   162 &  0.0 &  2.5 & 35.8 & 46.1 & 17.1 &  4.8 & 13.7 &  1.9 \\
\hline
\hline
e1T300 &   151 &  0.0 &  3.3 & 32.5 & 40.4 & 21.9 &  4.9 & 10.8 &  2.4 \\
 &   177 &  0.0 &  7.9 & 33.3 & 45.0 & 21.9 &  4.7 & 13.8 &  2.2 \\
\hline
e5T300 &   135 &  0.0 &  0.7 & 27.4 & 39.3 & 23.7 &  5.2 &  9.4 &  2.8 \\
 &   171 &  0.6 &  1.2 & 30.4 & 47.4 & 28.9 &  5.1 & 12.4 &  2.5 \\
\hline
e10T300 &   142 &  0.0 &  0.7 & 28.2 & 45.8 & 19.7 &  5.0 &  9.7 &  2.7 \\
 &   175 &  0.6 &  2.9 & 30.9 & 52.8 & 21.8 &  4.9 & 12.7 &  2.4 \\
\hline
\hline
e1T100Q &   122 &  0.0 &  0.8 & 56.6 & 28.7 & 12.3 &  4.6 &  9.1 &  2.4 \\
 &   166 &  0.0 &  0.6 & 54.8 & 41.0 & 17.2 &  4.6 & 12.1 &  2.3 \\
\hline
\end{tabular}
}
\end{table}

\subsection{Time development of relaxations}

After the growth process, when there was no more injected bombarding
atoms over the network, the films relaxed with full dynamics for
20~ps.  The deposited networks, which have a temperature gradient in
the $z$ direction released a considerable amount of heat. This is
attributable to the influence of the substrate kept at a constant
temperature.  Anomalous relaxation has been observed during quenching.
Temperature versus time function of this non-equilibrium process shows
a stretched-exponential form, $T_{film}(t) = c + \exp \left(a
t^{\beta}+b\right)$~K.

\begin{figure}
\centerline{\epsfig{file=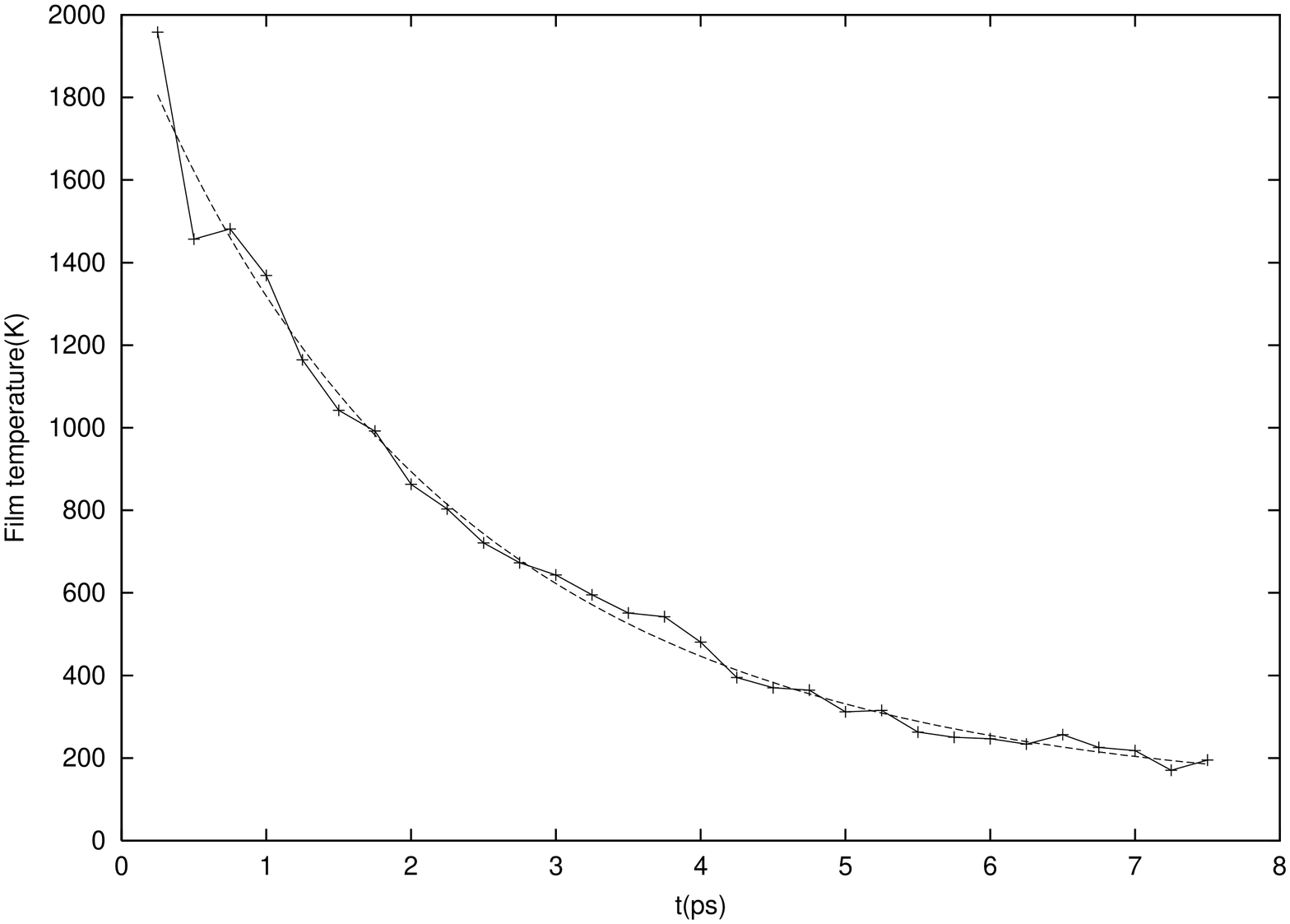,height=8.0truecm,angle=-90}}
\centerline{\epsfig{file=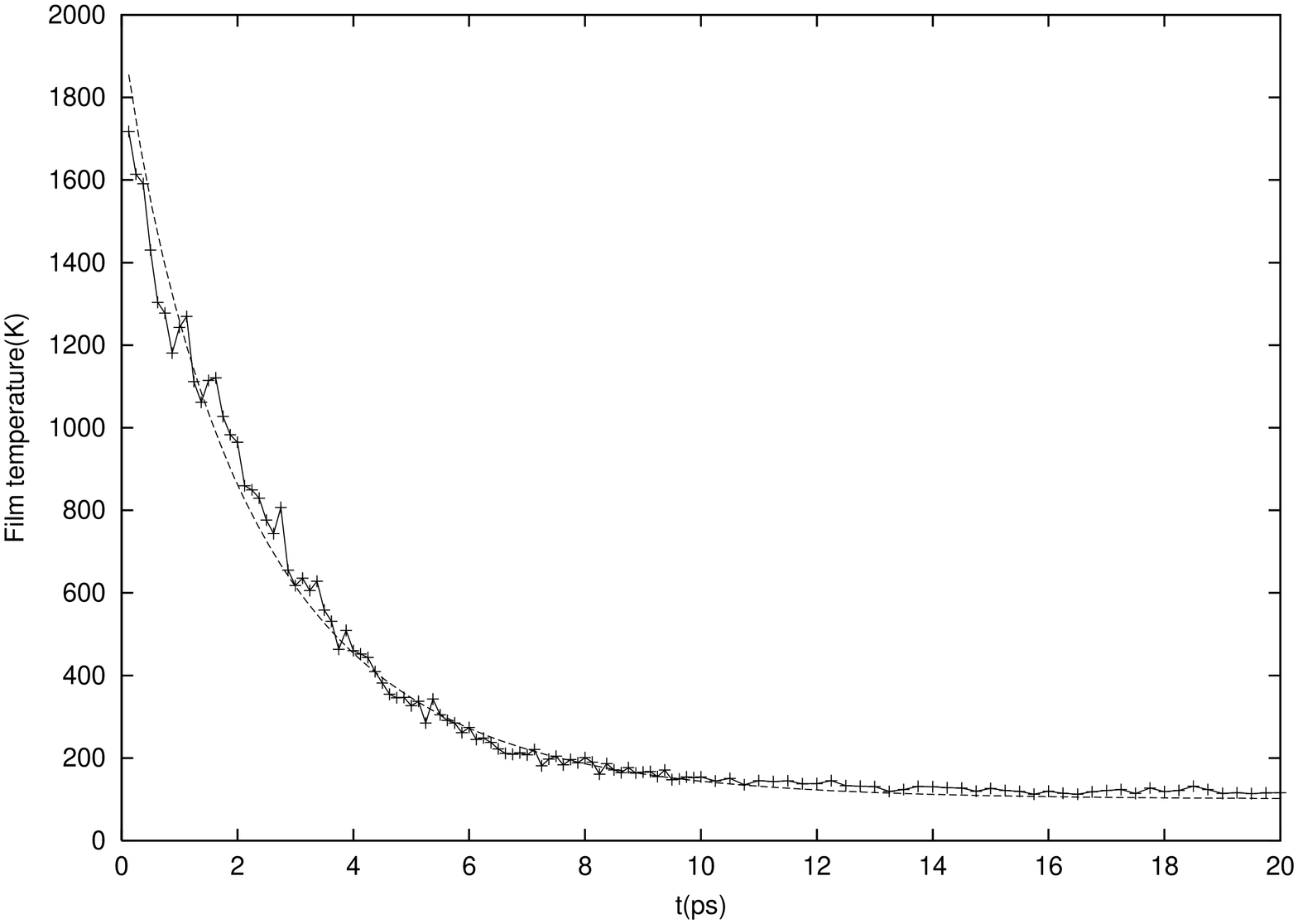,height=8.0truecm,angle=-90}}
\centerline{\epsfig{file=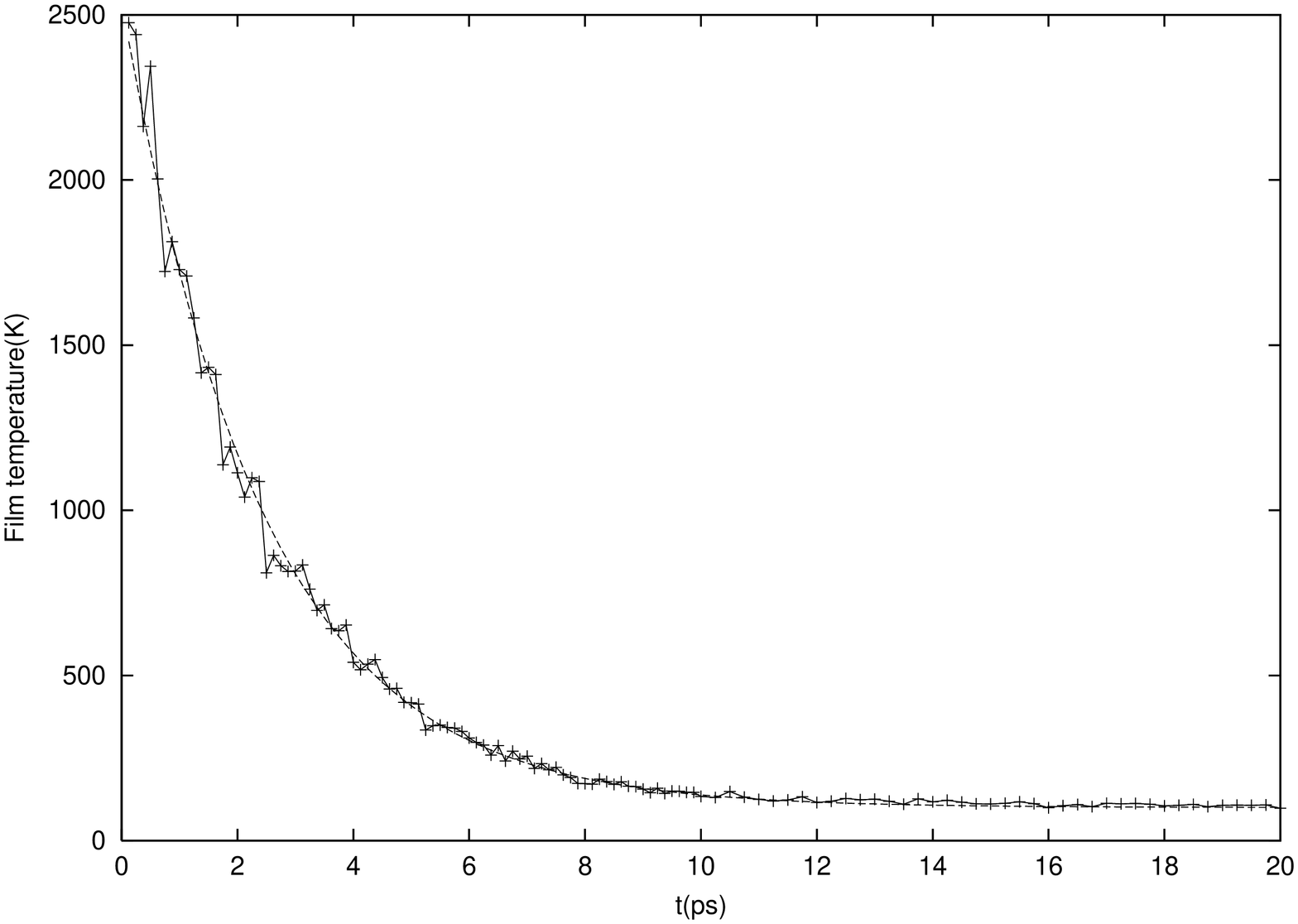,height=8.0truecm,angle=-90}}
\medskip
\caption{
\small
The observed stretched exponential functions for temperature
relaxations in films e1T100Q~(top), e1T100R20~(middle) and
e5T100R20~(bottom).  The best fit was retrieved according to
$T_{film}(t) = c + \exp \left(a t^{\beta}+b\right)$~K Fitting
parameters are $a$ = -0.46, -0.49, -0.41~fs$^{-1}$; $b$ = 7.57,
7.54, 7.79; $\beta$ = 0.95, 0.88, 1.00 and $c$ = 100~K, for e1T100Q,
e1T100R20 and e5T100R20, respectively. The $\beta \le$ 1 in all cases,
producing stretched exponential relaxation.
}
\label{fig3_Si}
\end{figure}

In the case of the e1T100Q model the fitting parameters are $a$ =
-0.46 (t is measured in fs) and $b$ = 7.57 and $c$ = 100~K (substrate
temperature).  The $\beta$ is equal to 0.95~($<$~1), producing
stretched exponential relaxation.  For the e1T100R20 model, if the fit
is done in the [0:10] ps interval, we obtain similar fitting
parameters; $\beta$ = 0.88, $a$ = -0.49, $b$ = 7.54, and for e5T100R20
$\beta$ = 1.00, $a$ = -0.41, $b$ = 7.79.  The fits are shown in Fig.\
\ref{fig3_Si}.  We note here that the search for the best fit is not a
well-defined problem in our case. A slight modification in the
parameter list can give qualitatively almost the same fit.

\subsection{Density}

To ignore the effect of the rough surface on the top side of the grown
film we identified two different cells: bulk and total sample~(see
section\ \ref{aC} and \cite{Kohary01}). The bottom of the latter was
0.77~\AA \ lower than the lowest $z$ coordinate; and the top of the
cell was the largest $z$ coordinate, among silicon atoms in the
amorphous network. For the bulk, this definition was slightly
modified, by simply decreasing the $z$ coordinate of the top of the
cell by 3~\AA.  The $x$ and $y$ size of the cell was determined
according to two-dimensional periodic boundary conditions.  Table\
\ref{table1_Si} contains the densities of different models.  Each row
is divided further into two rows. The upper ones always refer to the
bulk and the second rows refer to the total sample.  For realistic
density calculations one should consider only the "bulk densities",
which are always larger than for the total sample.  At T$_{sub}$ =
100~K bulk densities are between 2.3$-$2.5~g/cm$^3$, while in the
quick energy dissipation models~(T$_{sub}$ = 300~K) the structures
possess higher densities~(2.4$-$2.8~g/cm$^3$). For crystalline silicon
the density is equal to 2.33~g/cm$^3$, which is lower than the values
we obtained for a$-$Si, i.e.\ our tight-binding molecular dynamics
simulation provided more dense structures.  There was no significant
changing in density during relaxations.

\subsection{Ring statistics}

A surprising result was found in ring statistics. We defined a ring as
a closed path, which starts from a given atom walking only on the
first neighbor bonds. Every atom is visited only once and bonds can
exist only between adjacent atoms. The size of the ring is the number
of atoms forming a closed path. We considered only ring sizes smaller
or equal to eight but there are not any eightfold rings in the
networks.  The ring statistics are displayed in Table\
\ref{table2_Si}.  Networks prepared by our models have a significant
number of three membered rings, i.e.\ triangles are present in the
atomic arrangement.  Furthermore, squares can also be found.  Most of
the theoretical models for a$-$Si do not contain such fractions of the
structure. It seems to be an important result although a  neutron
diffraction measurement carried out on a pure evaporated amorphous
silicon sample, and evaluated by reverse Monte Carlo method had a
similar conclusion \cite{Kugler93}.  Nevertheless, the systematic
analysis of the structural data of the Cambridge Structural Database
suggests that the equilateral triangles and near planar squares can
also be natural local configurations inside the atomic arrangements of
a$-$Si \cite{Kugler01}.  All of the three independent pieces of
evidence support the existence of triangles in a$-$Si structures.

Significant differences can be found in ring statistics between
samples constructed by rapid quenching  and by atom-by-atom
deposition.  From Table\ \ref{table2_Si} it can be seen that for the
model e1T100Q~(rapid quenching) the number of rings is much lower than
for e1T100R20 and for e1T100Q, i.e.\ models prepared by different
techniques have different medium range order.  We can also conclude
that for the latter model the number of rings has decreased compared
to the initial e1T100R20 model.  With the exception of the ring
statistics which were different, all the other structural parameters
of these models were similar.

\begin{table}
\caption{
Ring statistics, for different first neighbor shells
2.6, 2.7, 2.82 and 3.0~\AA, respectively.
\label{table2_Si}
}
\vspace*{0.5truecm}
\centerline{
\begin{tabular}{|c||c|c|c|c|c|}
\hline
Model& \multicolumn{5}{|c|}{Ring statistics (2.59~\AA)}\\
\hline
& 3 & 4 & 5 & 6 & 7 \\
\hline
\hline
e1T100R5 &    12 &    10 &    50 &    86 &    95 \\
\hline
e1T100R10 &    25 &    10 &    61 &    85 &    93 \\
\hline
e1T100R20 &     9 &    14 &    68 &    89 &   102 \\
\hline
e5T100R5 &    11 &     9 &    51 &    45 &    65 \\
\hline
e5T100R10 &    14 &    21 &    74 &    84 &    96 \\
\hline
e5T100R20 &    18 &    20 &    78 &   103 &   116 \\
\hline
e1T300 &    13 &     9 &    33 &    36 &    46 \\
\hline
e5T300 &    13 &     8 &    44 &    31 &    33 \\
\hline
e10T300 &     7 &     8 &    16 &    19 &    12 \\
\hline
e1T100Q &     6 &    13 &    63 &    72 &    86 \\
\hline
\hline
Model& \multicolumn{5}{|c|}{Ring statistics (2.7~\AA)}\\
\hline
& 3 & 4 & 5 & 6 & 7 \\
\hline
\hline
e1T100R5 &    39 &    26 &    83 &   111 &   157 \\
\hline
e1T100R10 &    53 &    20 &    91 &   134 &   155 \\
\hline
e1T100R20 &    53 &    24 &    95 &   142 &   188 \\
\hline
e5T100R5 &    38 &    36 &   119 &   120 &   199 \\
\hline
e5T100R10 &    38 &    35 &   118 &   133 &   173 \\
\hline
e5T100R20 &    53 &    28 &   107 &   130 &   165 \\
\hline
e1T300 &    37 &    26 &    61 &    67 &    84 \\
\hline
e5T300 &    56 &    33 &    78 &    91 &   131 \\
\hline
e10T300 &    34 &    25 &    48 &    33 &    41 \\
\hline
e1T100Q &    35 &    26 &    75 &    98 &   130 \\
\hline
\end{tabular}
\begin{tabular}{|c||c|c|c|c|c|}
\hline
Model& \multicolumn{5}{|c|}{Ring statistics~(2.82~\AA)}\\
\hline
& 3 & 4 & 5 & 6 & 7 \\
\hline
\hline
e1T100R5 &    63 &    35 &   115 &   138 &   202 \\
\hline
e1T100R10 &    73 &    37 &   122 &   169 &   232 \\
\hline
e1T100R20 &    66 &    35 &   121 &   160 &   248 \\
\hline
e5T100R5 &    62 &    53 &   151 &   163 &   247 \\
\hline
e5T100R10 &    75 &    58 &   152 &   184 &   265 \\
\hline
e5T100R20 &    81 &    36 &   128 &   162 &   229 \\
\hline
e1T300 &    63 &    45 &    84 &    87 &    97 \\
\hline
e5T300 &    92 &    55 &   102 &   132 &   169 \\
\hline
e10T300 &    54 &    43 &    71 &    60 &    82 \\
\hline
e1T100Q &    75 &    29 &   109 &   129 &   181 \\
\hline
\hline
Model& \multicolumn{5}{|c|}{Ring statistics (3.0~\AA)}\\
\hline
& 3 & 4 & 5 & 6 & 7 \\
\hline
\hline
e1T100R5 &    80 &    45 &   120 &   174 &   235 \\
\hline
e1T100R10 &    91 &    50 &   135 &   181 &   273 \\
\hline
e1T100R20 &    82 &    56 &   137 &   177 &   276 \\
\hline
e5T100R5 &    95 &    84 &   169 &   230 &   323 \\
\hline
e5T100R10 &   102 &    66 &   170 &   197 &   307 \\
\hline
e5T100R20 &    95 &    58 &   157 &   188 &   293 \\
\hline
e1T300 &    81 &    55 &    87 &    95 &   118 \\
\hline
e5T300 &   123 &    72 &   136 &   158 &   277 \\
\hline
e10T300 &    72 &    47 &    85 &    69 &    92 \\
\hline
e1T100Q &    96 &    41 &   140 &   147 &   222 \\
\hline
\end{tabular}
}

\end{table}

\subsection{Coordination number}

It is known from diffractions that the first neighbor coordination
number in amorphous silicon is approximately four.  We have found that
this number can only be provided in our models if we choose 2.6~\AA \
for the first neighbor shell.  Nevertheless, the densities of our
models are bigger than in crystalline silicon.  This means, our
amorphous models do not have big voids, indicating that the model
potential shows more or less properties of amorphous and liquid
silicon systems at the same time.  In the case of taking the first
neighbor shell to 2.82 and 3.0~\AA, we received coordination numbers
close to five, which is between the average coordination number in
amorphous~($\langle Z \rangle \approx$ 4) and liquid~($\langle Z
\rangle \approx$ 6) silicon.

Figure\ \ref{fig1_Si} illustrates the first neighbor distances. It is
clear that in the case of 2.6~\AA \ first neighbor shells, the lowest
first neighbor distances are dominated by Si$_{Z=4}$$-$Si$_{Z=4}$
distances~(dashed line). The distances between other coordinated
silicon atoms only contribute to larger distances than the interatomic
separation in crystalline silicon~(2.36~\AA) and do not contribute to
lower distances, which can be seen in the experimental curve~(dotted
line) in Fig.\ \ref{fig1_Si}.

To our present knowledge, there is no silicon tight-binding model
potential, which can more accurately describe the amorphous phase.
Vink {\it et al.} proposed a modified Stillinger-Weber classical
empirical potential very recently \cite{Vink01}, which they believe
can be used to describe the structure of amorphous silicon.

\subsection{Radial distribution and bond angle distribution functions}

In Fig.\ \ref{fig1_Si} the first and second neighbor peaks of e1T100Q
model can be seen. All the other models show similar radial
distribution functions. The tight-binding potential applied in our
computer experiments overestimates the first neighbor distance.
However, if the potential would provide only silicon atoms with a
coordination equal to four, the position of the peak for the first
neighbor shell would be in the right position. 

The main contribution to the bond angle distribution arises from
angles between 95$^\circ$ and 125$^\circ$. We found a large amount of
triangles in our samples, there is therefore a significant
contribution from bond angles close to 60$^\circ$. We believe this
causes the average bond angle to be smaller than 109$^\circ$, which is
the tetrahedral angle in diamond structures, in our systems.

\section{Summary}

We have developed a tight-binding molecular dynamics computer code to
simulate the preparation procedure of amorphous silicon, which are
grown by a vapor deposition technique. Nine structures were simulated
by this method. An additional tenth model~(e1T100Q) was prepared by
rapid cooling in order to make a comparison between the atom-by-atom
deposition on a substrate and melt-quenching preparation techniques.

The most important difference we have found between the models
prepared in varied conditions, is in the ring statistics, i.e.\ in
medium range order. In our simulations triangles are present in the
atomic arrangement. Two other independent pieces of evidence support
the existence of triangles in a$-$Si structures.  Nevertheless, the
fragments like equilateral triangles have never been considered at
electronic density of states calculations or with unsolved problems
such as the breaking of weak bonds by prolonged illumination, which is
the major mechanism for the light-induced defect creation
\cite{Shimakawa95}.

The tight-binding model potential used in our computer experiments
slightly overestimates the bond distances, giving values which were
observed previously in liquid forms \cite{Ishimaru96}. A more
important problem is that the TB potential provides overcoordinated
atomic sites (five- even sixfold coordinated atoms) during the
amorphous silicon simulations. The sixfold coordinated atoms we
obtained can be found in the liquid phase of silicon instead of in the
amorphous structure. The well-known Stillinger-Weber classical
empirical potential for amorphous silicon had a similar difficulty,
yet this has recently been successfully modified \cite{Vink01}.
A~possible future aim is to improve this TB potential by a similar
process.

\chapter{One-dimensional hopping in disordered organic solids}
\markboth{1D HOPPING IN DISORDERED ORGANIC SOLIDS}{1D HOPPING IN DISORDERED ORGANIC SOLIDS}
\label{hopping}

\section{Introduction}

Transport of charge carriers in disordered organic solids, such as
molecularly doped  polymers, conjugated polymers and organic glasses
has been the subject of intensive experimental and theoretical study
for about 20 years \cite{Baessler93}. In recent years particular
attention has been devoted to the columnar discotic
liquid--crystalline glasses due to their potential technical
applications for electrophotography and as transport materials in
light-emitting diodes \cite{Bacher97, Christ97}. Dielectric
measurements have clarified that charge transport in most of such
materials is extremely anisotropic \cite{Boden95} so that it is
reasonable to describe the transport of charge carriers as a
one--dimensional process \cite{Bleyl99}.  Moreover, experimentally
observed dependences of the conductivity on the frequency, on the
strength of the applied electric field and on the temperature evidence
that an incoherent hopping process is the dominant transport mechanism
in such materials \cite{Boden95, Bleyl99, Ochse99a, Ochse99b}.  It has
been found reasonable to describe the transport of charge carriers as
a hopping process via molecules arranged in spatially ordered
one--dimensional chains \cite{Bleyl99}.  The energy levels which are
responsible for charge transport are usually characterized by a
Gaussian density of states (DOS) \cite{Baessler93, Dunlap96, Bleyl99,
Ochse99a}:

\begin{equation}
g(\epsilon) = \frac{1}{\sqrt{2\pi\sigma^2}}
\exp{\left(-\frac{\epsilon^2}{2\sigma^2}\right)}.
\label{1}
\end{equation}
The origins of the energy disorder characterized by parameter $\sigma$
could be fluctuations in the polarization energy as well as dipolar
interactions or impurity molecules \cite{Bleyl99}. Such systems
represent an excellent field to test various theoretical approaches.
These approaches differ from each other, for example, with respect to
the choice of transition rates for single hopping events. The choice
of transition rates is of course limited as they must fulfill the
detailed balance condition to be physically reasonable in thermal
equilibrium.  The most plausible transition probabilities for the
hopping of charge carriers via localized states were suggested by
Miller and Abrahams \cite{Miller60, Baessler93, Bleyl99}: the jump
rate from an occupied localized state $i$ to an empty state $j$
separated by distance $R_{i,j}$ is the product of  a prefactor
$\Gamma_{1}$, overlap of the wave functions of states $i$ and $j$, and
the Boltzmann factor for jumps upward in energy \cite{Miller60},

\begin{equation}
\Gamma_{i,j}=\Gamma_{1}\exp{\left(-\frac{2R_{i,j}}{\alpha}\right)}
\exp{\left(-\frac{\epsilon_{j}- \epsilon_{i} + 
\mid\epsilon_{j}-\epsilon_{i}\mid}{2kT}\right)}.  
\label{2}
\end{equation}
Here $\alpha$ is the decay length of the carrier wave function in the
localized states and $k$ is the Boltzmann constant.  Allowing only
hops to nearest neighbors and assuming that chains of localized states
are spatially regular we omit the $R$-dependence of the transition
probabilities and use the Miller-Abrahams transition rates in the form

\begin{equation}
\Gamma_{i,j}=\nu_0\exp{\left(-\frac{\epsilon_j-
\epsilon_i + \mid\epsilon_j-\epsilon_i\mid}{2kT}\right)}  
\label{3}
\end{equation}
with prefactor $\nu_0=\Gamma_1\exp{(-2b/\alpha)}$, $b$ being the
distance between the neighboring sites on the conduction chain.
Nearest--neighbor hopping within the model described by Eq.\ (\ref{1})
and Eq.\ (\ref{3}) has been recently studied by computer simulations
\cite{Bleyl99}.  Other forms of transition probabilities have also
been considered in the literature. For example, a symmetric form was
recently used in order to analyze analytically the field and
temperature dependences of the hopping conductivity in
one--dimensional systems \cite{Dunlap96}:

\begin{equation}
\Gamma_{i\pm1,i} = \nu_0
\exp{\left(-\frac{\epsilon_{i\pm1}-\epsilon_i}{2kT}\right)}.  
\label{4}
\end{equation}

Two models have been discussed in the literature with respect to the
space-energy correlations of localized states responsible for
transport. In the so-called Gaussian disorder model (GDM), such
correlations are neglected \cite{Baessler93, Bleyl99}. In the
correlated disorder model (CDM), it is assumed that energy
distributions for spatially close sites are correlated, for example,
due to dipole or quadrapole interactions \cite{Dunlap96, Gartstein95,
Novikov98a, Novikov98b}. 

The drift mobility of carriers in organic disordered media usually
demonstrates exponential dependences on temperature $T$ and electric
field $E$ in the form 

\begin{equation}
\mu \propto \exp{\left[-\left(\frac{T_{0}}{T}\right)^2\right]}
\exp{\left[\sqrt{\frac{E}{E_{0}}}\right]}, 
\end{equation}
where $T_{0}$ and $E_{0}$ are parameters.  Much attention has been
paid in recent years to the dependence of the mobility on electric
field \cite{Dunlap96, Gartstein95, Novikov98a, Novikov98b} and it has
been claimed that in three--dimensional (3D) systems the observed
field dependence in a wide range of electric fields can be explained
only by the CDM.  We check below whether this is true for a
one--dimensional (1D) system. 

As far as the dependence of the mobility on temperature is concerned,
the situation is widely believed to have been cleared up many years
ago.  Computer simulations \cite{Schoenherr81} and analytical
calculations \cite{Gruenewald84, Movaghar86} in 3D case were carried
out giving the dependence 

\begin{equation}
\mu\propto\exp{\left[-\left(C\frac{\sigma}{kT}\right)^2\right]}
\label{5}
\end{equation}
with $C\approx2/3$. This expression is widely believed to be universal
for the above model and it is often used to determine disorder
parameter $\sigma$ of assumed Gaussian DOS in various materials from
the experimental measurements \cite{Ochse99a, Nemeth-Buhin96}.
However, it is necessary to emphasize that there is no agreement among
researchers concerning the magnitude of the coefficient $C$ for
one--dimensional systems.  Ochse \etal\cite{Ochse99a} used the 3D
value $C=2/3$ for columnar systems with one--dimensional transport,
while Bleyl \etal \cite{Bleyl99} obtained $C\approx0.9$ in their
computer simulations for 1D case with asymmetric transition
probabilities. The analytic calculations of Dunlap \etal
\cite{Dunlap96} for symmetric probabilities predict $C=1$ in 1D case. 

In this chapter the temperature dependence in Eq.\ (\ref{5}) will be
studied in order to clarify this dependence for the GDM and the CDM.
This chapter concentrates on the contribution of the author in that
field, i.e.  in computer simulations \cite{kohary_prb01_hop}. However,
the work was done together with the analytical work performed by Dr.
Harald Cordes \cite{cordes_prb01}.  The analytical results will be
summarized in section \ref{analytic_work} and may occasionally be
mentioned together with the computer simulation results as well. In
section \ref{mc_simulations} the results of Monte Carlo simulations
will be gathered.

\section{Analytic theory}
\label{analytic_work}

The steady--state drift velocity of carriers in one--dimensional
periodic systems can be written exactly using the general solution
found by Derrida \cite{Derrida83},

\begin{equation}
v=\frac{Nb\left[1-
\prod\limits_{k=0}^{N-1}\frac{\Gamma_{k+1,k}}{\Gamma_{k,k+1}}\right]}{
\sum\limits_{k=0}^{N-1}\frac{1}{\Gamma_{k,k+1}}\left[1+\sum\limits_{i=1}^{N-1}
\prod\limits_{j=1}^i\frac{\Gamma_{k+j,k+j-1}}{\Gamma_{k+j,k+j+1}}\right]},
\label{8}
\end{equation}
where $N$ is the periodicity of the system and $b$ is the distance
between nearest neighbor sites.  This formula was recently used
by Dunlap \etal \cite{Dunlap96} to study the field dependence of the
drift mobility in a one--dimensional system with symmetric transition
rates \cite{Dunlap96}.  The drift mobility $\mu$ is related to the
steady--state velocity as

\begin{equation}
\mu=\frac{v}{E}.
\label{12}
\end{equation}
However in experiments, the drift mobility is usually obtained by the
time--of--flight technique. Here  charge carriers pass only once
through a sample of finite thickness \cite{Bleyl99, Ochse99a} and
drift mobility is calculated by dividing the sample length $Nb$ by the
mean transit time $\bar{t}_{0N}$ and electric field $E$:

\begin{equation}
\mu=\frac{Nb}{\bar{t}_{0N}E}.
\label{9}
\end{equation}
Murthy and Kehr \cite{Murthy89} have derived the analytical expression
for the mean transit time of carriers through a given chain of $N+1$
sites

\begin{equation}
\bar{t}_{0N}=
\sum\limits_{k=0}^{N-1}\frac{1}{\Gamma_{k,k+1}}+
\sum\limits_{k=0}^{N-2}\frac{1}{\Gamma_{k,k+1}}
\sum\limits_{i=k+1}^{N-1}\prod\limits_{j=k+1}^i
\frac{\Gamma_{j,j-1}}{\Gamma_{j,j+1}}.
\label{10}
\end{equation}
This formula gives the transit time for a charge carrier that starts
on site $0$ and finishes on site $N$.  The time is averaged over many
transit times through the same chain with given values of transition
probabilities $\Gamma_{ij}$.  

One should be cautious with the application of Eqs.\ (\ref{9}) and
(\ref{10}) at low electric fields. Even without an applied electric
field carriers starting at site $0$ will pass through the system
solely due to a diffusion process. At low fields, diffusion dominates
the motion of carriers and it would be wrong to use Eq.\ (\ref{9}).
In such cases it would be more appropriate to estimate the mobility via
the diffusion formula

\begin{equation}
\mu=\frac{e}{kT}\frac{b^2 N^2}{2\bar{t}_{0N}}.
\label{11}
\end{equation}
which presumes the validity of the Einstein relation. 

Note that Eqs.\ (\ref{8}), (\ref{12}) and Eqs.\ (\ref{9}), (\ref{10})
are valid for any type of nearest--neighbor hopping of
noninteracting carriers in one--dimensional systems.

\subsection{Drift mobility in the random-barrier model}
\markboth{1D HOPPING IN DISORDERED ORGANIC SOLIDS}{DRIFT MOBILITY IN RBM}

We start our analysis with a simple random-barrier model (RBM).  In
the RBM, all sites have equal energies being separated by energy
barriers.  Transition probabilities in the presence of electric field
$E$ are given by expressions 

\begin{equation}
\Gamma_{k,k+1}=\nu_0\exp\left(-\frac{\Delta_k-\frac{1}{2}ebE+
|\Delta_k-\frac{1}{2}ebE|}{2kT}\right)
\label{13a}
\end{equation}
and

\begin{equation}
\Gamma_{k+1,k}=\nu_0\exp\left(-\frac{\Delta_k+\frac{1}{2}ebE+
|\Delta_k+\frac{1}{2}ebE|}{2kT}\right),
\label{13b}
\end{equation}
where $\Delta_k$ is the energy barrier between sites $k$ and $k+1$. 

\begin{figure}
\centerline{\epsfig{file=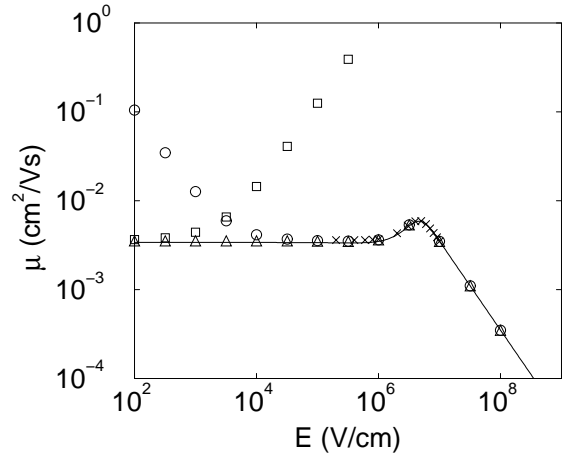,height=6.0truecm}}
\medskip
\caption{
\small
Field dependence of the carrier mobility in the RBM for Gaussian
distribution of barrier heights with the width $\sigma=$ 50~meV at
temperature $kT$ = 25~meV. The solid line represents exact solution
for the infinite chain. Data shown by circles and squares were
calculated via Eq.\ (\ref{10}) using drift and diffusion relation,
respectively.  Data shown by triangles were calculated via Eq.\
(\ref{8}), while those shown by crosses were obtained by Monte Carlo
simulations presented in the section \ref{mc_simulations}. The chain
length $N$ was 500 and averaging was performed over $c=1000$ chains.
}
\label{fig1_hop}
\end{figure}

In the limit of long periods (long chains), $N\gg 1$, calculation of
the drift mobility via Eqs.\ (\ref{8}), (\ref{12}) or Eqs.\ (\ref{9}),
(\ref{10}) for any chain with a given distribution of transition rates
is equivalent to the averaging of Eq.\ (\ref{8}) or Eq.\ (\ref{10})
over all possible distributions of transition rates determined by
disorder.  In the following we will call this "averaging over
disorder".  Within the RBM, the averaging over disorder is given by
\cite{Derrida83, Murthy89} 

\begin{equation}
v=b\left[1-\left\langle\frac{\Gamma_{k+1,k}}{\Gamma_{k,k+1}}\right\rangle
\right]\Bigg/\left\langle\frac{1}{\Gamma_{k,k+1}}\right\rangle ,
\label{14}
\end{equation}
where $\langle \dots \rangle$ denotes the averaging over the distribution
of barrier heights $p(\Delta)$. Note that Eq.\ (\ref{14}) can be
derived by both equations, Eq.\ (\ref{8}) and Eq.\ (\ref{10}). The
explicit expressions for the averaging in the RBM are 

\begin{equation}
\left\langle\frac{1}{\Gamma_{k,k+1}}\right\rangle=
\nu_0^{-1}\int\limits_0^{ebE/2} d\Delta\ p(\Delta)+\nu_0^{-1}
\int\limits_{ebE/2}^\infty d\Delta\ p(\Delta) e^{\frac{\Delta-ebE/2}{kT}} 
\label{15a}
\end{equation}
and

\begin{equation}
\left\langle\frac{\Gamma_{k+1,k}}{\Gamma_{k,k+1}}\right\rangle=
\int\limits_0^{ebE/2} d\Delta\ p(\Delta)
e^{-\frac{\Delta+ebE/2}{kT}} +
\int\limits_{ebE/2}^\infty d\Delta\ p(\Delta) e^{-\frac{ebE}{kT}}. 
\label{15b}
\end{equation}
Assuming a Gaussian distribution of barriers,

\begin{equation}
p(\Delta)=\frac{2}{\sqrt{2\pi\sigma^2}}
\exp{\left(-\frac{\Delta^2}{2\sigma^2}\right)},
\label{7}
\end{equation}
one obtains for the drift mobility

\begin{equation}
\mu=\frac{b\nu_0}{E}\ \frac{1-e^{\frac{\sigma^2}{2(kT)^2}-\frac{ebE}{2kT}}\left[
\mbox{erf}\left(\frac{ebE}{\sqrt{8}\sigma}+\frac{\sigma}{\sqrt{2}kT}\right)-
\mbox{erf}\left(\frac{\sigma}{\sqrt{2}kT}\right)
\right]-e^{-\frac{ebE}{kT}}\left[1-\mbox{erf}\left(\frac{ebE}{\sqrt{8}\sigma}\right)\right]
}{
\mbox{erf}\left(\frac{ebE}{\sqrt{8}\sigma}\right)+e^{\frac{\sigma^2}{2(kT)^2}-\frac{ebE}{2kT}}
\left[1-\mbox{erf}\left(\frac{ebE}{\sqrt{8}\sigma}-\frac{\sigma}{\sqrt{2}kT}\right)\right]
},
\label{16}
\end{equation}
where $\mbox{erf}(x)=\frac{2}{\sqrt{\pi}}\int_0^x \exp(-y^2) dy$ is the error
function.  In Fig.\ \ref{fig1_hop} the calculated field dependences of
the drift mobility are shown for parameters $b$ = 3.6~\AA, $\sigma$ =
50~meV, $kT$ = 25~meV. The solid line represents the exact solution
for infinite chain given by Eq.\ (\ref{16}). All points in the figure
correspond to mobilities for finite chains of $N=500$ sites averaged
over $1000$ different chains. Circles were obtained via Eqs.\
(\ref{9}), (\ref{10}) while squares via Eqs.\ (\ref{10}), (\ref{11}).
Results of our Monte Carlo computer simulation are shown by crosses to
demonstrate the excellent agreement of the simulation results with the
analytical theory.  At low fields, drift approximation (Eqs.\ (\ref{9}),
(\ref{10})) for finite systems leads to the increase of the calculated
mobility with decreasing field strength. Similar results are obtained
for all considered models of disorder and various forms of the
transition probabilities between neighboring sites. This result simply
reflects the fact that charge carriers can penetrate through a finite
system via diffusion motion even in the absence of an external
electric field.  By using Eq.\ (\ref{9}) one overestimates the
mobility at low fields. The same happens with using Eq.\ (\ref{11})
at higher fields. Comparison of Eqs.\ (\ref{9}) and (\ref{11}) reveals
the strength of the electric field at which a transition from the
diffusion-controlled to the drift-controlled transit times takes
place: $E\simeq 2kT/eNb$. 

At low fields, transition time $\bar{t}_{0N}$ does not depend on the
strength of the electric field $E$ being determined mostly by
diffusion. Using Eq.\ (\ref{9}) one artificially obtains an increasing
drift mobility at decreasing field $E$. In numerous publications
experimental results were reported that evidence the decrease of the
drift mobility with increasing electric field \cite{Peled88, Schein92,
Abkowitz92}. This was always observed at high temperatures at which
diffusion can really dominate the motion of charge carriers.
Experimental evidence for decreasing mobility with increasing field is
usually obtained using equations similar to Eq.\ (\ref{9}), where the
drift mobility is determined via the measured transit time by dividing
the sample length by the value of this time and by the value of the
field strength. We believe that this calculation is not appropriate
and a diffusion equation similar to Eq.\ (\ref{11}) should be used at
low fields and high temperatures.  In 3D case, this equation should be
slightly modified, though the conclusion is the same: decreasing of
the drift mobility with increasing field strength at low fields is an
artifact caused by using the wrong equation [similar to Eq.\
(\ref{9})] for evaluation of the mobility on the basis of the measured
transit time.  In our 1D calculations we illustrate this idea by using
Eq.\ (\ref{11}) instead of Eq.\ (\ref{9}) for finite systems at low
electric fields. The result is shown by squares  in Fig.\
\ref{fig1_hop}.  The excellent agreement with the exact calculation
for the infinite system at low fields confirms our conclusion.  For
the chosen parameters, the linear transport regime with the mobility
independent of the electric field is valid up to the field strength of
approximately $10^6$ V/cm, where a nonlinear regime starts with the
mobility increasing with electric field. The latter regime is valid in
a rather narrow range. At higher electric fields it is replaced by the
trivial regime in which electric field eliminates the energy barriers
between the sites of the chain and the transit time becomes field
independent. In such a case mobility decreases proportionally to
$1/E$. 

\begin{figure}
\centerline{\epsfig{file=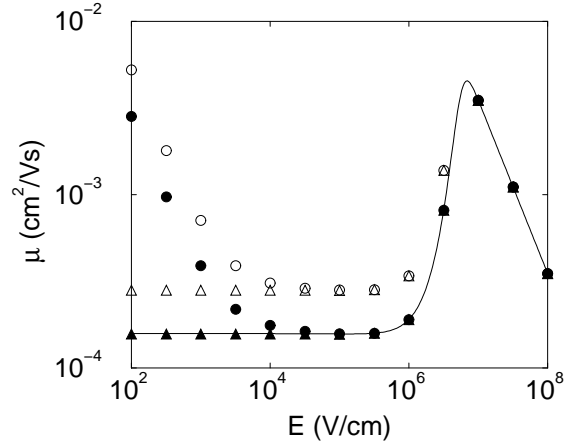,height=6.0truecm}}
\medskip
\caption{
\small
Field dependence of the carrier mobility in the RBM. Chain length
$N=500$, number of chains $c=1000$, width of barrier height
distribution $\sigma$ = 50~meV, temperature $kT$ = 15~meV. Open
circles and open triangles show the data obtained by averaging over
chain mobilities calculated by Eq.\ (\ref{10}) and Eq.\ (\ref{8}),
respectively.  Solid symbols give the data obtained by the
corresponding averaging over the inverse mobilities.
}
\label{fig2_hop}
\end{figure}

In Fig.\ \ref{fig2_hop} the field dependence of the drift mobility is
shown for a lower temperature $kT$ = 15~meV for the same system as in
Fig.\ \ref{fig1_hop} with $\sigma$ = 50~meV and $b$ = 3.6~\AA. Clear
mesoscopic effects are seen in this figure as evidenced by the
difference in the results obtained by averaging the mobilities
(open symbols) and those obtained by averaging the transit times or
the inverse mobilities (solid symbols). In all cases, averagings were
performed over $1000$ different chains, each chain consisting of $500$
sites. The difference in the results of the two different averaging
procedures reflects the presence of "fast" and "slow" chains.
Averaging of times corresponds to the successive arrangement of chains
where slow electron transitions over untypically high energy barriers
determine the transport coefficients. On the contrary, the averaging
of the mobilities corresponds to the parallel arrangement of chains,
in which the fastest chains with untypically low energy barriers
provide fast transit times of charge carriers. In the next section, we
show that qualitatively the same effects are also inherent for the
more realistic random-energy model.       

\subsection{Drift mobility in the random-energy model without correlations}
\markboth{1D HOPPING IN DISORDERED ORGANIC SOLIDS}{DRIFT MOBILITY IN GDM}

In this section we consider a random--energy model with a Gaussian DOS
described by Eq.\ (\ref{1}), presuming that there are no correlations
between spatial positions of chain sites and their energies. This
model is called in the literature a Gaussian disorder model (GDM).  It
has recently been suggested for the transport of charge carriers in
discotic columnar liquid crystalline glasses with transition rates
described by Eq.\ (\ref{3}) and it was studied by computer simulations
\cite{Bleyl99}. The exact analytic results for this model
based on Eqs.\ (\ref{8}) and (\ref{10}) will be presented below. 

In Eq.\ (\ref{3}), site energies $\epsilon_{k}$ depend on the strength
of the electric field $E$. They are related to the zero field site
energies $\phi_{k}$ as $\epsilon_{k}=\phi_{k}-ekbE$.  In accordance
with Eq.\ (\ref{3}), the ratio of forward $\Gamma_{k,k+1}$ and
backward $\Gamma_{k+1,k}$ hopping rates for any pair of neighboring
sites is  

\begin{equation}
\frac{\Gamma_{k+1,k}}{\Gamma_{k,k+1}}=
\exp\left(\frac{\epsilon_{k+1}-\epsilon_k}{kT}\right)
\label{17}
\end{equation}
equivalent to the condition of the detailed balance, which recently
\cite{Kehr96} has been used to determine the diffusion coefficient in
thermal equilibrium for any form of the transition rates.

Inserting relation (\ref{17}) into Eq.\ (\ref{8}), one obtains

\begin{equation}
v=\frac{bN\left[1-\exp\left(-\frac{NebE}{kT}\right)\right]}{
\sum\limits_{k=0}^{N-1}\frac{1}{\Gamma_{k,k+1}}+
\sum\limits_{k=0}^{N-1}\exp\left(-\frac{\phi_k}{kT}\right)
\sum\limits_{i=1}^{N-1}\frac{\exp\left(-\frac{iebE}{kT}\right)}
{\exp\left(-\frac{\phi_{k+i}}{kT}\right)\Gamma_{k+i,k+i+1}}}.
\label{18}
\end{equation}
The only random variables in Eq.\ (\ref{18}) are the zero field site
energies $\phi_k$ and the forward transition rates
$\Gamma_{k+i,k+i+1}$, which in turn depend only on neighboring site
energies $\phi_{k+i}$ and $\phi_{k+i+1}$. Since in all products
appearing in the second term of the denominator, $\phi_k$ is not
correlated with $\phi_{k+i}$ or $\Gamma_{k+i,k+i+1}$, averaging over
disorder can be carried out for $\exp(-\phi_k/kT)$ and
$\exp(\phi_{k+i}/kT)\Gamma_{k+i,k+i+1}^{-1}$ separately.  Therefore
the steady-state velocity and the mean transit time of carriers
through long chains, $N\gg 1$ are given by

\begin{equation}
v=b \left(
\left\langle\frac{1}{\Gamma_{k,k+1}}\right\rangle+
\left\langle\frac{\left\langle\exp\left(-\frac{\phi_k}{kT}\right)\right\rangle}
{\exp\left(-\frac{\phi_{k}}{kT}\right)\Gamma_{k,k+1}}\right\rangle
\frac{\exp\left(-\frac{ebE}{kT}\right)}{1-\exp\left(-\frac{ebE}{kT}\right)}
\right)^{-1},
\label{19}
\end{equation}

\begin{equation}
\frac{\bar{t}_{0N}}{N}=\left\langle\frac{1}{\Gamma_{k,k+1}}\right\rangle+
\left\langle\frac{
\left\langle\exp\left(-\frac{\phi_k}{kT}\right)\right\rangle}
{\exp\left(-\frac{\phi_{k}}{kT}\right)\Gamma_{k,k+1}}\right\rangle
\frac{\exp\left(-\frac{ebE}{kT}\right)}
{1-\exp\left(-\frac{ebE}{kT}\right)}\, .
\label{21}
\end{equation}
The relation $v=bN/\bar{t}_{0N}$ provides the equivalence of Eq.\
(\ref{21}) and (\ref{19}) in long chains as should be expected.  Note
that Eqs. (\ref{19}) and (\ref{21}) are exact for all kinds of
nearest--neighbor hopping rates fulfilling Eq.\ (\ref{17}). Thus they
can as well be applied to the symmetric transition rates given by Eq.\
(\ref{4}).

\begin{figure}
\centerline{\epsfig{file=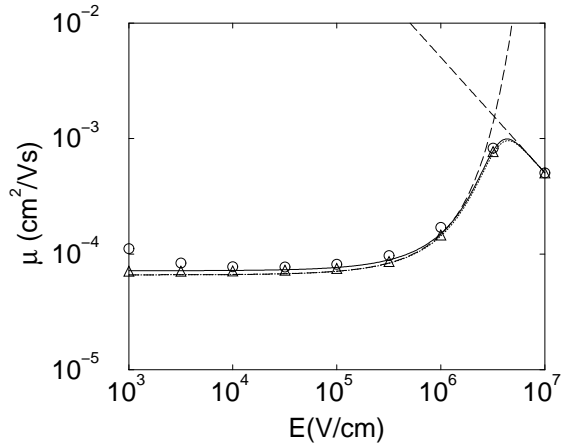,height=6.0truecm}}
\medskip
\caption{
\small
Field dependence of the carrier mobility for nearest neighbor hopping
with Miller--Abrahams rates between sites with Gaussian energy
distribution.  The solid line shows the exact solution for an infinite
chain. Dashed lines correspond to the low-field and high-field
approximations, while the dotted line gives the interpolation between
them. Circles show the averaged mobilities of $c=1000$ chains with
$N=2000$ sites each calculated according to Eq.\ (\ref{10}) and Eq.\
(\ref{9}).  Triangles represent the results obtained by averaging 
the inverse mobilities for the same chains calculated by Eq.\
(\ref{18}) and Eq.\ (\ref{12}).  Temperature and the width of the
energy distribution were chosen as $kT$ = 25~meV and $\sigma$ = 50~meV.
}
\label{fig3_hop}
\end{figure}

\begin{figure}
\centerline{\epsfig{file=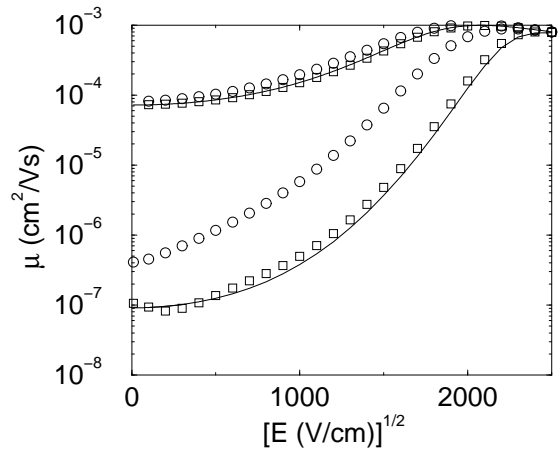,height=6.0truecm}}
\medskip
\caption{
\small
Poole--Frenkel plots for carrier mobilities at $kT$ = 25~meV (upper
curves) and $kT$ = 15~meV (lower curves). Circles show the averaged
mobilities calculated by Eq.\ (\ref{18}), squares show the
corresponding results obtained by averaging of inverse mobilities.
Number of chains was $c=10000$ with $N=500$ sites each.
}
\label{fig4_hop}
\end{figure}

The results of the averaging over Gaussian disorder for various terms
in Eqs.\ (\ref{19}), (\ref{21}) are:

\begin{equation}
\left\langle e^{-\frac{\phi_k}{kT}}\right\rangle
=\exp\left[\frac{1}{2}\left(\frac{\sigma}{kT}\right)^2\right],
\end{equation}

\begin{equation}
\left\langle\frac{1}{\Gamma_{k,k+1}}\right\rangle
=
\frac{1}{2\nu_0}\left\{
1+\mbox{erf}\left(\frac{ebE}{2\sigma}\right)
+\exp\left[\left(\frac{\sigma}{kT}\right)^2\right]\exp\left(-\frac{ebE}{kT}\right)
\left[1+\mbox{erf}\left(\frac{\sigma}{kT}-\frac{ebE}{2\sigma}\right)\right]\right\}
\end{equation}
and

\begin{equation}
\left\langle\frac{1}
{\exp\left({-\frac{\phi_{k}}{kT}}\right)\Gamma_{k,k+1}}\right\rangle
=
\frac{e^{\frac{\sigma^2}{2 (kT)^2}}}{2\nu_0}\left\{
1+\mbox{erf}\left(\frac{\sigma}{2kT}+\frac{ebE}{2\sigma}\right)+
e^{-\frac{ebE}{kT}}
\left[1+\mbox{erf}\left(\frac{\sigma}{2kT}-\frac{ebE}{2\sigma}\right)\right]
\right\}.
\end{equation}
Thus in the one--dimensional GDM the drift mobility of carriers on an
infinite  chain is 

\begin{eqnarray}
\mu&=&\frac{2\nu_0b}{E}\Bigg(
1+\mbox{erf}\left(\frac{ebE}{2\sigma}\right)+
e^{\frac{\sigma^2}{(kT)^2}-\frac{ebE}{kT}}
\left[1+\mbox{erf}\left(
\frac{\sigma}{kT}-\frac{ebE}{2\sigma}\right)
\right]+\nonumber\\&&
\frac{e^{\frac{\sigma^2}{(kT)^2}-\frac{ebE}{kT}}}
{1-e^{\frac{ebE}{kT}}}
\left\{e^{-\frac{ebE}{kT}}
\left[1+\mbox{erf}\left(\frac{\sigma}{2kT}-\frac{ebE}{2\sigma}\right)\right]+
1+\mbox{erf}\left(\frac{\sigma}{2kT}+\frac{ebE}{2\sigma}\right)
\right\}
\Bigg)^{-1}.
\label{26}
\end{eqnarray}
At low electric fields this exact expression can be approximated by

\begin{equation}
\mu\simeq\frac{\nu_0b}{2E}\exp\left(-\frac{\sigma^2}{(kT)^2}\right)
\left[\exp\left(\frac{ebE}{kT}\right)-1\right],
\label{27}
\end{equation}
while at high fields $\mu\simeq \nu_0b/E$.  Interpolation of both
approximations gives

\begin{equation}
\mu\simeq\frac{\nu_0b}{E}\left[1+\frac{2\exp\left(\frac{\sigma^2}{(kT)^2}\right)}
{\exp\left(\frac{ebE}{kT}\right)-1}\right]^{-1}.
\label{29}
\end{equation}
Field dependences of the drift mobility are shown in Fig.\
\ref{fig3_hop} for $b$ = 3.6~\AA, $\sigma$ = 50~meV, $kT$ = 25~meV.
The solid line represents the exact solution for an infinite chain
given by Eq.\ (\ref{26}). Dashed lines show approximations via Eqs.\
(\ref{27})  and $\mu\simeq \nu_0b/E$, respectively. The dotted line
illustrates the interpolation formula [ Eq.\ (\ref{29})] which appears
to be in excellent agreement with the exact solution.  Circles and
triangles in the figure show the calculated results for finite systems
of $2000$ sites averaged over $1000$ different chains once using the
averaging of inverse transit times (circles) and once using the
averaging of inverse velocities (triangles).  Qualitatively, the
results for the random-energy model in Fig.\ \ref{fig3_hop} resemble
those for the random-barrier model from Figs.\ \ref{fig1_hop} and
\ref{fig2_hop} and we refer the reader to the corresponding parts of
the previous section, where the latter results were discussed. In
Fig.\ \ref{fig4_hop}, the field dependences of the drift mobility are
shown in the Poole-Frenkel representation ($\ln\mu$ versus $\sqrt E$)
for two different temperatures and two different averaging procedures.
The figure clearly shows that in the chosen model the analytic solution of
the carrier mobility hardly can be described by the Poole-Frenkel law
$\ln\mu\propto\sqrt E$. This conclusion is in agreement with previous
studies of Gartstein and Conwell \cite{Gartstein95}, Dunlap \etal
\cite{Dunlap96} and of Novikov \etal \cite{Novikov98a, Novikov98b}.
Following these authors we consider in the next section the field
dependence of the drift mobility in the model of correlated disorder
(CDM). In the rest of this section we concentrate on the temperature
dependence of the drift mobility at low electric fields in the
Gaussian disorder model (GDM).

\begin{figure}
\centerline{\epsfig{file=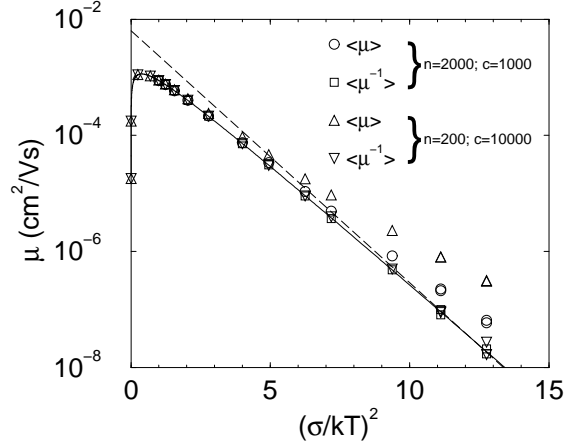,height=6.0truecm}}
\medskip
\caption{
\small
Temperature dependence of the low-field mobility for $\sigma$ =
50~meV.  Solid line represents the solution for the infinite chain
given in Eq.\ (\ref{34}).  Circles and squares show the results
obtained by averaging of mobilities and averaging of inverse
mobilities calculated by Eq.\ (\ref{32}), respectively, with averaging
over $c=1000$ chains of $N=2000$ sites. Upward and downward triangles
are the corresponding values for $c=10000$ and $N=200$. The dashed
line indicates for illustration the temperature dependence with the
slope equal to $-1$.
}
\label{fig5_hop}
\end{figure}

The low field mobility for a finite chain is given by Eqs.\ (\ref{18})
and (\ref{12}) in the limit of $E\rightarrow 0$:

\begin{equation}
\mu=\frac{eb^2}{kT}\left[
\frac{1}{N}
\sum\limits_{k=0}^{N-1}\exp\left(-\frac{\phi_k}{kT}\right)
\ \times\ \frac{1}{N}\sum\limits_{k=0}^{N-1}\frac{1}
{\exp\left(-\frac{\phi_{k}}{kT}\right)\Gamma_{k,k+1}}
\right]^{-1}\, .
\label{30}
\end{equation}
For the infinite chain this expression is equal to

\begin{equation}
\mu=\frac{eb^2}{kT}\Bigg/\left\langle\frac{\left\langle\exp\left(-\frac{\phi_k}{kT}\right)\right\rangle}
{\exp\left(-\frac{\phi_{k}}{kT}\right)\Gamma_{k,k+1}}\right\rangle.
\label{31}
\end{equation}
With hopping rates described by Eq.\ (\ref{3}) low field mobility in
the GDM is exactly given by

\begin{equation}
\mu=\frac{\nu_0eb^2}{kT}\exp\left[-\left(\frac{\sigma}{kT}\right)^2\right]
\left[1+\mbox{erf}\left(\frac{\sigma}{2kT}\right)
\right]^{-1}.
\label{32}
\end{equation}
This equation is valid for the infinite chain. The corresponding
temperature dependence of the low--field mobility is shown by the
solid line in Fig.\ \ref{fig5_hop}.  The slope of $\ln\mu$ vs.
$(\sigma/kT)^2$ in this curve differs slightly from $-1$, due to the
temperature dependence of the preexponential factor in Eq.\
(\ref{32}). This result is in good agreement with the recent computer
simulations of Bleyl \etal \cite{Bleyl99} and with the analytic
calculations of Dunlap \etal \cite{Dunlap96}, although the latter
analytic calculations have been carried out for correlated systems
with symmetric transition rates described by Eq.\ (\ref{4}). This
shows that in 1D systems Eq.\ (\ref{5}) with $C\simeq1$ is correct
and even stable against the choice of the form of transition rates.

The temperature dependences of the low--field drift mobility in finite
systems calculated according to Eq.\ (\ref{30}) are also shown in
Fig.\ \ref{fig5_hop}.  These results are really striking. They clearly
show that the temperature dependence of the mobility is related to the
size of the system. This is a manifestation of the mesoscopic
character of the hopping transport problem which is mostly pronounced
in 1D systems. As has been already discussed with respect to the RBM,
the averaging of transit times or inverse mobilities over various
finite chains corresponds to the calculation of the mobility in a long
system consisting of all these chains connected sequentially one after
another. On the contrary, the averaging of the mobilities over various
finite chains corresponds to the calculation of the carrier mobility
in the system in which these chains are arranged in parallel to each
other. In the latter case, the "fast" chains with untypically short
transit times strongly contribute to the averaged value of the carrier
mobility and they dominate the transport of carriers. This is the very
essence of the mesoscopic effects \cite{Raikh91}. Having in mind an
application of our results to discotic organic disordered systems
where many quasi--1D current channels are connected in parallel being
perpendicular to the sample surface, one should conclude that the
temperature dependence of the electrical current in such systems does
change with the thickness of the samples. For example, a comparison
between the data obtained by the averaging of mobility values for
chains with $N=2000$ sites and chains with $N=200$ sites suggests that
for shorter chains and hence for the thinner samples the temperature
dependence of the low--field drift mobility should be weaker than that
for thicker samples.  Of course, it would be desirable to verify
our prediction experimentally.

\begin{figure}
\centerline{\epsfig{file=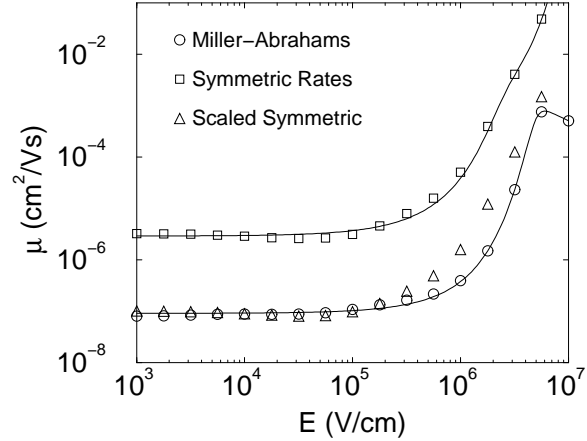,height=6.0truecm}}
\medskip
\caption{
\small
Influence of hopping rates on the field dependence of the carrier
mobility.  Circles show the results calculated by Eq.\ (\ref{18}) with
Miller--Abrahams rates defined in Eq.\ (\ref{3}). Squares display the
mobilities calculated with symmetric hopping rates defined in Eq.\
(\ref{4}). Above data were obtained for 1000 chains each containing
500 sites. Solid lines show the corresponding analytic solutions for
infinite chains. For a better comparison, triangles are plotted
indicating the mobilities with symmetric hopping rates scaled by
factor $\frac{1}{2}\exp[{\sigma^2}/{(2kT)^2}]$.
}
\label{fig6_hop}
\end{figure}

We now turn to the question how decisive for our results is the choice
of the Miller-Abrahams transition probabilities described by Eq.\
(\ref{3}) and used for the above calculations.  For comparison, one
can insert into described mathematical apparatus symmetrical
transition probabilities described by Eq.\ (\ref{4}). One
straightforwardly comes via Eqs.\ (\ref{19}) and (\ref{12}) to the
result      

\begin{equation}
\mu=\frac{\nu_0b}{E}\times\frac{e^{\frac{ebE}{2kT}}
e^{-\frac{\sigma^2}{4(kT)^2}}}{
1+e^{\frac{\sigma^2}{2(kT)^2}}\left[e^{\frac{ebE}{kT}}-1\right]^{-1}}
\label{33}
\end{equation}
In Fig.\ \ref{fig6_hop} we compare the field dependence of the
mobility given by Eq.\ (\ref{33}) with that for Miller-Abrahams rates.
After an appropriate scaling of the results for symmetric rates the
mobilities in both cases (symmetric and Miller-Abrahams hopping
probabilities) are the same at low electric fields. This confirms the
conclusion of Dunlap \etal \cite{Dunlap96} that the form of transition
probabilities is not essential for the field dependence of the drift
mobility at low fields, provided the detailed balance is fulfilled.
Dunlap \etal \cite{Dunlap96} obtained this result for correlated
systems (CDM), while we confirm their conclusion for the uncorrelated
Gaussian model (GDM). In the non-linear regime at higher fields
($E\sim10^6$ V/cm) deviations between the results for symmetric and
for Miller-Abrahams hopping probabilities are seen in Fig.\
\ref{fig6_hop}.  

\subsection{Drift mobility in the random-energy model with correlated disorder}
\markboth{1D HOPPING IN DISORDERED ORGANIC SOLIDS}{DRIFT MOBILITY IN CDM}

So far we have considered the hopping transport of charge carriers
along regular chains of sites with Gaussian distribution of site
energies and without correlations between the spatial positions of
sites and their energies. Such a model has recently been suggested by
Bleyl \etal \cite{Bleyl99} for transport of charge carriers in
columnar discotic liquid--crystalline glasses. Bleyl \etal \ treated
this model by computer simulations and obtained results similar to our
exact analytic results from the previous section. We then analyzed
whether or not the assumption on the uncorrelated distribution of site
energies is essential for these results. Such a question was raised by
Gartstein and Conwell \cite{Gartstein95}, by Dunlap \etal
\cite{Dunlap96} and by Novikov \etal \cite{Novikov98a, Novikov98b}.
Using a model similar to that of Gartstein and Conwell, we present
below, exact analytic solutions for the drift mobility of charge
carriers in systems with correlated disorder (CDM).

\begin{figure}
\centerline{\epsfig{file=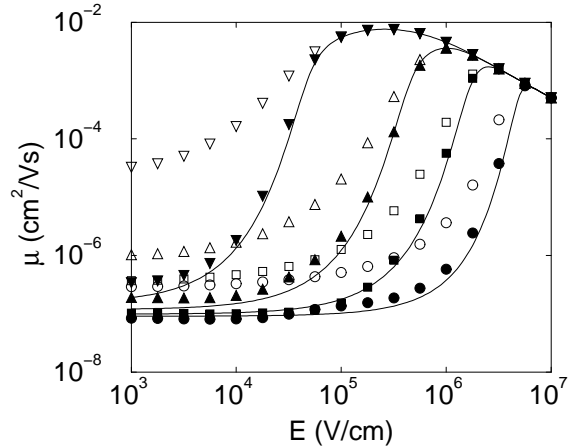,height=6.0truecm}}
\medskip
\caption{
\small
Influence of the space-energy correlations on the field dependence of
the carrier mobility at $kT$ = 15~meV and $\sigma$ = 50~meV.  The
different correlation lengths are $\lambda=1$~(circles),
$\lambda=3$~(squares), $\lambda=11$~(upward triangles) and
$\lambda=101$~(downward triangles). Open symbols show the averaged
mobilities calculated by Eq.\ (\ref{18}). Solid symbols show the
corresponding results obtained by averaging of the inverse mobilities
for ($N=1000, c=1000$).  Solid lines show the solutions for the
infinite chain given by Eq.\ (\ref{43}).  
}
\label{fig7_hop}
\end{figure}

In order to construct a model with correlated disorder (CDM) we first
generate provisional site energies $\psi_{k}$.  The zero-field energy
$\phi_{k}$ of a charge carrier on site $k$ is then determined by the
arithmetic average of provisional energies $\psi$ of neighboring
sites:  

\begin{eqnarray}
\phi_k=\lambda^{-1}\sum_{j=-m}^{m}\psi_{k+j}
\label{34}
\end{eqnarray}
Here $\lambda=2m+1$ is the correlation length in units of the
intersite separation $b$.  To ensure that the resulting site energies
$\phi_k$ have a Gaussian distribution with variance $\sigma$ we
distribute energies $\psi$ according to

\begin{equation}
g(\psi)=\frac{1}{\sqrt{2\pi\lambda\sigma^2}}\exp\left(-\frac{\psi^2}{2\lambda\sigma^2}\right).
\label{35}
\end{equation}
Due to the correlations, two site energies at zero field $\phi_k$ and
$\phi_{k+i}$ are not independent as long as $i\le\lambda$. This
prevents the possibility to carry out the averaging over disorder in
the style as it was done above on the way from Eq.\ (\ref{18}) to Eq.\
(\ref{19}). Instead we have to split the sum in the second term of the
denominator in Eq.\ (\ref{18}) into two terms:

\begin{equation}
v=\frac{bN\left[1-\exp\left(-\frac{NebE}{kT}\right)\right]}{
\sum\limits_{k=0}^{N-1}\frac{1}{\Gamma_{k,k+1}}+
\sum\limits_{k=0}^{N-1}e^{-\frac{\phi_k}{kT}}
\sum\limits_{i=\lambda}^{N-1}\frac{e^{-\frac{iebE}{kT}}}
{e^{-\frac{\phi_{k+i}}{kT}}\Gamma_{k+i,k+i+1}}+
\sum\limits_{k=0}^{N-1}e^{-\frac{\phi_k}{kT}}
\sum\limits_{i=1}^{\lambda-1}\frac{e^{-\frac{iebE}{kT}}}
{e^{-\frac{\phi_{k+i}}{kT}}\Gamma_{k+i,k+i+1}}}.
\label{36}
\end{equation}
In the second term of the denominator the site energies $\phi_k$ are
not correlated with site energies $\phi_{k+i}$ or with transition
rates $\Gamma_{k+i,k+i+1}$. Hence here the averaging of
$\exp(-\phi_k/kT)$ and $[\exp(-\phi_{k+i}/kT)\Gamma_{k+i,k+i+1}]^{-1}$
over disorder can be carried out separately.  Inserting Eq.\
(\ref{34}) into the last term of the denominator one obtains

\begin{eqnarray}
\sum\limits_{k=0}^{N-1}\sum\limits_{i=1}^{\lambda-1}\frac{e^{-\frac{iebE}{kT}}}{\Gamma_{k+i,k+i+1}}
e^{\frac{\phi_{k+i}-\phi_k}{kT}}=
\sum\limits_{k=0}^{N-1}\sum\limits_{i=1}^{\lambda-1}\frac{e^{-\frac{iebE}{kT}}}{\Gamma_{k+i,k+i+1}}
\prod_{j=1}^i e^{\frac{\psi_{k+m+j}-\psi_{k-m-1+j}}{\lambda kT}}.
\label{37}
\end{eqnarray}
With our choice of site energy correlation, Eq.\ (\ref{34}), the
energy difference between neighboring sites $k+i$ and $k+i+1$ is
$\lambda^{-1}(\psi_{k+i+1+m}-\psi_{k+i-m})$. The transition rates
$\Gamma_{k+i,k+i+1}$ are uncorrelated with $\psi_{k+m+j}$ and
$\psi_{k-m+1+j}$ for all $j$ and the averaging for the corresponding
terms can be performed independently.  Note that for long-range
correlations one has to average the entire product in Eq.\ (\ref{37}).
In the latter case Eq.\ (\ref{36}) cannot be used and one has to solve
Eq.\ (\ref{18}) instead.  This may prevent an exact solution.  For
correlations given by Eq.\ (\ref{34}) we get 

\begin{eqnarray}
\mu=\frac{b}{E}\left[
\left\langle\frac{1}{\Gamma_{k,k+1}}\right\rangle\left(1+
\sum_{i=1}^{\lambda-1}\left\langle e^{\frac{\psi_k}{\lambda kT}}\right\rangle^i
\left\langle e^{-\frac{\psi_k}{\lambda kT}}\right\rangle^i
e^{-\frac{iebE}{kT}}\right)+ \right.\nonumber\\\left.
\left\langle\frac{\left\langle\exp\left(-\frac{\phi_k}{kT}\right)\right\rangle}
{\exp\left(-\frac{\phi_{k}}{kT}\right)\Gamma_{k,k+1}}\right\rangle
\frac{\exp\left(-\frac{\lambda ebE}{kT}\right)}{1-\exp\left(-\frac{ebE}{kT}\right)}
\right]^{-1}.
\label{38}
\end{eqnarray}
Averaging over disorder with using Gaussian distribution of site
energies $g(\psi)$ given by Eq.\ (\ref{35}) leads to the following
result for the drift mobility of carriers in an infinite chain within
the CDM 


\begin{eqnarray}
\mu &=& \frac{2\nu_0b}{E}\left(
\frac{1-e^{\frac{\sigma^2}{(kT)^2}-\frac{\lambda ebE}{kT}}}
{1-e^{\frac{\sigma^2}{\lambda (kT)^2}-\frac{ebE}{kT}}}
\left\{1+\mbox{erf}\left(\frac{\sqrt{\lambda}ebE}{2\sigma}\right)
\right.\right.+\nonumber\\ &+& \left. \left.
e^{\frac{\sigma^2}{\lambda (kT)^2}-\frac{ebE}{kT}}
\left[1+\mbox{erf}\left(
\frac{\sigma}{\sqrt{\lambda}kT}-\frac{\sqrt{\lambda}ebE}{2\sigma}\right)
\right]\right\}+\right.\nonumber\\&+&\left.
\frac{e^{\frac{\sigma^2}{(kT)^2}-\frac{\lambda ebE}{kT}}}
{1-e^{\frac{ebE}{kT}}}
\left\{
1+\mbox{erf}\left(\frac{\sigma}{\sqrt{\lambda}2kT}+\frac{\sqrt{\lambda}ebE}{2\sigma}\right)
\right.\right.+\nonumber\\ &+& \left. \left.
e^{-\frac{ebE}{kT}}
\left[1+\mbox{erf}\left(\frac{\sigma}{\sqrt{\lambda}2kT}-\frac{\sqrt{\lambda}ebE}{2\sigma}\right)\right]
\right\}
\right)^{-1}.
\label{43}
\end{eqnarray}
In Fig.\ \ref{fig7_hop} the corresponding field dependences are
plotted for different correlation lengths $\lambda$ for finite chains
(symbols) as calculated via Eqs.\ (\ref{18}), (\ref{12}) and for the
infinite chain [Eq.\ (\ref{43})]. The particular shape of the field
dependence of the drift mobility in the CDM for the infinite system
has been already discussed in detail by Gartstein and Conwell
\cite{Gartstein95}, by Dunlap \etal \cite{Dunlap96} and by Novikov
\etal \cite{Novikov98a, Novikov98b}. We would like to focus our
attention here on the other aspect of the phenomenon, namely, on the
increasing difference between the results obtained by averaging of
mobilities and those obtained by averaging of inverse mobilities in
finite systems with increasing correlation length. This trend is
clearly related to the smaller number of high barriers for charge
carriers in systems with longer correlation lengths. The mesoscopic
effects become more pronounced in systems with longer correlation
length.  The increase of the correlation length in the CDM is
analogous to the decrease of the total number of sites in the chain in
the GDM.

\subsection{Concluding remarks}
\markboth{1D HOPPING IN DISORDERED ORGANIC SOLIDS}{1D HOPPING IN DISORDERED ORGANIC SOLIDS}

Analytic calculations were presented for the hopping drift mobility of
charge carriers in a one--dimensional regular chain of localization
sites with energetic disorder. Exact results were obtained for
temperature and field dependences of the mobility in systems with
uncorrelated disorder (GDM) and those with a correlated disorder (CDM)
for both asymmetrical and symmetrical transition probabilities. These
dependences are shown to be determined by the length of the system.
Therefore even the exact solutions usually obtained for infinite
systems might be of low value for thin samples which are used in real
devices. Focusing on the description of the transport properties of the
discotic liquid--crystalline glasses, one should take into account
that the real device samples usually contain only 50 to 100 sites
\cite{Christ97b}. For such systems one should use the theoretical
results obtained in this report for finite systems and not those
derived for infinite chains. 

In this section we considered only transitions of charge carriers to
the nearest sites on a conducting chain. In the next section (section\
\ref{mc_simulations}) it is shown by a Monte Carlo computer simulation
that in a correlated system (CDM) transitions to more distant
neighbors do not play an essential role while for uncorrelated systems
such distant transitions can be important. Hence in the CDM our
analytical results can be definitely considered as exact. In the case of
uncorrelated disorder (GDM), longer electron hops can modify the
results for the field and the temperature dependences of the carrier drift
mobility as shown in section \ref{mc_simulations}.

\section{Monte Carlo simulations}
\label{mc_simulations}

In section \ref{analytic_work} analytic calculations were carried
out for the steady-state drift mobility of charge carriers in
one-dimensional systems with taking into account hopping transitions
only between the nearest localized states. Special attention was
paid to the temperature and field dependences of the mobility for
various hopping models. We check here the corresponding results by a
Monte Carlo computer simulation and show that analytic theory works
perfectly.  However, this theory is unable to take into account
hopping events to localized states situated further in space than the
nearest-neighboring sites.  We will show below that electron
transitions to more distant sites decisively influence the most
fundamental transport properties, such as, for example, the
temperature dependence of the drift mobility.  This dependence at low
electric fields is described by Eq.\ (\ref{5}).  Such a dependence is
observed experimentally and it has also been obtained in computer
simulations and analytic calculations.  In the three-dimensional~(3D)
case, coefficient $C$ in Eq.\ (\ref{5})  is believed to have the
magnitude \cite{Baessler93} $C \approx 2/3$, while in the
one--dimensional~(1D) case the value $C \approx 1$ has been obtained
for symmetric transition rates determined by Eq.\ (\ref{4})
\cite{Dunlap96}.  Equation\ (\ref{5}) is widely used to determine
parameter $\sigma$ of the density of states (DOS) in Eq.\ (\ref{1})
for various materials from experimental measurements of the $\ln(\mu)$
vs. $(1/T)^2$ dependences, in particular for the 1D hopping transport
\cite{Ochse99a, Bleyl99}. Therefore, the magnitude of the coefficient
$C$ is really important. This coefficient $C$ in Eq.\ (\ref{5}) cannot
be universal. Equation\ (\ref{2}) shows that both the energy
difference between localized states involved in the carrier jump and
the distance between these states determine the hopping probability.
In such processes, characteristic temperature $T_0$ should depend on
the concentration of localized states $N$ and on the decay length of
the carrier wave function in the localized states $\alpha$ in the
combination $N\alpha$ in the 1D case and $N\alpha^3$ in the 3D case
\cite{shklovskii_springer84}.  In the computer simulations
\cite{Schoenherr81} this parameter was taken as $N\alpha^3 = 0.001$
and it is not surprising that some particular value for coefficient
$C$ was obtained.  A very close value $C\approx 2/3$ was also obtained
by the effective-medium theory in the 3D case \cite{Gruenewald84,
Movaghar86}. It is not correct to use this value for 1D systems.
Concentrating on the 1D transport we emphasize that the value of the
coefficient $C = 1$ obtained in the analytic calculation for
Miller-Abrahams transition rates and also known from the literature
for symmetric rates \cite{Dunlap96} can also be wrong. This value was
obtained only for the nearest-neighbor hopping.  Below we show that in
systems with uncorrelated energetic and spatial distributions of
localized states, hopping transitions to more distant sites than the
nearest neighbors influence the value of the coefficient $C$, which is
decisive for determination of the disorder parameter $\sigma$ of the
DOS for various materials from experimental measurements of the
$\ln(\mu)$ vs.  $(1/T)^2$.

\subsection{Numerical algorithm and simulation details}
\label{sim_det}

In our simulations a Gaussian density of states (DOS) was assumed
following the data from the literature \cite{Baessler93, Dunlap96,
Ochse99a, Bleyl99}.  In the so-called Gaussian disorder model~(GDM),
the energy distribution was described by Eq.\ (\ref{1}) and no
correlations between energy and space positions of localized states
were allowed.  In the so-called correlated disorder model~(CDM)
\cite{Dunlap96, Gartstein95, Novikov98a}, the site energies
were correlated with energies of neighboring sites. We used
correlations described by Eqs.\ (\ref{34}, \ref{35}) with m equal to 1.
In the simulation procedure, $c$ independent regular chains of
localization sites, each chain containing $N$ sites were studied
simultaneously.  Only one particle was considered on each chain.  The
simulation algorithm was the following.  A charge carrier starts its
motion on the beginning site of a chain (site number 1).  The
simulations were stopped after 90\% of carriers reached the final
sites of the chains (sites number $N$). The rate $\Gamma_{ij}$ of a
hopping transition from an occupied site $i$ to an empty site $j$ were
determined by Miller-Abrahams equation (Eq.\ \ref{2}). The lifetime of
a carrier at a site $i$ was calculated as 

\begin{equation} 
\label{eq_lifetime} 
t_i = \frac{\ln(x)}{\sum\limits_j{\Gamma_{ij}}}
\end{equation} 
where $x$ is a random number from the uniform random-number
distribution between 0 and 1.  The sum in Eq.\ (\ref{eq_lifetime}) was
taken over sites with numbers $j = i \pm 1$ for nearest-neighbor
hopping and $j = i \pm 1, i \pm 2, i \pm 3, i \pm 4, i \pm 5, i \pm 6$
for simulations with tunneling to further neighbors.

For the fastest 10\% of particles two different methods to calculate
the mobility were used. In the first method, transition times for
different chains were averaged and the obtained value $\langle t
\rangle_{10\%}$ was used to calculate the carrier mobility via
relation $\mu_{10\%}^{tav} = \frac{d}{E\langle t\rangle_{10\%}}$,
where $d$ is the length of the chains and $E$ is the strength of the
applied electric field.  The other method was based on the calculation
of carrier mobilities for the fastest 10\% of carriers and on
averaging the mobility values.  The latter method in our simulations
is equivalent to the averaging of the inverse transit times $\langle
1/t\rangle_{10\%}$.  The averaged mobility in this method was
calculated as $\mu_{10\%}^{mobav} =
\frac{d\langle1/t\rangle_{10\%}}{E}$.  The same averaging procedures
were performed for the 90\% of charge carriers with finding the
corresponding quantities $\mu_{90\%}^{tav} = \frac{d}{E\langle t
\rangle_{90\%}}$, and $\mu_{90\%}^{mobav} =
\frac{d\langle1/t\rangle_{90\%}}{E}$.

\begin{figure}
\centerline{\epsfig{file=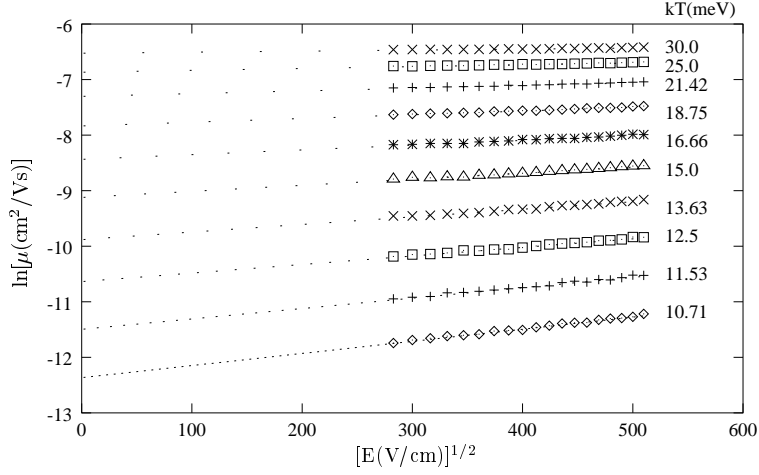,height=7.0truecm}}
\medskip
\caption{
\small
Field dependences of the carrier mobility at various temperatures
in the Poole-Frenkel representation for the GDM with $\sigma$ =
30~meV.
}
\label{fig8_hop}
\end{figure}

Simulations were carried out for various magnitudes of the electric
field $E$ and temperature $T$.  In order to obtain the value of the
zero-field mobility $\mu_{zf}$, the procedure suggested by Bleyl \etal
\cite{Bleyl99} has been used.  The data were plotted in the
Poole-Frenkel (PF) representation 

\begin{equation}
\label{eq_zf_fielddep} 
\ln(\mu) = \ln(\mu_{zf}) + \gamma \sqrt{E}
\end{equation} 
with the aim of finding $\mu_{zf}$ via extrapolation to zero field.  To
illustrate this extrapolation procedure, we show some of our results
in Fig.\ \ref{fig8_hop}.  These particular results were obtained for
the GDM with the energy scale of the DOS $\sigma$ = 30~meV and the
temperature range corresponds to $kT$ between 10~meV and 30~meV. The
mobility values were obtained by averaging of the inverse transit
times for 90\% of charge carriers.  After having found $\mu_{zf}$ for
various temperatures, the $T$ dependence of this quantity was plotted
as 

\begin{equation} 
\label{eq_zf_tempdep}
\mu_{zf}\left(\frac{\sigma}{kT}\right)= 
\mu_0 \exp\left[-\left(C\frac{\sigma}{kT}\right)^2\right] 
\end{equation}
where  $\mu_0$ is a temperature-independent prefactor and $C$ is a
constant to be determined from such plots. 

\begin{figure}
\centerline{\epsfig{file=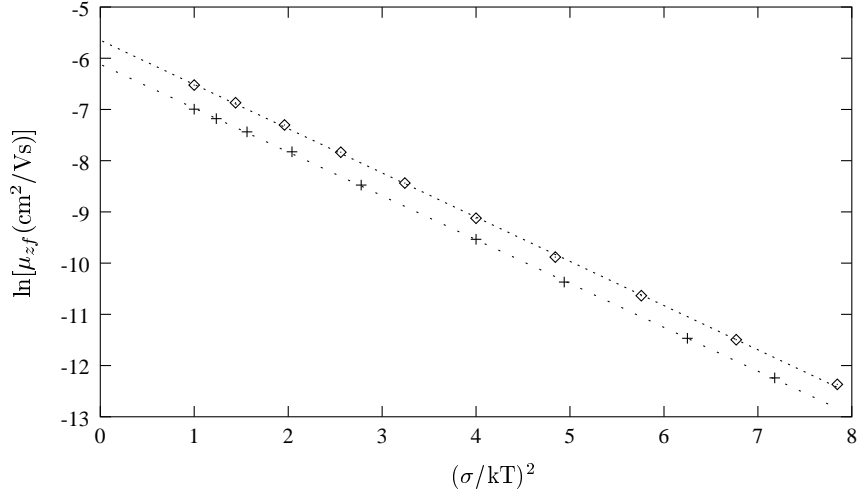,height=7.0truecm}}
\medskip
\caption{
\small
Temperature dependence of the zero-field drift mobility according
to Eq.\ (\ref{eq_zf_tempdep}) for the GDM with $\sigma$ = 30~meV
(diamonds) and $\sigma$ = 50~meV (crosses).  Dashed and dotted
lines represent Eq.\ (\ref{eq_zf_tempdep}) with $C$ = 0.928 and
$C$ = 0.924 respectively.  
}
\label{fig9_hop}
\end{figure}

Obtained temperature dependence described by Eq.\
(\ref{eq_zf_tempdep}) is plotted in Fig.\ \ref{fig9_hop} illustrating
some of our data obtained for the nearest-neighbor hopping in the GDM.
A similar Monte Carlo simulation was carried out recently by Bleyl
\etal \cite{Bleyl99}, who obtained very similar results.  Our data
confirm the coefficient $C \approx 0.9$ in Eq.\ (\ref{eq_zf_tempdep})
obtained earlier by Bleyl \etal \ However, computer simulations of the
nearest-neighbor hopping are not of considerable interest, as exact
results for such processes can be obtained analytically for finite and
infinite systems as previously shown in the analytical work~(section
\ref{analytic_work}).  Hopping processes to further neighbors than the
nearest ones can hardly be treated exactly by analytic theories.  In
this section we use computer simulations to study whether such distant
hopping transitions are important for transport coefficients, like the
drift mobility, or not.

In order to answer this question, we performed computer simulations of
hopping processes to distant neighbors for both GDM and CDM and
compared the results with those obtained only for hopping to the
nearest neighbors.  The results for the latter are given in 
section \ref{nn_hop}, while the results for distant hops are presented in section
\ref{further_hop}. Also the question will be addressed on the significance of the
space-energy correlations for transport processes as given by the
comparison of the results for the GDM and CDM. The latter question has
recently been studied with respect to the dependence of the carrier
drift mobility on the electric field \cite{Dunlap96, Gartstein95,
Novikov98a, Novikov98b} and it was shown that the space-energy
correlations play an essential role for the field dependence of the
mobility.  Below we mainly consider the temperature dependence of the
carrier mobility.  The quantity of our main interest will be the
coefficient $C$ in Eq.\ (\ref{eq_zf_tempdep}).  Before presenting the
results, we would like to note that the extrapolation procedure according
to Eq.\ (\ref{eq_zf_fielddep}) cannot be exact. We have found that the
results for $\mu_{zf}$ can slightly depend (by less than 10\%) on the
chosen range of field strengths used for extrapolation.  We always
used the interval $9\times10^4 \le E \le 2.5\times10^5$ V/cm where the
PF dependence seems to be valid as shown in Fig.\ \ref{fig8_hop}.
Moreover, the values of the determined coefficient $C$ (by
less than 10\%) depend slightly on the way how parameter $\sigma$/$kT$ was
changed -- by changing $T$ with fixed $\sigma$, or by changing $\sigma$
at fixed $T$.  This is caused by the extrapolation procedure based on
Eq.\ (\ref{eq_zf_fielddep}), because $T$ and $\sigma$ determine the
field dependence of the carrier drift mobility not in the combination
$\sigma$/$kT$ which is essential for the temperature dependence of the
zero-field mobility as has been clearly shown by Dunlap \etal
\cite{Dunlap96}. Therefore, not the absolute values of the coefficient
$C$, but its trends with changing the simulation parameters will be of
most interest with respect to our simulation results.

\subsection{Nearest neighbor hopping}
\label{nn_hop}

\begin{figure}
\centerline{\epsfig{file=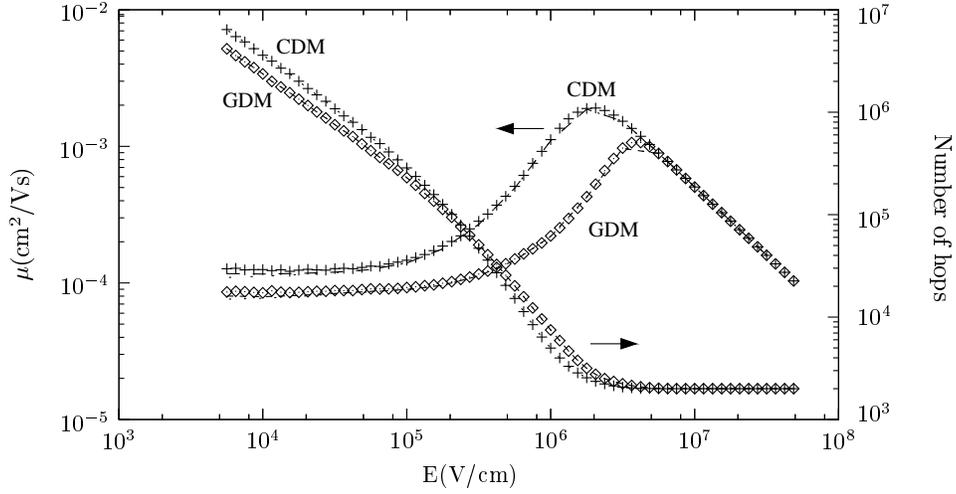,height=7.0truecm}}
\medskip
\caption{
\small
Left axis: dependences of the carrier mobility  $\mu_{90\%}^{tav}$
on the electric field for the GDM and CDM. The lines show the
results of the exact analytic solutions for system
parameters $\sigma$ = 50~meV and $kT$ = 25~meV; right axis:
average numbers of forward steps as functions of electric field.
The chain length was equal to 2000 sites in all simulation runs.
}
\label{fig10_hop}
\end{figure}

All simulation results given below were averaged over $c$ = 1000
chains, each containing $N$ = 2000 sites.  Following Bleyl \etal
\cite{Bleyl99} only Miller--Abrahams transition rates determined by
Eq.\ (\ref{3}) were used.  Parameters $\Gamma_1 = 1.875\cdot10^{14}
s^{-1}$, $b$ = 3.6~\AA, $\alpha$ = 1~\AA \ were chosen, where $b$ is
the distance between the neighboring sites. As long as only the
nearest-neighbor hopping is considered, the choice of parameters $b$
and $\alpha$ cannot influence the temperature dependence of the
mobility which is of prime interest in our study. In order to
illustrate the simulation results, we plot in Fig.\ \ref{fig10_hop}
the field dependence of the drift mobility at $kT$ = 25~meV and
$\sigma$ = 50~meV for both GDM and CDM.  Results of our computer
simulation are shown by crosses for the CDM and by diamonds for the
GDM. Dashed lines represent the results of the exact analytic
calculations for the particular chains used in the simulation
procedure based on the methods described in section
\ref{analytic_work}.  Excellent agreement with the exact analytic
results confirms the validity of our simulation procedure.  In
agreement with previous calculations \cite{Dunlap96, Gartstein95,
Novikov98a}, one can see in Fig.\ \ref{fig10_hop} that the mobility
values are higher and the field dependence starts at lower fields for
the CDM than for the GDM.  We also present in Fig.\ \ref{fig10_hop}
the statistics for the average number of hops necessary for a carrier
to penetrate through a chain containing 2000 sites. At low fields, the
motion is diffusive and the number of hops is enormously larger than
the number of sites indicating that charge carriers often move back
and forth.  This supports the conclusion of analytical
calculations~(section \ref{analytic_work}) that at low fields one
should use the diffusion equation [Eq.\ (\ref{11})] rather than the
drift approach [Eq.\ (\ref{9})]. Erroneous application of Eq.\ (\ref{9})
at low fields leads to the artificial increase of mobility values with
decreasing field strength. Unfortunately, this artifact is often
treated as a true physical effect.  At very high fields, the carriers
mostly hop in the direction prescribed by the field as indicated by
the equality between the number of hops necessary for charge carrier
to penetrate through the chain and the number of chain sites ($N$ =
2000). In such extreme conditions, the drift velocity does not depend
on the field strength and Eq.\ (\ref{12}) leads to the relation $\mu
\propto 1/E$ for high fields. Performing the extrapolation and fitting
procedure described in section \ref{sim_det}, we obtained the values
for the coefficient $C$ in various conditions.  These values were
obtained by the four different averaging methods described in section
\ref{sim_det}. In the following, $C_{10\%}^{tav}$ for example, will
denote the coefficient $C$ corresponding to fastest 10\% of charge
carriers (i.e., "fastest" 10\% chains) and "$tav$" means that the
values were obtained by averaging of arrival times. Values for the
coefficients $\mu^{mobav}$ were correspondingly obtained via the
averaging of mobilities.  Obtained results are gathered in Table\
\ref{table1_hop}, where different horizontal lines correspond to
simulation results at different values of the disorder parameter
$\sigma$ in Eq.\ (\ref{1}).  Temperature values were changed so that
parameter $\sigma$/$kT$ had magnitudes between 1 and 3.  In all cases,
the values of the coefficient $C$ are between 0.9 and 1, in good
agreement with the results of previous computer simulations
\cite{Bleyl99} and analytical calculations \cite{Dunlap96} (see also
the analytical solution in section \ref{analytic_work}).  No striking
difference concerning the temperature dependence of the carrier drift
mobility has been found between the GDM and CDM as can clearly be seen from
Table\ \ref{table1_hop}.

\begin{table}
\caption{
Coefficients $C$ in Eq.\ (\ref{eq_zf_tempdep}) for the
nearest-neighbor hopping.
}
\vspace*{1.0truecm}
\label{table1_hop}
\centerline{
\begin{tabular}{|c|c|c|c|c|c|}
\hline
model&parameter&$C_{10\%}^{tav}$&$C_{10\%}^{mobav}$&$C_{90\%}^{tav}$&$C_{90\%}^{mobav}$\\
\hline
&$\sigma = 30 meV$&0.89&0.89&0.95&0.93\\
\cline{2-6}
GDM&$\sigma = 40 meV$&0.90&0.89&0.95&0.93\\
\cline{2-6}
&$\sigma = 50 meV$&0.90&0.90&0.94&0.92\\
\hline
\hline
&$\sigma = 28.867 meV$&0.92&0.92&0.99&0.96\\
\cline{2-6}
CDM&$\sigma = 40 meV$&0.92&0.92&0.98&0.96\\
\cline{2-6}
&$\sigma = 50 meV$&0.91&0.91&0.96&0.94\\
\hline
\end{tabular}
}
\end{table}

\subsection{Hopping with tunneling to further neighbors}
\label{further_hop}

In the previous section as well as in the analytical
calculations~(section \ref{analytic_work}), only transitions of charge
carriers to the nearest-neighboring sites were considered.  In such
conditions, transition rates are determined by Eq.\ (\ref{3}). Now we
would like to check whether and at what conditions, if at all, the
tunneling hops to further sites than the nearest neighbors influence
the transport coefficients. In order to answer this question we
repeated all simulations described in section \ref{nn_hop}, but this
time allowing the tunneling of charge carriers to further sites. For
such distant hops explicit $R-$dependent transition probabilities
described by Eq.\ (\ref{2}) should be taken into account instead of
the simplified probabilities described by Eq.\ (\ref{3}). Moreover, we
carried out all simulations with changing the values of parameter
$\alpha$ which determines the decay length of the carrier wave
function in the localized states. The intersite distance was kept
constant $b$ = 3.6~\AA, the values of $\alpha$ were changed between
1~\AA \ and 3~\AA.  Transitions to six neighbors in each direction
were allowed in the simulation procedure.  We have checked that
tunneling to further states than these 12 neighbors does not play any
role at all sets of parameters used in the simulation runs.  Results
obtained for coefficient $C$ in Eq.\ (\ref{eq_zf_tempdep}) at various
values of $\alpha$ and $\sigma$ and for various averaging procedures
are gathered in Table\ \ref{table2_hop}. The main conclusion one can
draw from this table is the following.  For the CDM, i.e., for a
system with correlated disorder, tunneling to further sites than the
nearest neighbors do not play any essential role, while for the GDM,
i.e., for uncorrelated systems these distant hopping transitions
slightly influence the transport coefficients.  The insignificance of
distant transitions for systems with correlated disorder is
unsurprising. In such systems, energies of neighboring sites are close
to each other due to correlation effects. Thus the variable-range
hopping is not significant for such systems, because the closest in
energy and space states are chosen by carriers in their motion and
these states are the nearest neighbors.  The quantitative confirmation
of this qualitatively clear statement is of great importance because
it implies that for systems with correlated disorder one can use
analytic theories described in section \ref{analytic_work} and one can
be sure that the restriction of the analytic theory which considers
only the nearest-neighbor hopping is not at all essential for the
obtained results.

\begin{table}
\caption{
Coefficients $C$ in Eq.\ (\ref{eq_zf_tempdep}) when taking into 
account further hopping transitions than those to the 
nearest-neighboring sites.  
\label{table2_hop}
}
\vspace*{1.0truecm}
\centerline{
\begin{tabular}{|c|c|c|c|c|c|c|}
\hline
model &parameter &$\alpha$(\AA) &
$C_{10\%}^{tav}$ & $C_{10\%}^{mobav}$ & $C_{90\%}^{tav}$ & $C_{90\%}^{mobav}$\\
\hline
&&1.0&0.86&0.86&0.92&0.89\\
\cline{3-7}
&$\sigma = 30 meV$&2.0&0.79&0.79&0.86&0.82\\
\cline{3-7}
&&3.0&0.75&0.75&0.83&0.79\\
\cline{2-7}
\cline{2-7}
&&1.0&0.87&0.87&0.92&0.90\\
\cline{3-7}
GDM&$\sigma = 40 meV$&2.0&0.79&0.79&0.84&0.82\\
\cline{3-7}
&&3.0&0.74&0.74&0.79&0.77\\
\cline{2-7}
\cline{2-7}
&&1.0&0.85&0.84&0.91&0.87\\
\cline{3-7}
&$\sigma = 50 meV$&2.0&0.77&0.77&0.82&0.80\\
\cline{3-7}
&&3.0&0.73&0.73&0.78&0.76\\
\hline
\hline
&&1.0&0.93&0.93&1.00&0.97\\
\cline{3-7}
&$\sigma = 28.867 meV$&2.0&0.91&0.91&0.98&0.96\\
\cline{3-7}
&&3.0&0.88&0.88&0.96&0.93\\
\cline{2-7}
\cline{2-7}
&&1.0&0.92&0.92&0.98&0.95\\
\cline{3-7}
CDM&$\sigma = 40 meV$&2.0&0.90&0.89&0.96&0.94\\
\cline{3-7}
&&3.0&0.87&0.87&0.93&0.91\\
\cline{2-7}
&&1.0&0.92&0.92&0.97&0.95\\
\cline{3-7}
&$\sigma = 50 meV$&2.0&0.89&0.89&0.95&0.93\\
\cline{3-7}
&&3.0&0.87&0.87&0.92&0.90\\
\hline
\end{tabular}
}
\end{table}

For systems with uncorrelated disorder the situation is also rather
optimistic with respect to the exact analytic solutions~(see section
\ref{analytic_work}).  From Table\ \ref{table2_hop} it can be seen that
even for rather high value of the decay constant $\alpha$ = 2~\AA \
which makes the distant hops very favorable, the value of coefficient
$C \approx 0.8$ is not that dissimilar from its value $C \approx
0.9$ for the nearest-neighbor hopping. Thus even for systems with
uncorrelated disorder, exact analytic solutions~(see section
\ref{analytic_work}) taking into account only the
nearest-neighbor hopping give reasonable values for transport
coefficients.

\subsection{Concluding remarks}

Computer simulations of charge carrier transport in disordered 1D
systems show that for systems with correlated space-energy disorder
the variable-range hopping does not play an essential role, implying
that exact analytic solutions~(section \ref{analytic_work})
based on the nearest-neighbor hopping are valid for such systems
without any restrictions. For systems with uncorrelated disorder the
analytic results for transport coefficients from section \ref{analytic_work}
can be considered as an approximation. For example, taking into
account the more distant hops than hops to the nearest neighbors one
obtains the value of the coefficient $C$ in Eq.\
(\ref{eq_zf_tempdep}) by approximately 10\% different to that for only
the nearest-neighbor hopping.  With such an accuracy one can even use
analytic results from section \ref{analytic_work} for systems with
uncorrelated disorder.


\addcontentsline{toc}{chapter}{~~~~Summary}
\chapter*{Summary}
\markboth{SUMMARY}{SUMMARY}


The study of different physical properties of disordered systems by
means of computer simulations was the aim of the present thesis.
The~incredibly huge and fast growing advances in information
technology in the last decades, has allowed physicists to use computer
resources more effectively. The~extensive use of computers has enabled
the development of a new and extensive part of physics, lying
somewhere between experimental and theoretical physics.

In order to describe the physical properties of amorphous materials it
is very important that one has knowledge of the atomic arrangement.
The structure of disordered solids usually depends on their
preparation method. It is therefore important that the preparation
method is modeled first, following this, the received structures need
to be studied. In chapter\ \ref{aC}, low energy molecular dynamics
simulations of atomic beam growth of amorphous carbon on a diamond
[111] surface were carried out.  Our aim was to construct large scale
amorphous carbon models over 100 atoms by the quantum-mechanical
treatment of the interatomic potential.  The interactions were
described by a well-tested and widely applied tight-binding potential.
The structures were grown by atom-by-atom deposition onto substrates.
The growth process was simulated as in experiments without any
artificial model of energy dissipation.  We used two different average
bombarding energies E$_{beam}$ = 1 and 5~eV, and two different
substrate temperatures T$_{sub}$ = 100 and 300~K for 30~ps simulation
time.  The simulations were prolonged by 15~ps at T$_{sub}$ = 100~K to
obtain larger structures.  Six networks were prepared by the
deposition technique with periodic boundary conditions in two
dimensions. The most important structural parameters, like densities,
radial distribution functions, coordination numbers, bond angle
distributions, and ring statistics were then analyzed. The deposition
rate was much higher than it is in experiments. On the basis of our
time development investigation we are certain that much lower deposit
rates would not cause a significant difference in our models.  Our
lower substrate temperatures, compared to experiments, provide a
quicker energy dissipation which slightly compensates for the high
deposition rate.

In chapter\ \ref{CxHy}, we simulated the atomistic simulation of the
bombardment process during the bias enhanced nucleation~(BEN) phase of
diamond chemical vapor deposition~(CVD). Special attention was paid to
the study of different hydrocarbon~(\cxhy) projectile bombardment at
varying impact energies. The simulations showed that larger kinetic
energies produce deeper subplantations and that it is more
probable that the species will be dissociated.  For acetylene~(\ctht)
the carbon-carbon bond is almost never broken at low energies~(20,
40~eV) and it is broken in 50 to 90\% of cases with larger
energies~(80~eV). The computer experiments manifested that
approximately 60\% of the projectile hydrogen-carbon bonds are broken
during the subplantation.  Usually, one of the hydrogen-carbon bonds
is not broken for methyl~(\cht) and the other two hydrogens form H$_2$
molecules.  At the typical 200~V bias voltage, which is optimal for
diamond nuclei formation in BEN, the \ctht \ projectiles have 80~eV,
and \cht \ ion species have 40~eV kinetic energy.  There are
significant differences in the structural rearrangements in the film
caused by these two projectiles at this voltage. For the \ctht \
species, the penetration is deeper and these projectiles cause changes
in the deeper region than is the case with the \cht \ species.  The
\cht \ projectiles do not contribute to the deep variations in the
film and cause structural rearrangements only in the surface regions.
The structural changes in the film showed that higher kinetic energy
ionic species under the average deposition depth generate a thicker
$sp^3$ rich film and the surface mostly consists of atoms with $sp^2$
hybrids. The border between these two regimes move downwards with
larger deposition energies.  The results support the subplantation
model. In the future it would be a very interesting task to do
simulations over a silicon substrate and study the effect of this
crystalline layer on the formation of diamond nuclei in the amorphous
carbon film.

Chapter\ \ref{Si} deals with the simulation of amorphous silicon
structures.  The same atomic deposition program code, which had been
developed for studying the amorphous carbon structure growth and was
reported in chapter\ \ref{aC}, was applied here for silicon.  We used
a well tested tight-binding Hamiltonian to describe the interatomic
interaction between silicon atoms.  A similar tight-binding potential
was developed by the same group for carbon systems and we successfully
applied it for the description of the amorphous carbon growth~(see
chapter\ \ref{aC}). Nine different amorphous silicon structures were
obtained by our deposition computer method.  An additional tenth model
was prepared by rapid cooling in order to check the difference between
atom-by-atom deposition on substrate and melt-quenching preparation
techniques.  The most significant difference we have found between the
models prepared in different conditions, is in the ring statistics,
i.e. in the medium range order.  Triangles are also present in the
atomic arrangement, supporting three other independent pieces of
evidence into the existence of triangles in amorphous silicon
structures.  Nevertheless, the fragments like equilateral triangles
have never been considered at electronic density of states
calculations or at an unsolved problems such as the breaking of weak
bonds by prolonged illumination, which is the major mechanism for the
light-induced defect creation. The tight-binding model potential used
in our computer experiments slightly overestimates the bond distances,
by giving values which were observed previously in liquid silicon
forms. The potential gives overcoordinated atomic sites (fivefold, and
even sixfold coordinated atoms) during the amorphous silicon structure
simulations.  The sixfold coordinated atoms can be found in the liquid
phase of silicon instead of in the amorphous structure.  The
well-known Stillinger-Weber classical empirical potential for
amorphous silicon had a similar difficulty and has recently been
successfully modified.  A~possible future aim would be to improve this
tight-binding potential by a similar process.

Finally, in chapter\ \ref{hopping}, calculations for the hopping drift
mobility of charge carriers in a one--dimensional regular chain of
localization sites with energetic disorder were performed.  In the
analytic calculations, exact results were obtained for temperature and
field dependences of the mobility in systems with uncorrelated
disorder~(Gaussian Disordered Model) and those with a correlated
disorder~(Correlated Disorder Model) for both asymmetrical and
symmetrical transition probabilities. These dependences are shown to
be determined by the length of the system. Therefore even the exact
solutions usually obtained for infinite systems might be insignificant
for thin samples which are used in real devices. Focusing on the
description of the transport properties of the discotic
liquid--crystalline glasses, one should take into account that the
real device samples usually contain only 50 to 100 sites. For such
systems one should use the theoretical results obtained in section\
\ref{analytic_work} for finite systems and not those derived for
infinite chains.  In analytic calculations we considered only
transitions of charge carriers to the nearest sites on a conducting
chain.  The results of Monte Carlo computer simulation reveal that for
systems with uncorrelated space-energy disorder, the variable-range
hopping plays an important role. Longer electron hops can modify the
results for the field and temperature dependences of the carrier drift
mobility as presented in section \ref{mc_simulations}.  For such
systems the analytic results for transport coefficients can be
considered as an approximation.  Taking into account, for example, the
more distant hops than hops to the nearest neighbors one obtains the
value of the coefficient $C$, which describes the temperature
dependence of mobility value in Eq.\ (\ref{eq_zf_tempdep}), which is
approximately 10\% different to that for only the nearest-neighbor
hopping.  With such accuracy one can even use analytic results (from
section~\ref{analytic_work}) for systems with an uncorrelated
disorder.  However, in the case of correlated space-energy disorder,
this effect is much lower, implying that exact analytic solutions
based on the nearest-neighbor hopping are valid for such systems
without any restrictions.

\addcontentsline{toc}{chapter}{~~~~Zusammenfassung (in German)}
\chapter*{Zusammenfassung}
\markboth{ZUSAMMENFASSUNG}{ZUSAMMENFASSUNG}


Ziel dieser Dissertation war die Untersuchung verschiedener
physikalischer Eigenschaften von ungeordneten Systemen mit Hilfe von
Computer Simulationen. Die \"uberw\"altigend schnell gewachsenen
Entwicklungen in der Informationstechnologie in den letzen Jahrzehnten
hat es erm\"oglicht, Computer Ressourcen effektiver als je zuvor zu
nutzen. Die ausgdehnte Nutzung von Computern hat eine neues und
umfangreiches Gebiet der Physik hervorgebracht, das irgendwo zwischen
experimenteller und theoretischer Physik liegt.

Die Kenntnis der atomaren Struktur ist sehr wichtig um die
physikalischen Eigenschaften amorpher Materialien zu verstehen. Die
Strukturen ungeordneter Festk\"orper h\"angen normalerweise von ihren
Pr\"aparationsmethoden ab.  Daher is es wichtig, die Pr\"aparation
zuerst zu modellieren. Dadurch k\"onnen die resultierenden Strukturen
ermittelt werden. In Kapitel\ \ref{aC} wurden Niedrigenergie
Molecular-Dynamic Simulationen von Atomstrahlwachstum von amorphem
Kohlenstoff auf einer Diamant [111] Oberfl\"ache durchgef\"uhrt.
Unser Ziel war es, Modelle von amorphem Kohlenstoff mit \"uber 100
Atomen mit einer quantenmechanischen Behandlung des interatomaren
Potentials zu erhalten.  Als Wechselwirkung wurde ein wohlgetestetes
und weitbenutztes tight-binding Potential verwendet. Die Strukturen
wurden durch atomweise Anlagerung auf dem Substrat gewachsen.  Der
Wachstumsprozess wurde dem Experiment entsprechend ohne die Annahme
einer k\"unstlichen Energiedissipation simuliert. Wir haben zwei
unterschiedliche mittlere Atomstrahlenergien E$_{beam}$ = 1 und 5~eV
benutzt, und auch zwei veschiedene Substrattemperaturen T$_{sub}$ =
100 und 300~K f\"ur eine Simulationszeit von 30~ps.  Die Simulationen
wurden um  15~ps bei T$_{sub}$ = 100~K verl\"angert um gr\"o\ss ere
Strukturen herzustellen.  Es wurden sechs Netzwerke in zwei
Dimensionen mit periodischen Randbedingungen durch diese
Depositionstechnik erzeugt. Anschlie\ss end wurden die wichtigsten
strukturellen Parameter, wie die Dichten, radialen
Verteilungsfunktionen, Koordinationszahlen, die Verteilungsfunktionen
der Bindungswinkel und Ringstatistiken  analysiert. Die
Depositionsgeschwindigkeit war gr\"o\ss er als in den Experimenten.
Auf Grund unserer Untersuchung zur zeitlichen Entwicklung sind wir
sicher, da\ss \ eine niedrigere Geschwindigkeit der Anlagerung keine
wesentlichen Unterschiede in unseren Modellen verursachen w\"urde.
Unsere niedrigeren Substrattemperaturen verglichen mit den
Experimenten stellen ein schnellere Energiedissipation sicher, die die
hohe Depositionsgeschwindingkeit einigerma\ss en kompensiert.

In Kapitel\ \ref{CxHy}, haben wir die atomistische Simulation des
Chemical-Vapor-Deposi\-tionsprozesses w\"ahrend der Phase der
bias-enhanced nucleation~(BEN) von Diamant  durchgef\"uhrt.
Besonderes Augenmerk wurde auf die Untersuchung des
Projektilbeschusses mit verschiedenen Hydrokarbonaten~(\cxhy) bei
unterschiedlichen Energien gerichtet. Die Simulationen zeigten, da\ss
\ gro\ss ere kinetische Energien tiefere Subplantationen zur Folge
haben und da\ss \ es wahrscheinlicher wird, da\ss \ diese Spezies
dissoziert werden.  F\"ur Azetylen~(\ctht) wird die
Kohlenstoff-Kohlenstoff Bindung fast nie bei niedrigen Energien~(20,
40~eV) gebrochen. Sie wird andererseits in 50 bis 90\% der F\"alle
f\"ur gr\"o\ss ere Energien~(80~eV) gebrochen.  Die Computer
Experimente offenbarten, da\ss \ ungef\"ahr 60\% der
Wasserstoff-Kohlenstoff Bindungen der Projektile w\"ahrend der
Subplantation gebrochen werden. Normaleweise wird bei Methyl~(\cht)
eine der Wasserstoff-Kohlenstoff Bindungen nicht gebrochen und die
anderen beiden formen H$_2$ Molek\"ule. Bei einer typischen
bias-Spannung von 200~V, die optimal f\"ur die Formation von
Diamantkeimen in BEN ist, haben die \ctht \ Projektile 80~eV und die
\cht \ Ionspezies 40~eV kinetische Energie. Es gibt bedeutensvolle
Unterschiede in den strukturellen Umordungen der Filme, die von den
beiden Projektilen bei dieser Spannung verursacht wurden. F\"ur die
\ctht \ Spezies ist die Durchdringung tiefer und diese Projektile sind
verantwortlich f\"ur die \"Anderungen in tiefern Regionen als es der
Fall mit den \cht \ Spezies ist.  Die \cht \ Projektile tragen nicht zu
\"Anderungen in tiefen Lagen der Filme bei und verursachen
strukturelle \"Anderungen nur an der Oberfl\"ache. Die strukturellen
\"Anderungen in den Filmen deuten darauf hin, da\ss \ Spezies mit
hoher kinetischer Energie einen dickeren $sp^3$-reichen Film unterhalb
der mittleren Depositionstiefe erzeugen und da\ss \ die Oberfl\"ache
aus Atomen mit $sp^2$ Hybriden besteht.  Die Grenze zwischen diesen
beiden Regionen bewegt sich mit gr\"o\ss erer Depositionsenergie nach
unten. Diese Ergebnisse unterst\"utzen das Modell der Subplantation
und deuten darauf hin, da\ss \ $sp^3$ Hybride unter der
Durchdringungschwelle geordnet werden k\"onnen. Zuk\"unftig w\"are es
eine sehr interessante Aufgabe, Simulationen zur Siliziumstruktur zu
unternehmen sowie die Auswirkungen dieser kristallinen Schicht auf die
Formation von Diamantkeimen in den amorphen Kohlenstoffilmen zu
untersuchen.

Kapitel\ \ref{Si}  handelt von der Simulation von amorphen
Siliziumstrukturen. Genau das gleiche Programm f\"ur die Deposition,
wie es in Kapitel\ \ref{aC} f\"ur Kohlenstoff entwickelt wurde, fand
hierbei Anwendung.  Wir haben einen gut getesteten tight-binding
Hamiltonian benutzt um die interatomare Wechselwirkung zwischen
Siliziumatomen zu beschreiben.  Neun verschiedene amorphe
Siliziumstrukturen wurden durch unsere numerische Depositionstechnik
generiert.  Zus\"atzlich wurde ein zehntes Modell durch schnelle
Abk\"uhlung pr\"apariert, um den Unterschied zwischen atomarer
Deposition und Abk\"uhlen aus der Schmelze zu untersuchen.  Der wohl
bedeutungsvollste Unterschied zwischen diesen Modellen lag in den
resultierenden Ringstatistiken, beziehungsweise in der Medium-Range
Order. Selbst Dreiecke sind in der atomaren Struktur sichtbar. Dies
unterst\"utzt drei andere unabh\"angige Hinweise auf die Existenz von
Dreiecken in amorphen Siliziumstrukturen.  Dennoch wurden in der
Literatur strukturelle Einheiten wie gleichseitige Dreiecke weder bei
Rechnungen zur elektronischen Zustandsdichte noch bei der Theorie des
Aufbrechens schwacher Bindungen ber\"ucksichtigt. Letzterer ist der
dominante Mechanismus bei der lichtinduzierten Defekterzeugung.  Das
tight-binding Modellpotential in unseren Computer Experimenten
\"ubersch\"atzt etwas die Bindungsl\"angen, indem es Werte f\"ur
Silizium in der fl\"ussigen Phase liefert.  Es erzeugt
\"uberkoordinierte Atome (f\"unffach und sogar sechsfach koordinierte
Atome) w\"ahrend der Simulation der amorphen Struktur. Die sechsfach
koordinierten Atome finden sich in der fl\"ussigen Phase von Silizium,
nicht jedoch in der amorphen Phase.  Das wohlbekannte klassische,
empirische Stillinger-Weber Potential f\"ur amorphes Silizium hatte
das gleiche Problem und wurde k\"urzlich modifiziert. Es w\"are
interessant, unser tight-binding Potential auf \"ahnliche Weise zu
verbessern.

In Kapitel\ \ref{hopping} schlie\ss lich wurden Rechnungen f\"ur die
Hopping-Driftbeweglichkeit von Ladungstr\"agern in einer
ein-dimensionalen Kette mit lokalisierten elektronischen Zust\"anden
mit energetischer Unordnung ausgef\"uhrt. In den analytischen
Rechnungen wurden exakte Ergebnisse wurden f\"ur Abh\"angigkeit der
Beweglichkeit von der Temparatur und dem Feld f\"ur  Systeme mit
unkorrelierter Unordnung (Gaussches Unordnungsmodell) erzielt und
soche f\"ur korrelierte  Unordnug (korreliertes Unordnungsmodell)
f\"ur asymmetrische und symmetrische \"Ubergangswahrscheinlichkeiten.
Es zeigte sich, da\ss \ diese Abh\"angigkeiten von der L\"ange des
Systems abh\"angen.  Daher k\"onnten exakte L\"osungen, wie sie
\"ublicherweise f\"ur unendliche Systeme erzielt werden, f\"ur
realistische d\"unne Systeme, wie sie in  Anwendungen vorkommen, nicht
relevant sein.  In Bezug auf die Beschreibung der
Transporteigenschaften von diskotischen fl\"ussigkristallinen
Gl\"asern sollte man daran erinnern, da\ss \ die echten Device-Proben
normalerweise nur 50 bis 100 molekulare Sites in der Stapelrichtung
aufweisen.  Hierf\"ur sind die in Abschnitt\ \ref{analytic_work} f\"ur
endliche Systeme erzielten Ergebnisse relevant.  In den analytischen
Rechnungen haben wir nur \"Uberg\"ange der Ladungstr\"ager zu den
Nachbarsites auf der Kette betrachtet. Die Resultate der Monte-Carlo
Computersimulationen zeigen, da\ss \ dagegen f\"ur Systeme mit
unkorrelierter r\"aumlich-energetischer Unordnung
Variable-Range-Hopping eine wichtige Rolle spielt. L\"angere
Elektronenspr\"unge k\"onnen die Resultate f\"ur die Feld- und
Temperaturabh\"angigkeiten der Beweglichkeit der Ladungstr\"ager
\"andern. F\"ur solche Systeme k\"onnen die analytischen Ergebnisse
f\"ur die Transportkoeffizienten (aus Teil\ \ref{analytic_work}) als
eine Approximiation betrachtet werden.  Nimmt man zum Beispiel
Spr\"unge \"uber mehrere Nachbarn an, so bekommt man  Werte f\"ur die
Koeffizienten $C$, die die Temparaturabh\"angigkeit der Mobilit\"at
beschreiben (\ref{eq_zf_tempdep}), die ungef\"ahr 10\% von denjenigen
abweichen, die man f\"ur ausschlie\ss liche Spr\"unge zu den n\"achsten
Nachbarn erh\"alt.  Mit einer solchen Genauigkeit kann man sogar
analytische Ergebnisse (aus Teil~\ref{analytic_work}) f\"ur Systeme
mit unkorrelierter Unordnung benutzen. Jedoch ist im Fall korrelierter
r\"aumlich-energe\-tischer Unordnung dieser Effekt viel weniger
ausgepr\"agt. Dies impliziert, da\ss \ die exakten analytischen
Ergebnisse f\"ur H\"upfen zum n\"achsten Nachbarn f\"ur solche Systeme
ohne Einschr\"ankung g\"ultig sind.
\addcontentsline{toc}{chapter}{~~~~\"Ossze\-fog\-la\-l\'as (in Hun\-ga\-ri\-an)}
\chapter*{\"Osszefoglal\'as}
\markboth{\"OSSZEFOGLAL\'AS}{\"OSSZEFOGLAL\'AS}


Ez~a dol\-go\-zat ren\-de\-zet\-len rend\-sze\-rek k\"u\-l\"on\-b\"o\-z\H o fi\-zi\-kai pa\-ra\-m\'e\-te\-re\-i\-nek
sz\'a\-m\'{\i}\-t\'o\-g\'e\-pes m\'od\-szer\-rek\-kel va\-l\'o vizs\-g\'a\-la\-t\'a\-val fog\-lal\-ko\-zik.
Az~ut\'ob\-bi \'ev\-ti\-ze\-dek\-ben az in\-for\-m\'a\-ci\-\'os tech\-no\-l\'o\-gi\-\'a\-ban a hi\-het\-le\-n\"ul
in\-te\-z\'{\i}v \'es gyors fej\-l\H o\-d\'es le\-he\-t\H o\-v\'e tet\-te a fi\-zi\-ku\-sok sz\'a\-m\'a\-ra a
sz\'a\-m\'{\i}\-t\'o\-g\'e\-pek na\-gyon ha\-t\'e\-kony hasz\-n\'a\-la\-t\'at. A~sz\'eles\-k\"o\-r\H u al\-kal\-ma\-z\'a\-sok
ki\-fej\-lesz\-tet\-tek egy \'uj \'es ma m\'ar sz\'e\-les k\"or\H{u} r\'e\-sz\'et a
fi\-zi\-k\'a\-nak, amely va\-la\-hol az el\-m\'e\-le\-ti \'es k\'{\i}\-s\'er\-le\-ti fi\-zi\-ka k\"o\-z\"ot\-ti
ha\-t\'ar\-mezs\-gy\'en ta\-l\'al\-ha\-t\'o.

Az~amorf anya\-gok fi\-zi\-kai tu\-laj\-don\-s\'a\-ga\-i\-nak me\-g\'er\-t\'e\-s\'e\-hez na\-gyon
fon\-tos az ato\-mi el\-ren\-de\-z\'es is\-me\-re\-te.  A~rendezetlen rend\-sze\-rek
szer\-ke\-ze\-te \'al\-ta\-l\'a\-ban f\"ugg az el\-k\'e\-sz\'{\i}\-t\'e\-s\"uk m\'od\-sze\-r\'e\-t\H ol. Ez\'ert fon\-tos,
hogy ezt a fi\-zi\-kai fo\-lya\-ma\-tot mo\-del\-le\-z\"uk el\H o\-sz\"or, majd ut\'a\-na
vizs\-g\'al\-juk a k\'esz szer\-ke\-ze\-te\-ket. A~\ref{aC}. fe\-je\-zet\-ben, amorf sz\'en
[111] szubszt\-r\'a\-tum fe\-lett va\-l\'o mo\-le\-ku\-l\'a\-ris di\-na\-mi\-kai n\"o\-vesz\-t\'e\-s\'e\-nek
ki\-se\-ner\-gi\-\'as, ato\-mi bom\-b\'a\-z\'a\-sos m\'od\-szer se\-g\'{\i}t\-s\'e\-g\'e\-vel vizs\-g\'al\-tuk. Az~volt
a c\'e\-lunk, hogy t\"obb mint 100 r\'e\-szecs\-k\'et tar\-tal\-ma\-z\'o, ato\-mi sk\'a\-l\'an
nagy\-m\'e\-re\-t\H u, amorf sz\'en mo\-del\-le\-ket \'al\-l\'{\i}t\-sunk el\H o a po\-ten\-ci\-\'al
kvan\-tum\-me\-cha\-ni\-kai le\-\'{\i}\-r\'a\-s\'a\-val.  A~k\"ol\-cs\"on\-ha\-t\'ast egy ala\-po\-san tesz\-telt
\'es sz\'e\-les k\"or\-ben al\-kal\-ma\-zott szo\-ros k\"ot\'es\H o~(tight-bin\-ding)
po\-ten\-ci\-\'al\-lal mo\-del\-lez\-t\"uk. A~szerkezetek az ato\-mok a szubszt\-r\'a\-tum\-ra
va\-l\'o egy\-m\'as ut\'a\-ni p\'a\-ro\-log\-ta\-t\'a\-sos n\"o\-vesz\-t\'e\-s\'e\-vel k\'e\-sz\"ul\-tek. A~n\"ove\-ke\-d\'es
a k\'{\i}\-s\'er\-le\-tek\-nek meg\-fe\-le\-l\H o\-en volt szi\-mu\-l\'al\-va.  K\'et k\"u\-l\"on\-b\"o\-z\H o \'at\-la\-gos
bom\-b\'a\-z\'a\-si ener\-gi\-\'at, E$_{beam}$ = 1 \'es 5~eV, \'es k\'et k\"u\-l\"on\-b\"o\-z\H o
szubszt\-r\'at h\H o\-m\'er\-s\'ek\-le\-tet T$_{sub}$ = 100 \'es 300~K al\-kal\-maz\-tunk a 30~ps
hossz\'u szi\-mu\-l\'a\-ci\-\'os id\H o alatt. A~szimul\'aci\-\'ok 15~ps-mal meg vol\-tak
hossza\-b\'{\i}t\-va T$_{sub}$ = 100~K ese\-t\'en, hogy na\-gyobb m\'eret\H{u}
szer\-ke\-ze\-te\-ket kap\-junk. Hat da\-rab mo\-dell k\'e\-sz\"ult el ez\-zel a
p\'a\-ro\-log\-ta\-t\'a\-sos m\'od\-szer\-rel, k\'et$-$di\-men\-zi\-\'os pe\-ri\-\'o\-di\-kus ha\-t\'ar\-fel\-t\'e\-telt
al\-kal\-maz\-va.  A~legfontosabb szer\-ke\-ze\-ti pa\-ra\-m\'e\-te\-re\-ket, mint a
s\H u\-r\H u\-s\'e\-get, ra\-di\-\'a\-lis elosz\-l\'as f\"ugg\-v\'enyt, ko\-or\-di\-n\'a\-ci\-\'os sz\'a\-mot,
sz\"o\-ge\-losz\-l\'ast, \'es gy\H u\-r\H us\-ta\-tisz\-ti\-k\'at ele\-mez\-t\"uk. A~p\'aro\-log\-ta\-t\'a\-si
r\'a\-ta sok\-kal na\-gyobb volt mint a k\'{\i}\-s\'er\-le\-tek\-ben. A~rendszer
id\H o\-fej\-l\H o\-d\'e\-s\'et k\"o\-ve\-t\H o vizs\-g\'a\-la\-tunk alap\-j\'an azon\-ban azt v\'ar\-juk, hogy
sok\-kal ki\-sebb p\'a\-ro\-log\-ta\-t\'a\-si r\'a\-ta ese\-t\'en sem kap\-n\'ank je\-len\-t\H os
v\'al\-to\-z\'a\-so\-kat a mo\-dell\-je\-ink\-ben. Az~\'al\-ta\-lunk hasz\-n\'alt, k\'{\i}\-s\'er\-le\-tek\-hez
k\'e\-pest sok\-kal ala\-cso\-nyabb szubszt\-r\'a\-tum h\H o\-m\'er\-s\'ek\-let gyor\-sabb ener\-gia
disszi\-p\'a\-ci\-\'ot ered\-m\'e\-nyez, \'es kis\-s\'e kom\-pen\-z\'al\-ja a nagy p\'a\-ro\-log\-ta\-t\'a\-si
r\'a\-t\'at.

A~\ref{CxHy}. fe\-je\-zet\-ben, gy\'e\-m\'ant, k\'e\-mi\-ai g\H oz\-f\'a\-zis le\-v\'a\-lasz\-t\'a\-s\'u
(Che\-mi\-cal Va\-por De\-po\-si\-ti\-on - CVD), elekt\-ro\-mos el\H o\-fe\-sz\'{\i}\-t\'e\-ses nuk\-le\-\'a\-ci\-\'os
f\'a\-zi\-s\'a\-nak (Bi\-as En\-han\-ced Nuc\-le\-a\-ti\-on - BEN) ato\-mi szin\-t\H u mo\-del\-le\-z\'e\-se
van \"ossze\-fog\-lal\-va. K\"u\-l\"o\-n\"os fi\-gyel\-met for\-d\'{\i}\-tot\-tunk k\"u\-l\"on\-b\"o\-z\H o
sz\'enhidr\'at~(\cxhy) gy\"o\-k\"ok, m\'as-m\'as ki\-ne\-ti\-kus ener\-gi\-\'a\-val va\-l\'o
bom\-b\'a\-z\'a\-s\'a\-ra. Az ered\-m\'e\-nyek azt mu\-tat\-j\'ak, hogy na\-gyobb ki\-ne\-ti\-kus
ener\-gia m\'e\-lyebb szubp\-lan\-t\'a\-ci\-\'ot okoz \'es a gy\"o\-k\"ok disszo\-ci\-\'a\-ci\-\'o\-ja is
gya\-ko\-ribb. Az~acetil\'en~(\ctht) ese\-t\'en a sz\'en-sz\'en k\"o\-t\'es na\-gyon rit\-k\'an
bom\-lik fel ala\-csony energ\'as~(20, 40~eV) bom\-b\'a\-z\'as ese\-t\'en, vi\-szont az
ese\-tek 50$-$90\%-ban fel\-bom\-lik na\-gyobb ener\-gi\-\'ak eset\'en~(80~eV).
A~sz\'am\'{\i}\-t\'o\-g\'e\-pes k\'{\i}\-s\'er\-le\-tek ki\-mu\-tat\-t\'ak, hogy k\"o\-r\"ul\-be\-l\"ul az ese\-tek
60\%-ban a sz\'en-hid\-ro\-g\'en k\"o\-t\'es fel\-bom\-lik a szubp\-lan\-t\'a\-ci\'o so\-r\'an.
A~metil~(\cht) ese\-t\'en \'al\-ta\-l\'a\-ban az egyik sz\'en-hid\-ro\-g\'en k\"o\-t\'es meg\-ma\-rad,
\'es a m\'a\-sik k\'et hid\-ro\-g\'en H$_2$ mo\-le\-ku\-l\'at al\-kot.  A~tipikus 200~V-os
el\H o\-fe\-sz\'{\i}\-t\'es ese\-t\'en, ami op\-ti\-m\'a\-lis a gy\'e\-m\'ant nuk\-le\-usz k\'ep\-z\H o\-d\'e\-s\'e\-hez a
BEN so\-r\'an, a \ctht \ gy\"o\-k\"ok 80~eV-os, m\'{\i}g a \cht \ ion gy\"o\-k\"ok 40~eV-os
ki\-ne\-ti\-kus ener\-gi\-\'a\-val ren\-del\-kez\-nek.  En\-n\'el az el\H o\-fe\-sz\'{\i}\-t\'es\-n\'el, je\-len\-t\H os
k\"u\-l\"onb\-s\'e\-gek fi\-gyel\-he\-t\H o\-ek meg a k\'et gy\"ok \'al\-tal oko\-zott amorf sz\'en
v\'e\-kony\-r\'e\-teg\-be\-li szer\-ke\-ze\-ti \'at\-ren\-de\-z\H o\-d\'e\-s\'e\-ben.  A \ctht \ gy\"ok ese\-t\'en, a
be\-ha\-to\-l\'as m\'e\-lyebb \'es a gy\"ok m\'e\-lyebb tar\-to\-m\'a\-nyok\-ban okoz v\'al\-to\-z\'a\-so\-kat
mint a \cht \ gy\"ok.  A~\cht \ l\"o\-ve\-d\'ek nem j\'a\-rul hoz\-z\'a m\'ely
v\'al\-to\-z\'a\-sok\-hoz az amorf r\'e\-teg\-ben \'es csak a fe\-l\"u\-let k\"o\-ze\-l\'e\-ben okoz
struk\-t\'u\-r\'a\-lis \'ata\-la\-ku\-l\'a\-so\-kat.  A~v\'ekony\-r\'e\-teg\-ben meg\-fi\-gyelt szer\-ke\-ze\-ti
v\'al\-to\-z\'a\-sok ki\-mu\-tat\-t\'ak, hogy na\-gyobb ki\-ne\-ti\-kus ener\-gi\-\'a\-j\'u gy\"o\-k\"ok az
\'at\-la\-gos be\-ha\-to\-l\'a\-si m\'ely\-s\'eg alatt egy vas\-tag, $sp^3$ hib\-ri\-dek\-ben gaz\-dag
r\'e\-te\-get hoz\-nak l\'et\-re \'es a fe\-l\"u\-let pe\-dig $sp^2$ hib\-ri\-dek\-kel ren\-del\-ke\-z\H o
ato\-mok\-b\'ol \'all.  A~hat\'ar ek\"o\-z\"ott a k\'et r\'e\-teg k\"o\-z\"ott lej\-jebb to\-l\'o\-dik
na\-gyobb bom\-b\'a\-z\'a\-si ener\-gia ese\-t\'en. Az ered\-m\'e\-nye\-ink me\-ge\-r\H o\-s\'{\i}\-tik a
szubp\-lan\-t\'a\-ci\-\'os mo\-dell al\-kal\-ma\-z\'a\-s\'at.  Egy igen \'er\-de\-kes fe\-la\-dat
le\-het a j\"o\-v\H o\-ben me\-gis\-m\'e\-tel\-ni a szi\-mu\-l\'a\-ci\-\'os k\'{\i}\-s\'er\-le\-te\-ket egy szi\-l\'{\i}\-ci\-um
szubszt\-r\'a\-tum fe\-lett, \'es vizs\-g\'al\-ni en\-nek a kris\-t\'a\-lyos r\'e\-teg\-nek a
ha\-t\'a\-s\'at az amorf sz\'en v\'e\-kony\-r\'e\-teg\-ben a gy\'e\-m\'ant nuk\-le\-u\-szok
ki\-a\-la\-ku\-l\'a\-s\'a\-hoz vezet\H{o} fo\-lya\-ma\-t\'at.

A~\ref{Si}. fe\-je\-zet az amorf szi\-l\'{\i}\-ci\-um szer\-ke\-ze\-t\'e\-nek szi\-mu\-l\'a\-ci\-\'o\-j\'a\-val
fog\-lal\-ko\-zik. Ugya\-nazt az ato\-mi p\'a\-ro\-log\-ta\-t\'a\-si prog\-ram\-cso\-ma\-got
hasz\-n\'al\-tuk, mint ame\-lyet ki\-fej\-lesz\-tet\-t\"unk az amorf sz\'en n\"o\-vesz\-t\'e\-s\'e\-nek
vizs\-g\'a\-la\-t\'a\-ra \'es a~\ref{aC}. fe\-je\-zet\-ben mu\-tat\-tunk be. Egy sz\'e\-les k\"or\-ben
tesz\-telt, szo\-ros k\"ot\'es\H u~(tight-bin\-ding) Ha\-mil\-tont for\-ma\-liz\-must
al\-kal\-maz\-tunk a szi\-l\'{\i}\-ci\-um ato\-mok k\"o\-z\"ot\-ti k\"ol\-cs\"on\-ha\-t\'as le\-\'{\i}\-r\'a\-s\'a\-ra. Egy
ha\-son\-l\'o szo\-ros k\"o\-t\'e\-s\H o po\-ten\-ci\-\'alt fej\-lesz\-tett ki ugya\-nez a
ku\-ta\-t\'o\-cso\-port a sz\'en rend\-sze\-rek\-re \'es mi ezt al\-kal\-maz\-tuk ered\-m\'e\-nye\-sen
az amorf sz\'en v\'e\-kony\-r\'e\-te\-gek n\"o\-vesz\-t\'e\-s\'e\-nek le\'{\i}r\'as\'ara~(ld. \ref{aC}.
fe\-je\-zet).  Ki\-lenc k\"u\-l\"on\-b\"o\-z\H o amorf szi\-l\'{\i}\-ci\-um szer\-ke\-ze\-tet k\'e\-sz\'{\i}\-tet\-t\"unk
el a sz\'a\-m\'{\i}\-t\'o\-g\'e\-pes p\'a\-ro\-log\-ta\-t\'a\-si m\'od\-sze\-r\"unk\-kel. Egy to\-v\'ab\-bi ti\-ze\-dik
mo\-dell gyors\-h\H u\-t\'e\-ses tech\-ni\-k\'a\-val k\'e\-sz\"ult el, hogy \"ossze\-ha\-son\-l\'{\i}t\-suk az
ato\-mi p\'a\-ro\-log\-ta\-t\'a\-si \'es az ol\-va\-d\'ek h\H u\-t\'e\-ses m\'od\-sze\-rek k\"o\-z\"ot\-ti
k\"u\-l\"onb\-s\'e\-ge\-ket. Az~\'al\-ta\-lunk ta\-l\'alt leg\-fon\-to\-sabb k\"u\-l\"onb\-s\'eg a
gy\H u\-r\H us\-ta\-tisz\-ti\-k\'a\-ban fi\-gyel\-he\-t\H o meg, azaz a k\"o\-z\'ep\-t\'a\-v\'u rend\-ben.
H\'a\-rom\-sz\"o\-gek ugyan\-csak meg\-ta\-l\'al\-ha\-t\'o\-ak az ato\-mi el\-ren\-de\-z\'es\-ben,
me\-ge\-r\H o\-s\'{\i}t\-ve ez\-zel h\'a\-rom ko\-r\'ab\-bi f\"ug\-get\-len bi\-zo\-nyos\-s\'a\-g\'at a h\'a\-rom\-sz\"o\-gek
el\H o\-for\-du\-l\'a\-s\'a\-nak az amorf szi\-l\'{\i}\-ci\-um\-ban.  Min\-da\-mel\-lett, olyan
frag\-men\-sek, mint p\'el\-d\'a\-ul egyen\-l\H o ol\-da\-l\'u h\'a\-rom\-sz\"o\-gek so\-ha\-sem vol\-tak
fi\-gye\-lem\-be v\'e\-ve az elekt\-ron \'al\-la\-pot\-s\H u\-r\H u\-s\'eg sz\'a\-m\'{\i}\-t\'a\-s\'a\-n\'al, il\-let\-ve olyan
me\-gol\-dat\-lan prob\-l\'e\-m\'ak ese\-t\'en, mint a gyen\-ge k\"o\-t\'e\-sek fel\-bom\-l\'a\-sa hossz\'u
meg\-vi\-l\'a\-g\'{\i}\-t\'as ese\-t\'en, ami a f\H o me\-cha\-niz\-mu\-sa a f\'eny \'al\-tal in\-du\-k\'alt hi\-ba
kel\-t\'e\-s\'e\-nek.  A~szoros k\"o\-t\'e\-s\H u mo\-dell po\-ten\-ci\-\'al egy pi\-cit t\'ul\-be\-cs\"u\-li a
k\"o\-t\'es\-t\'a\-vol\-s\'a\-got a sz\'a\-m\'{\i}\-t\'o\-g\'e\-pes k\'{\i}\-s\'er\-le\-te\-ink\-ben, olyan \'er\-t\'e\-ke\-ket ad\-va,
ame\-lye\-ket ko\-r\'ab\-ban fo\-lya\-d\'ek szi\-l\'{\i}\-ci\-um\-ban fi\-gyel\-tek meg.  Az amorf
szi\-l\'{\i}\-ci\-um szer\-ke\-ze\-t\'e\-nek szi\-mu\-l\'a\-ci\-\'o\-ja so\-r\'an a po\-ten\-ci\-\'al t\'ul\-ko\-or\-di\-n\'alt
ato\-mo\-kat is ad~(\"ot\"os, s\H ot ha\-tos ko\-or\-di\-n\'alt\-s\'a\-g\'u ato\-mo\-kat is).  Ha\-tos
ko\-or\-di\-n\'alt\-s\'a\-g\'u ato\-mok, in\-k\'abb a szi\-l\'{\i}\-ci\-um fo\-lya\-d\'ek f\'a\-zi\-s\'a\-ban
ta\-l\'al\-ha\-t\'o\-ak, mint az amorf szi\-l\'{\i}\-ci\-um\-ban. Az~amorf szi\-l\'{\i}\-ci\-um ese\-t\'e\-ben a
j\'ol is\-mert Stil\-lin\-ger-We\-ber klasszi\-kus em\-pi\-ri\-kus po\-ten\-ci\-\'al is ha\-son\-l\'o
ne\-h\'e\-zs\'e\-gek\-kel k\"uz\-d\"ott, \'es mos\-ta\-n\'a\-ban ered\-m\'e\-nye\-sen lett m\'o\-do\-s\'{\i}t\-va.  Egy
j\"o\-v\H o\-be\-li c\'el le\-het a szo\-ros k\"ot\'es\H u~(tight-bin\-ding) po\-ten\-ci\-\'al ha\-son\-l\'o
m\'o\-don va\-l\'o ja\-v\'{\i}\-t\'a\-sa.

V\'e\-ge\-ze\-t\"ul, az \ref{hopping}. fe\-je\-zet\-ben, egy egy-di\-men\-zi\-\'os ren\-de\-zet\-len
lo\-ka\-li\-z\'a\-ci\-\'os ener\-gia \'al\-la\-po\-to\-kon ke\-resz\-t\"ul ha\-la\-d\'o t\"ol\-t\'es\-hor\-do\-z\'ok
moz\-g\'e\-kony\-s\'a\-g\'a\-nak sz\'a\-m\'{\i}\-t\'a\-sa ta\-l\'al\-ha\-t\'o meg.  Az~analitikus
sz\'a\-m\'{\i}\-t\'a\-sok\-ban, eg\-zakt ered\-m\'e\-nye\-ket kap\-tunk a moz\-g\'e\-kony\-s\'ag elekt\-ro\-mos
t\'e\-re\-r\H os\-s\'eg \'es h\H o\-m\'er\-s\'ek\-let f\"ug\-g\'e\-s\'e\-re mind nem\-kor\-re\-l\'alt ren\-de\-zet\-len
rend\-szer eset\'en~(Ga\-us\-si\-an di\-sor\-de\-red mo\-del - GDM) \'es mind kor\-re\-l\'a\-ci\'o
ese\-t\'en is~(cor\-re\-la\-ted di\-sor\-der mo\-del - CDM), asszi\-met\-ri\-kus \'es
szim\-met\-ri\-kus \'at\-me\-ne\-ti r\'a\-t\'ak ese\-t\'en egya\-r\'ant. Ezek az ered\-m\'e\-nyek a
rend\-szer m\'e\-re\-t\'e\-t\H ol f\"ug\-ge\-nek. Ez\'ert m\'eg az eg\-zakt me\-gol\-d\'a\-sok v\'eg\-te\-len
hossz\'u rend\-sze\-rek ese\-t\'en is \'er\-v\'e\-ny\"u\-ket veszt\-he\-tik olyan v\'e\-kony min\-t\'ak
ese\-t\'en, ame\-lye\-ket a va\-l\'o\-di al\-kal\-ma\-z\'a\-sok\-ban hasz\-n\'al\-nak. K\"o\-z\'ep\-pont\-ba
he\-lyez\-ve az \'ugy ne\-ve\-zett osz\-lo\-po\-san ko\-rong\-sze\-r\H u fo\-lya\-d\'ekk\-ris\-t\'a\-lyo\-kat,
fi\-gye\-lem\-be kell ven\-ni, hogy az al\-kal\-ma\-zott esz\-k\"o\-z\"ok \'al\-ta\-l\'a\-ban csak
50-100 "\'ep\'{\i}\-t\H o elem\-b\H ol" \'all\-nak. Ilyen rend\-sze\-rek ese\-t\'en az
\ref{analytic_work}.~alfejezetben ka\-pott el\-m\'e\-le\-ti ered\-m\'e\-nyek k\"o\-z\"ul a
v\'e\-ges m\'e\-re\-t\H u rend\-sze\-rek\-re ka\-pott ered\-m\'e\-nye\-ket kell al\-kal\-maz\-ni, a
v\'eg\-te\-len hossz\'u rend\-sze\-rek\-re le\-ve\-ze\-tett ered\-m\'e\-nyek he\-lyett.
Az~analitikus sz\'a\-m\'{\i}\-t\'a\-sok so\-r\'an a t\"ol\-t\'es\-hor\-do\-z\'ok hop\-ping\-j\'a\-n\'al csak
el\-s\H o\-szom\-sz\'ed \'at\-me\-ne\-te\-ket vet\-t\"unk fi\-gye\-lem\-be. A~Monte Car\-lo szi\-mu\-l\'a\-ci\-\'ok
ered\-m\'e\-nyei ki\-mu\-tat\-t\'ak, hogy azok\-ban a rend\-sze\-rek\-ben, ahol nincs
kor\-re\-l\'a\-ci\'o az ener\-gia \'es lo\-ka\-li\-z\'a\-ci\-\'os \'al\-la\-po\-tok t\'er\-be\-li el\-he\-lyez\-ke\-d\'e\-se
k\"o\-z\"ott, a v\'al\-to\-z\'o t\'a\-vol\-s\'a\-g\'u hop\-ping fon\-tos sze\-re\-pet j\'at\-szik \'es
ha\-t\'as\-sal van a t\"ol\-t\'es\-hor\-do\-z\'o moz\-g\'e\-kony\-s\'a\-g\'a\-nak t\'e\-re\-r\H os\-s\'eg \'es
h\H o\-m\'er\-s\'ek\-let f\"ugg\'esre~(ld.  \ref{mc_simulations}.~alfejezet). Ezek\-re a
rend\-sze\-rek\-re a transz\-port egy\"utt\-ha\-t\'ok\-ra ka\-pott ana\-li\-ti\-kus ered\-m\'e\-nyek
k\"o\-ze\-l\'{\i}\-t\'es\-k\'ent al\-kal\-maz\-ha\-t\'o\-ak.  P\'el\-d\'a\-ul, ha t\'a\-vo\-lab\-bi szom\-sz\'ed
hop\-pin\-got is fi\-gye\-lem\-be ve\-sz\"uk, ak\-kor a $C$ egy\"utt\-ha\-t\'o\-ra, ami a
moz\-g\'e\-kony\-s\'ag h\H o\-m\'er\-s\'ek\-let f\"ug\-g\'e\-s\'et \'{\i}r\-ja le az (\ref{eq_zf_tempdep})
egyen\-let\-ben, 10\%-os k\"u\-l\"onb\-s\'eg ta\-l\'al\-ha\-t\'o a csak el\-s\H o\-szom\-sz\'ed hop\-ping\-ra
le\-ve\-ze\-tett ered\-m\'e\-nyek\-hez k\'e\-pest. Ilyen hi\-ba mel\-lett, ak\'ar az
ana\-li\-ti\-kus eredm\'enyek~(ld.  \ref{analytic_work}.~alfejezet) is
m\'er\-va\-d\'o\-ak le\-het\-nek a kor\-re\-l\'a\-lat\-lan rend\-sze\-rek\-ben. Ugya\-nak\-kor, a
kor\-re\-l\'alt ren\-de\-zet\-len\-s\'eg ese\-t\'en, ez az ef\-fek\-tus sok\-kal ki\-sebb, utal\-va
ar\-ra, hogy az el\-s\H o\-szom\-sz\'ed hop\-ping\-ra ka\-pott eg\-zakt ana\-li\-ti\-kus
ered\-m\'e\-nyek k\"u\-l\"o\-n\"o\-sebb kor\-l\'a\-to\-z\'a\-sok n\'el\-k\"ul \'er\-v\'e\-nye\-sek ezek\-ben a
rend\-sze\-rek\-ben.

\addcontentsline{toc}{chapter}{~~~~Acknowledgments}
\chapter*{Acknowledgments}
\markboth{ACKNOWLEDGMENTS}{ACKNOWLEDGMENTS}

In Riezlern~(September, 1999) I was really fortunate, as after we had
dinner I could sit down and talk about my research with three very
clever people.  Basically, PD Dr. Sergei Baranovskii was explaining the
model of a physical problem to me that I was going to work on in the
following months. At that conversation Professor Peter Thomas said a
sentence, which I shall never forget. The sentence went like this {\it
"This problem is very easy to solve with computer simulations!"}.
First I thought that Professor Peter Thomas should be right.  However,
the first reaction of Dr. Sergei Baranovskii was to say, {\it "Wait,
maybe it is not that easy"}. By that time I had executed lots of
computer simulations, bad ones as well as good ones.  In the following
two years after that discussion further thousands of programs were
executed. This Ph.D. thesis tries to summarize how "{\it{easy}}" this
work was. Furthermore, according to the sentence of my Hungarian
supervisor Dr.  S\'andor Kugler {\it "The working hours of a Ph.D.
student is 168 hours weekly, plus overtime."}, I hope the thesis shows
how much work I did or I did not do.

The present Ph.D. work was done under the (patient) guidance of my two
supervisors, Dr. S\'andor Kugler and PD Dr. Sergei Baranovskii. Words
are not appropriate enough to express my gratitude to these two
people, who have done an excellent job in edifying guidance. I am
eternally grateful to Professor Peter Thomas for his valuable and
helpful discussions and for organizing my stays in Marburg.

During the three years of work I had the opportunity to be a member of
two scientific groups. The first being with Theoretical Physics
Department at the Budapest University of Technology and Economics, and
the second in the Semiconductor Group of the Institute of Physics at
the Philipps--Universit\"at Marburg.  I would like to thank all of the
members of those groups for the various help they supplied to me. Let
me name a few of them here; Dr. Istv\'an L\'aszl\'o for his valuable
help and discussions during the developments of the tight-binding
molecular dynamics program code; Dr.  Harald Cordes, with whom we
worked on the one--dimensional hopping transport problem in disordered
organic solids; and some of the Ph.D. and ex-Ph.D. students:
L\'aszl\'o Borda, Ralf Eichmann, L\'aszl\'o Kullmann, Dr. J\'anos
T\"or\"ok, and Tam\'as Unger.

I would also like to thank Dr.~P\'eter~De\'ak~(Department of Atomic
Physics, Budapest University of Technology and Economics) and
Professor Thomas Frauenheim~(Department of Theoretical Physics,
Universit\"at-GH Paderborn) for organizing my summer work in Paderborn
in 1999. In addition I would like to thank all the members of these
departments for their support and I would like to name among them Dr.
Sza\-bolcs K\'atai, who helped me to understand in more detail the
experimental results of the field I worked in and Dr. Zolt\'an Hajnal,
who helped me greatly in understanding the TBMD program code developed
by the group in Paderborn.

Finally, I would like to thank my family for their continuous support
and patience. Last, but not least, I would like to thank my girlfriend
Sophie for simply everything and for making me happy all the time.

\addcontentsline{toc}{chapter}{~~~~List of publications}
\chapter*{List of publications}
\markboth{LIST OF PUBLICATIONS}{LIST OF PUBLICATIONS}

\begin{itemize}

\item[1.)] Krisztian  Kohary, Sandor Kugler, Istvan Laszlo:
{\it "Molecular dynamics simulation of amorphous carbon
structures"},
J. Non-Cryst. Solids, {\bf 227-230} (1998) p594-596.

\item[2.)] Kriszti\'an Koh\'ary:
{\it "Molecular Dynamics Simulation of Amorphous Carbon Structures"},
(thesiswork written in Hungarian)

\item[3.)] S. Kugler, K. Kohary, I. Laszlo:
{\it "Microscopic Structure of Amorphous Carbon.
Tight-Binding Molecular Dynamics Study"},
Proceedings of N2M Workshop, Grenoble, France, 1998.
AIP Conference Proceedings 479. p64-69.

\item[4.)] S. Kugler, I. Laszlo, K. Kohary, K. Shimakawa:
{\it "Molecular dynamics simulation of amorphous carbon structures"},
Functional Materials {\bf 6}, No.3 (1999) p459.

\item[5.)] K. Kohary, S. Kugler:
{\it "Growth of amorphous semiconductors:
tight-binding molecular dynamics study"},
J. Non-Cryst. Solids, {\bf 266--269} (2000) p746-749.

\item[6.)] H. Cordes, S. Baranovskii, K. Kohary, P. Thomas, S. Yamasaki,
F. Hensel, and J$-$H. Wendorff:
{\it "One-dimensional hopping transport in disordered organic solids.
I.~Analytic calculations"},
Phys. Rev. B {\bf 63}, 094201 (2001).

\item[7.)] K. Kohary, H. Cordes, S. Baranovskii, P. Thomas, S. Yamasaki,
F. Hensel, and J$-$H. Wendorff:
{\it "One-dimensional hopping transport in disordered organic solids.
II.~Monte Carlo simulation"},
Phys. Rev. B {\bf 63}, 094202 (2001).

\item[8.)] K. Kohary, S. Kugler:
{\it "Growth of amorphous carbon.  Low energy Molecular Dynamics
simulation of atomic bombardment"},
Phys. Rev. B {\bf 63}, 193404 (2001).

\item[9.)] S.D. Baranovskii, H. Cordes, K. Kohary, and P. Thomas: 
{\it "On disorder-enhanced diffusion in condensed aromatic melts"},
Phil. Mag. B {\bf 81}, 955 (2001).

\item[10.)] K. Kohary and S. Kugler:
{\it "Time development during growth and relaxation of amorphous carbon
films preparation. Tight-binding molecular dynamics study."},
J. Non-Cryst. Solids (in print).

\item[11.)] K. Kohary, S. Kugler, Z. Hajnal, Th. K\"ohler, Th. Frauenheim, 
Sz. K\'atai, and P. De\'ak:
{\it "Atomistic simulation of the bombardment process during the BEN
phase of diamond CVD"},
(accepted for publication in Diam. Rel. Mat).

\item[12.)] K. Kohary and S. Kugler:
{\it "Growth of amorphous silicon. Low energy molecular dynamics
simulation of atomic bombardment II."},
(submitted).

\item[13.)] I.P. Zvaygin, S.D. Baranovskii, K. Kohary, H. Cordes, and P. Thomas: 
{\it "Hopping in Quasi-One-Dimensional Disordered Solids: Beyond the
Nearest-Neighbor Approximation"}
(submitted).

\end{itemize}
\addcontentsline{toc}{chapter}{~~~~Curriculum Vitae}
\chapter*{Curriculum Vitae}
\markboth{CURRICULUM VITAE}{CURRICULUM VITAE}

\begin{tabular}{ll}
1st February, 1975       & Born in Kaposv\'ar~(Hungary)\\
                         & Parents: Koh\'ary Istv\'an \\
\vspace{0.8truecm}
                         & and Koh\'aryn\'e Drajk\'o \'Agnes\\
1981-1986                & G\'ardonyi G\'eza Primary School,
                           Budapest~(Hungary)\\
1986-1988                & 29th Primary and Secondary School,
                           Kalinin~(U.S.S.R.)\\
\vspace{0.25truecm}
1988-1989                & G\'ardonyi G\'eza Primary School, 
                           Budapest~(Hungary)\\
\vspace{0.25truecm}
1989-1993                & Szent Istv\'an Secondary School, 
                           Budapest~(Hungary)\\ 
\vspace{0.25truecm}
11th June, 1993          & A-level exam\\
1993-1998                & Studies in Engineering Physics at Budapest University\\
\vspace{0.25truecm}			 
                         & of Technology and Economics~(Hungary)\\  
\vspace{0.25truecm}
16th June, 1998            & M.Sc. in Engineering Physics\\
From 1st September, 1998 & PhD student at Budapest University of\\
                         & Technology and Economics~(Hungary) and\\
			 & Philipps--Universit\"at Marburg~(Germany)\\
			 & (mutual education)\\

\end{tabular}

\end{document}